\documentclass[10pt,a4paper]{article}\usepackage[]{graphicx}\usepackage[]{color}
\makeatletter
\def\maxwidth{ %
  \ifdim\Gin@nat@width>\linewidth
    \linewidth
  \else
    \Gin@nat@width
  \fi
}
\makeatother

\definecolor{fgcolor}{rgb}{0.345, 0.345, 0.345}
\newcommand{\hlnum}[1]{\textcolor[rgb]{0.686,0.059,0.569}{#1}}%
\newcommand{\hlstr}[1]{\textcolor[rgb]{0.192,0.494,0.8}{#1}}%
\newcommand{\hlcom}[1]{\textcolor[rgb]{0.678,0.584,0.686}{\textit{#1}}}%
\newcommand{\hlopt}[1]{\textcolor[rgb]{0,0,0}{#1}}%
\newcommand{\hlstd}[1]{\textcolor[rgb]{0.345,0.345,0.345}{#1}}%
\newcommand{\hlkwb}[1]{\textcolor[rgb]{0.69,0.353,0.396}{#1}}%
\newcommand{\hlkwc}[1]{\textcolor[rgb]{0.333,0.667,0.333}{#1}}%
\newcommand{\hlkwd}[1]{\textcolor[rgb]{0.737,0.353,0.396}{\textbf{#1}}}%

\usepackage{framed}
\makeatletter
\newenvironment{kframe}{%
 \def\at@end@of@kframe{}%
 \ifinner\ifhmode%
  \def\at@end@of@kframe{\end{minipage}}%
  \begin{minipage}{\columnwidth}%
 \fi\fi%
 \def\FrameCommand##1{\hskip\@totalleftmargin \hskip-\fboxsep
 \colorbox{shadecolor}{##1}\hskip-\fboxsep
     \hskip-\linewidth \hskip-\@totalleftmargin \hskip\columnwidth}%
 \MakeFramed {\advance\hsize-\width
   \@totalleftmargin\z@ \linewidth\hsize
   \@setminipage}}%
 {\par\unskip\endMakeFramed%
 \at@end@of@kframe}
\makeatother

\definecolor{shadecolor}{rgb}{.97, .97, .97}
\definecolor{messagecolor}{rgb}{0, 0, 0}
\definecolor{warningcolor}{rgb}{1, 0, 1}
\definecolor{errorcolor}{rgb}{1, 0, 0}
\newenvironment{knitrout}{}{} % an empty environment to be redefined in TeX

\usepackage{alltt}
\usepackage[utf8x]{inputenc}
\usepackage{amsmath}
\usepackage{amsfonts}
\usepackage{amssymb}
\usepackage{graphicx}
\usepackage{subcaption}
\usepackage{array}
\usepackage{booktabs}

\usepackage{multirow}
\usepackage{booktabs} % for extra vertical spacing in tables

\usepackage{url}
\usepackage{fancyvrb}
\usepackage[unicode=true,pdfusetitle,
 bookmarks=true,bookmarksnumbered=true,bookmarksopen=true,bookmarksopenlevel=2,
 breaklinks=false,pdfborder={0 0 1},backref=false,colorlinks=false]
 {hyperref}
\hypersetup{
 pdfstartview={XYZ null null 1}}
\usepackage{breakurl}

\newcommand{\UBL}{\texttt{UBL}\ }

\newcommand{\pUBL}{package \texttt{UBL}\ }
\newcommand{\UBLp}{\texttt{UBL}\ package  }
\newcommand{\version}{0.0.5}

\author{Paula Branco, Rita P. Ribeiro and Luis Torgo\\FCUP - LIAAD/INESC Tec\\University of Porto\\
  \texttt{ \{paula.branco,rpribeiro,ltorgo\}@dcc.fc.up.pt}}
\title{UBL: an R Package for Utility-Based Learning}
\IfFileExists{upquote.sty}{\usepackage{upquote}}{}
\begin{document}

%$
\maketitle

\begin{abstract}
  
  This document describes the R \pUBL that allows the use of several methods for handling utility-based learning problems. Classification and regression problems that assume non-uniform costs and/or benefits pose serious challenges to predictive analytic tasks. In the context of meteorology, finance, medicine, ecology, among many other, specific domain information concerning the preference bias of the users must be taken into account to enhance the models predictive performance. To deal with this problem, a large number of techniques was proposed by the research community for both classification and regression tasks. 
  The main goal of \UBLp is to facilitate the utility-based predictive analytic task by providing a set of methods to deal with this type of problems in the R environment. It is a versatile tool that provides mechanisms to handle both regression and classification (binary and multiclass) tasks. Moreover, \UBLp allows the user to specify his domain preferences, but it also provides some automatic methods that try to infer those preference bias from the domain, considering some common known settings. 

\end{abstract}

% ====================================================================
\section{Introduction}

This document describes the methods available in \pUBL \footnote{This document was written for \UBLp version \version.} to deal with utility-based problems. \UBLp aims at providing a diverse set of methods to address predictive tasks where the user has a non-uniform preference bias across the domain. The package provides tools suitable for both classification and regression tasks. All the methods available in \UBLp were extended for being able to deal with multiclass problems and with regression problems possibly containing several relevant regions across the target variable domain.

Utility-based problems are defined in the context of predictive tasks where the user has a differentiated interest over the domain. This means that, in this type of problems, the user has non-uniform benefits for the correct predictions and/or assumes non-uniform costs for different errors. Many real world applications are utility-based learning problems because they encompass domain specific information which, if disregarded, may strongly penalize the performance of predictive models. This happens in the context of meteorology, finance, medicine, ecology, among many other, where specific domain information concerning the user preferences must be taken into account to enhance the models predictive performance.

In the utility-based learning framework we can frequently witness the conjugation of two important factors: i) an increased interest in some particular range(s)/class(es) of the target variable values and ii) a scarce representation of the examples belonging to that range(s)/class(es). This situation occurs in both classification and regression tasks and is usually known as the problem of imbalanced domains \cite{branco2015survey}.

Utility-based learning assumes non-uniform costs and/or benefits which are usually expressed through a cost/benefit matrix (in classification) or a cost/benefit surface (in regression). However, frequently this information is just not available, or is hard/expensive to obtain because it often requires the intervention of a domain expert. This means that for many domains, there is only an informal knowledge regarding which are the most costly mistakes and which are the most important classes/ranges of the target variable. In fact, considering the particular problem of imbalanced classes it is frequent to observe the assumption that ``the minority class is the most important one". This is an important information regarding the preferences of the user. However, it is stated in a very informal way, an no cost/benefit matrix is available in this situation. The approaches proposed in \pUBL are able to deal with these situations because they allow the use of both user specified preferences and automatic methods.

% The methods implemented in \UBL are pre-processing approaches which aim at altering the original data set to match the user preferences. This means that the methods for dealing with imbalanced domains are only relevant when the user preferences are focused on the least represented examples. In fact, if the most frequent cases are the most important, then the learning algorithms will naturally tend to focus on these examples, and therefore there is no need to change the original distribution. On the other hand, when the most relevant examples are scarcely represented any learning algorithm used will focus on the normal cases and will fail the most important predictions on the rare examples.

Several types of approaches exist for handling utility-based learning problems. These approaches were categorized into: pre-processing, change the learning algorithms, post-processing or hybrid \cite{branco2015survey}. The \textbf{pre-processing} approaches act before the learning stage by manipulating the examples distribution to match the user preferences. The methods that \textbf{change the learning algorithms} try to incorporate the user preference bias into the selected learning algorithm. There are also strategies that are applied as a \textbf{post-processing} step by changing the predictions made by a standard learner using the original data set. Finally there are \textbf{hybrid} approaches that combine some of the previous strategies.

In \pUBL we have focused on pre-processing strategies to address the problem of utility-based learning. These strategies change the original distribution of examples by removing or/and adding examples, i.e., by performing under-sampling or over-sampling. The under-sampling strategies may be random or focused. By focused under-sampling we mean that the discarded examples satisfy a given requirement, such as: are possibly noisy examples, are too distant from the decision border, are too close to the border, etc. Regarding the over-sampling methods there are two main options: over-sampling is accomplished by the introduction of replicas of examples or by the generation of new synthetic examples. For the strategies which include copies of existing examples, the cases may be selected randomly or in an informed fashion. Approaches that build synthetic cases differ among themselves in the generation process adopted. Several strategies combine under-sampling and over-sampling methods in different ranges/classes of the target variable.

This document is organized as follows. In Section \ref{sec:instal} some general installation guidelines for \UBLp are provided. Section \ref{sec:syntdata} briefly describes the two synthetic data sets provided with \UBLp. Section \ref{sec:example} presents two simple examples to show how \UBLp can be used both in classification and regression contexts and its impact on the models performance. Sections \ref{sec:methClass} and \ref{sec:methRegres} describe with detail each method currently implemented in \UBL for classification and regression tasks. Section \ref{sec:distFunc} describes the distance functions available in \pUBL which allow to asses the distance between examples in data sets containing nominal and/or numeric features. Finally, Section \ref{sec:conc} concludes this document.

% ====================================================================
\section{Package Installation Guidelines}\label{sec:instal}

The installation of any R package available on CRAN is performed as follows:
\begin{knitrout}\footnotesize
\definecolor{shadecolor}{rgb}{0.969, 0.969, 0.969}\color{fgcolor}\begin{kframe}
\begin{alltt}
\hlkwd{install.packages}\hlstd{(}\hlstr{"UBL"}\hlstd{)}
\end{alltt}
\end{kframe}
\end{knitrout}

This is mandatory, if you want to use the approaches available in \pUBL or even if you just want to try out the examples presented in the next sections. This installs the current stable version of \texttt{UBL} package which is version \version.

You may also install the development version of the package, that is available on the following GitHub Web page: \url{https://github.com/paobranco/UBL}. However, we strongly recommend the use of the CRAN stable version. If you still want to install the development version, you should do this with extreme care because this version is still being tested and therefore is more prone to bugs.
To install the development version from GitHub you should do the following in R:

\begin{knitrout}\footnotesize
\definecolor{shadecolor}{rgb}{0.969, 0.969, 0.969}\color{fgcolor}\begin{kframe}
\begin{alltt}
\hlkwd{library}\hlstd{(devtools)}
\hlkwd{install_github}\hlstd{(}\hlstr{"paobranco/UBL"}\hlstd{,}\hlkwc{ref}\hlstd{=}\hlstr{"development"}\hlstd{)}
\end{alltt}
\end{kframe}
\end{knitrout}

Further instructions may be found at the mentioned GitHub page. 
For reporting issues related with \UBLp you can use: \url{https://github.com/paobranco/UBL/issues}.

After installation using any of the above procedures, the package can be used as any other R package by doing:

\begin{knitrout}\footnotesize
\definecolor{shadecolor}{rgb}{0.969, 0.969, 0.969}\color{fgcolor}\begin{kframe}
\begin{alltt}
\hlkwd{library}\hlstd{(UBL)}
\end{alltt}
\end{kframe}
\end{knitrout}

%Further help and illustrations can be obtained through the many help pages of each function defined in the package that contain lots of illustrative examples. Again, these help pages can be accessed as any other R package, through R help system (e.g. running \texttt{help.start()} at R command line). If you just want to run some examples of a particular function you can simply use \texttt{example(<function name>)}.

\section{Synthetic Data for Classification and Regression Tasks}\label{sec:syntdata}

Package \texttt{UBL} includes two artificially generated data sets: one for classification (ImbC) and another for regression (ImbR). Both data sets were generated to depict situations of imbalanced domains. Thus, in both data sets, we assume the usual setting where the under-represented values of the target variable (either nominal or numeric) are the most important to the user. Table~\ref{tab:data} summarizes the main characteristics of these data sets.

\begin{table}[!hbt]
\centering
\resizebox{\textwidth}{!}{
\begin{tabular}{ccccccccccc}
    \toprule
    \multirow{2}{*}{Dataset}
    & \multirow{2}{*}{Task} &
    Total 
    & 
    \multicolumn{2}{c}{Relevant}
    & \multicolumn{5}{c}{Features}&\multirow{2}{*}{Target}\\
    \cline{6-10}
    &&Cases&\multicolumn{2}{c}{Cases}&name&type&min&mean&max\\
    \midrule
    \multirow{2}{*}{ImbC} & \multirow{2}{*}{Classif.} &\multirow{2}{*}{1000} &rare1&rare2&      X1&num.&-13.58 &-0.11&12.78& minority: rare1; rare2\\   
&&&10&131&X2&nom. &"cat" 300 &"fish" 300&"dog 400"& majority: normal\\
\midrule
    \multirow{3}{*}{ImbR} & \multirow{3}{*}{Regress.} &\multirow{3}{*}{1000} & \multicolumn{2}{c}{\multirow{3}{*}{50}}&X1&num.&0.37 &9.94&19.06&min: 10.00 \\
    &&&&&X2&num.&0.20 &10.08&19.47& mean: 10.98\\
    &&&&&&&&&&max: 23.17 \\
    \bottomrule     
\end{tabular}
}
\caption{Description of artificial data sets of \UBLp.}
\label{tab:data}

\end{table}

ImbC data consists of multiclass classification data set with 1000 cases and two features, $X1$ (numeric) and $X2$ (nominal). The target variable, denoted as $Class$, has two minority classes (\textit{rare1} and \textit{rare2}) and a majority class (\textit{normal}). The percentage of cases of classes \textit{rare1} and \textit{rare2} is 1\% and 13.1\%, respectively, while the \textit{normal} class has 85.9\% of the cases. This data set mimics the usual setting where the most relevant classes for the user are under-represented. This data set also simulates the existence of both class overlap and small disjuncts. Both issues are known for increasing the difficulty of dealing with imbalanced domains~\cite{lopez2013insight}. Figure~\ref{fig:OriginalC} shows this data set with some noise added to the nominal variable $X2$ to make the examples distribution more visible.

ImbC data set was generated as follows:
\begin{itemize}
  \item $X1 \sim \mathbf{N} \left(0, 4\right)$
  \item $X2$ labels "cat", "fish" and "dog" where randomly distributed with the restriction of having a   frequency of 30\%, 30\% and 40\% respectively.
  \item To obtain the target variable $Class$, we have define the following sets:
  \begin{itemize}
    \item $S_1=\{(X1, X2) : X1 > 9 \wedge (X2 \in \{"cat", "dog"\})\}$
    \item $S_2=\{(X1, X2) : X1 > 7 \wedge X2 = "fish" \}$
    \item $S_3=\{(X1, X2) :-1  <  X1 < 0.5\}$
    \item $S_4=\{(X1, X2) : X1 < -7 \wedge X2 = "fish"\}$
  \end{itemize}
  \item The following conditions define the target variable distribution of the ImbC synthetic data set:
  \begin{itemize}
    \item Assign class label "rare1" to: a random sample of 90\% of set $S_1$ and a random sample of 40\% of set $S_2$
    \item Assign class label "rare2" to: a random sample of 80\% of set $S_3$ and a random sample of 70\% of set $S_4$
    \item Assign class label "normal" to the remaing examples.
  \end{itemize}
\end{itemize}

\begin{knitrout}\footnotesize
\definecolor{shadecolor}{rgb}{0.969, 0.969, 0.969}\color{fgcolor}\begin{figure}

{\centering \includegraphics[width=0.5\textwidth]{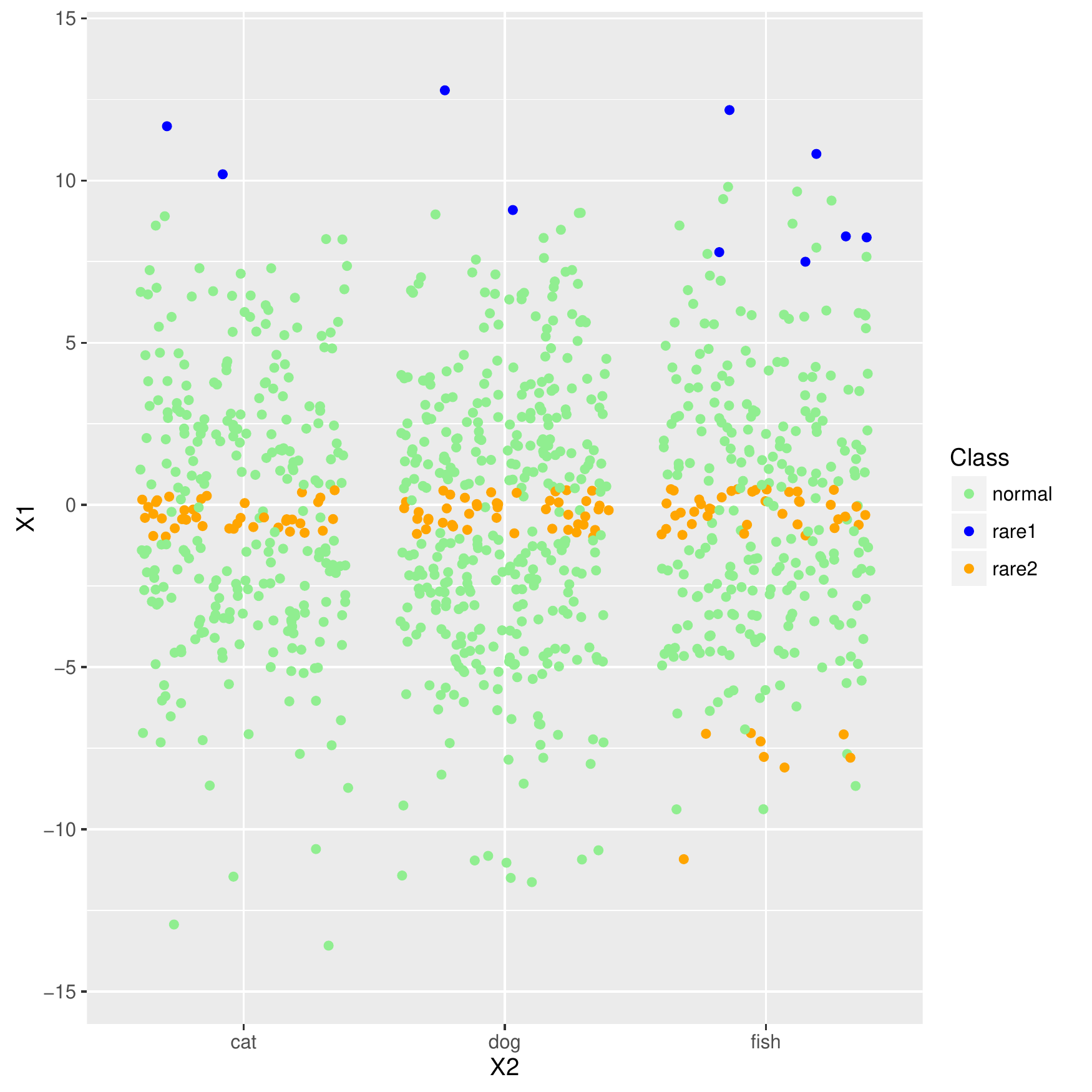} 

}

\caption[ImbC]{ImbC: artificial data set for classification.}\label{fig:OriginalC}
\end{figure}

\end{knitrout}

\begin{knitrout}\footnotesize
\definecolor{shadecolor}{rgb}{0.969, 0.969, 0.969}\color{fgcolor}\begin{figure}

{\centering \includegraphics[width=0.5\textwidth]{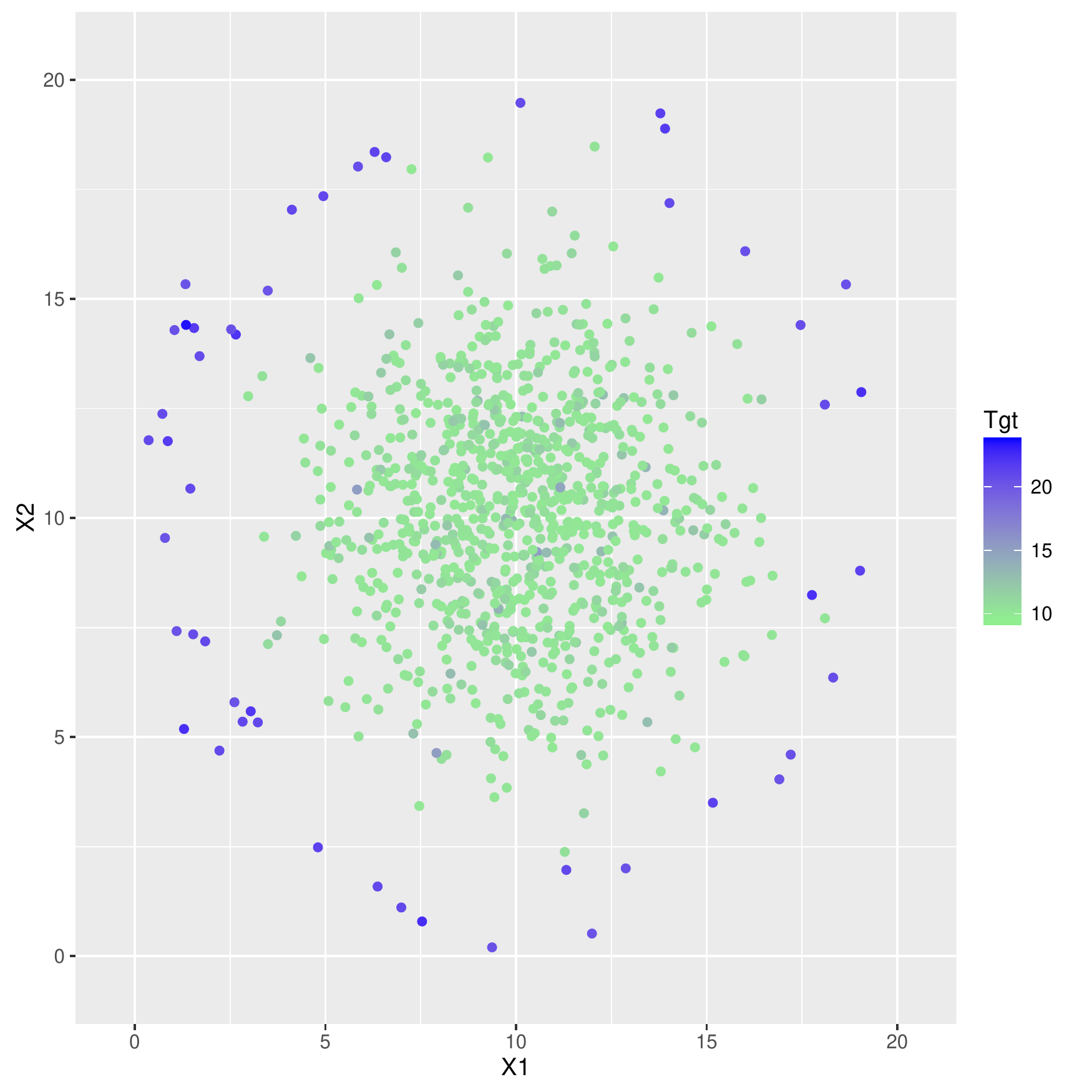} 

}

\caption[ImbR]{ImbR: artificial data set for regression.}\label{fig:OriginalR}
\end{figure}

\end{knitrout}
The regression data set ImbR has two numeric features ($X1$ and $X2$) and a continuous target variable $Tgt$. ImbR was generated in the following way: it includes 50 cases sampled from a circumference with white noise and the remaining 950 cases were sampled from a two dimensional normal distribution. Regarding the values of the continuous target variable ($Tgt$), they were obtained through a sample of two different Gamma distributions (one with higher values used for generating the target variable values of the examples in the circumference, and another with lower values used for the target variable values of the remaining examples). ImbR data simulates the usual setting in regression where the most relevant values are under-represented. In this case, we consider that the higher values of the target variable, whose predictor variables where sampled from a cirfumference, are the most important ones.

More formally, ImbR data was obtained as follows:
\begin{itemize}
  \item lower $Tgt$ values:
  \begin{itemize}
    \item $(X1, X2) \sim \mathbf{N}_{2} (\mathbf{10}_{2}, \mathbf{2.5}_{2})$
    \item $Tgt \sim \mathbf{\Gamma} \left( 0.5, 1 \right) + 10$
  \end{itemize}
  \item higher $Tgt$ values: 
  \begin{itemize} 
    \item $(X1, X2) \sim \left(\rho * cos(\theta) + 10, \rho * sin(\theta) + 10 \right)$, where $\rho \sim \mathbf{9}_{2}+\mathbf{N}_{2} \left(\mathbf{0}_{2}, \mathbf{I}_{2} \right)$ and $\theta \sim \mathbf{U}_{2} \left( \mathbf{0}_{2}, 2\pi \mathbf{I}_{2} \right)$ 
    \item $Tgt \sim \mathbf{\Gamma} \left( 1,1 \right) + 20$
    \end{itemize}
\end{itemize}

Figure~\ref{fig:OriginalR} shows the examples distribution of ImbR data set.
% ====================================================================
\section{Two Simple Illustrative Examples}\label{sec:example}

%Let us consider a classification task with 3 classes with different frequency. For illustration purposes, we will use the well-known iris data set with some examples removed to simulate a distribution with a rare class. Let us also suppose that the most important class for the user is this rare class (in our example, the virginica class). 
In this section we will show two simple examples of how to use the \UBLp. We will use the two data sets provided with the package (ImbC and ImbR) to illustrate a classification and a regression task.

Consider the ImbC synthetic data set. Let us suppose that the most important classes are the two minority classes, rare1 and rare2, and that we do not consider class normal relevant. Assuming this domain information, we begin by briefly observing the data set characteristics.

\begin{knitrout}\footnotesize
\definecolor{shadecolor}{rgb}{0.969, 0.969, 0.969}\color{fgcolor}\begin{kframe}
\begin{alltt}
\hlkwd{library}\hlstd{(UBL)}     \hlcom{# Loading our infra-structure}
\hlkwd{library}\hlstd{(e1071)}   \hlcom{# packge containing the svm we will use}
\hlkwd{data}\hlstd{(ImbC)}       \hlcom{# The synthetic data set we are going to use}
\hlkwd{summary}\hlstd{(ImbC)}    \hlcom{# Summary of the ImbC data  }
\end{alltt}
\begin{verbatim}
##        X1              X2         Class    
##  Min.   :-13.5843   cat :300   normal:859  
##  1st Qu.: -2.6930   dog :400   rare1 : 10  
##  Median : -0.1592   fish:300   rare2 :131  
##  Mean   : -0.1064                          
##  3rd Qu.:  2.4633                          
##  Max.   : 12.7836
\end{verbatim}
\begin{alltt}
\hlkwd{table}\hlstd{(ImbC}\hlopt{$}\hlstd{Class)}
\end{alltt}
\begin{verbatim}
## 
## normal  rare1  rare2 
##    859     10    131
\end{verbatim}
\end{kframe}
\end{knitrout}

Now we will obtain a random sample of 70\% of our data to train a svm.
Then, we observe the results on the remaining 30\% of data left for testing. We obtain the following:

\begin{knitrout}\footnotesize
\definecolor{shadecolor}{rgb}{0.969, 0.969, 0.969}\color{fgcolor}\begin{kframe}
\begin{alltt}
\hlkwd{set.seed}\hlstd{(}\hlnum{123}\hlstd{)}
\hlstd{samp} \hlkwb{<-} \hlkwd{sample}\hlstd{(}\hlnum{1}\hlopt{:}\hlkwd{nrow}\hlstd{(ImbC),} \hlkwd{nrow}\hlstd{(ImbC)}\hlopt{*}\hlnum{0.7}\hlstd{)}
\hlstd{train} \hlkwb{<-} \hlstd{ImbC[samp,]}
\hlstd{test} \hlkwb{<-} \hlstd{ImbC[}\hlopt{-}\hlstd{samp,]}

\hlstd{model} \hlkwb{<-} \hlkwd{svm}\hlstd{(Class}\hlopt{~}\hlstd{., train)}
\hlstd{preds} \hlkwb{<-} \hlkwd{predict}\hlstd{(model,test)}
\hlkwd{table}\hlstd{(preds, test}\hlopt{$}\hlstd{Class)} \hlcom{# confusion matrix }
\end{alltt}
\begin{verbatim}
##         
## preds    normal rare1 rare2
##   normal    256     3    41
##   rare1       0     0     0
##   rare2       0     0     0
\end{verbatim}
\end{kframe}
\end{knitrout}

Clearly, the model presents a poor performance on least represented classes. In effect, in this case, the model always predicts class normal.

Now, we can try to apply a pre-processing strategy for dealing with utility-based problems, and check again the models performance. In this case we selected the common strategy of balancing the data set classes and we used the SMOTE algorithm proposed by \cite{CBOK02}.

\begin{knitrout}\footnotesize
\definecolor{shadecolor}{rgb}{0.969, 0.969, 0.969}\color{fgcolor}\begin{kframe}
\begin{alltt}
\hlcom{# change the train data by applying the smote strategy}
\hlcom{# notice that we have to set the dist parameter to for instance "HEOM" }
\hlcom{# because the default distance (Euclidean) is not possible to use with nominal features}
\hlstd{newtrain} \hlkwb{<-} \hlkwd{SmoteClassif}\hlstd{(Class}\hlopt{~}\hlstd{., train,} \hlkwc{C.perc}\hlstd{=}\hlstr{"balance"}\hlstd{,} \hlkwc{dist}\hlstd{=}\hlstr{"HEOM"}\hlstd{)}


\hlcom{# generate a new model with the changed data}
\hlstd{newmodel} \hlkwb{<-} \hlkwd{svm}\hlstd{(Class}\hlopt{~}\hlstd{., newtrain)}
\hlstd{preds} \hlkwb{<-} \hlkwd{predict}\hlstd{(newmodel,test)}
\hlkwd{table}\hlstd{(preds, test}\hlopt{$}\hlstd{Class)}
\end{alltt}
\begin{verbatim}
##         
## preds    normal rare1 rare2
##   normal    104     0     4
##   rare1      12     3     0
##   rare2     140     0    37
\end{verbatim}
\end{kframe}
\end{knitrout}

We can observe that the least represented classes, rare1 and rare2, now present an improved result.
If the previous model was unable to correctly classify any examples of these classes, now most of those cases have a correct prediction. However, it is also important to highlight the increase in misclassification of class normal.

We can also observe the results obtained by applying a simple random over-sampling method. Again, we opted to balance the problem classes.

\begin{knitrout}\footnotesize
\definecolor{shadecolor}{rgb}{0.969, 0.969, 0.969}\color{fgcolor}\begin{kframe}
\begin{alltt}
\hlcom{# apply random over-sampling strategy}
\hlstd{newtrain2} \hlkwb{<-} \hlkwd{RandOverClassif}\hlstd{(Class}\hlopt{~}\hlstd{., train,} \hlkwc{C.perc}\hlstd{=}\hlstr{"balance"}\hlstd{)}

\hlcom{#generate a new model with the modified data set}
\hlstd{newmodel2} \hlkwb{<-} \hlkwd{svm}\hlstd{(Class}\hlopt{~}\hlstd{., newtrain2)}
\hlstd{preds} \hlkwb{<-} \hlkwd{predict}\hlstd{(newmodel2, test)}
\hlkwd{table}\hlstd{(preds, test}\hlopt{$}\hlstd{Class)}
\end{alltt}
\begin{verbatim}
##         
## preds    normal rare1 rare2
##   normal    129     0     4
##   rare1       9     3     0
##   rare2     118     0    37
\end{verbatim}
\end{kframe}
\end{knitrout}

Again, the pre-processing method applied allowed to improve the performance of the model on the least represented (and more important) class.

Let us now see how the pre-processing strategies can be applied on a regression task using the ImbR synthetic data set.

We start by loading the data and observe its main characteristics. Let us consider a random sample of 70\% of ImbR data for training a model. The remaining 30\% will be used as test set.

\begin{knitrout}\footnotesize
\definecolor{shadecolor}{rgb}{0.969, 0.969, 0.969}\color{fgcolor}\begin{kframe}
\begin{alltt}
\hlkwd{data}\hlstd{(ImbR)}
\hlkwd{summary}\hlstd{(ImbR)}
\end{alltt}
\begin{verbatim}
##        X1                X2              Tgt       
##  Min.   : 0.3654   Min.   : 0.201   Min.   :10.00  
##  1st Qu.: 8.2821   1st Qu.: 8.246   1st Qu.:10.06  
##  Median : 9.9811   Median :10.129   Median :10.22  
##  Mean   : 9.9418   Mean   :10.078   Mean   :10.98  
##  3rd Qu.:11.7202   3rd Qu.:11.903   3rd Qu.:10.72  
##  Max.   :19.0565   Max.   :19.474   Max.   :23.17
\end{verbatim}
\begin{alltt}
\hlkwd{set.seed}\hlstd{(}\hlnum{123}\hlstd{)}
\hlstd{samp} \hlkwb{<-} \hlkwd{sample}\hlstd{(}\hlnum{1}\hlopt{:}\hlkwd{nrow}\hlstd{(ImbR),} \hlkwd{as.integer}\hlstd{(}\hlnum{0.7}\hlopt{*}\hlkwd{nrow}\hlstd{(ImbR)))}
\hlstd{trainD} \hlkwb{<-} \hlstd{ImbR[samp,]}
\hlstd{testD} \hlkwb{<-} \hlstd{ImbR[}\hlopt{-}\hlstd{samp,]}
\end{alltt}
\end{kframe}
\end{knitrout}

It is required that the user states which are the cases that he considers more/less relevant. The \UBLp includes an automatic method for defining a relevance of the examples based on the data distribution. This method, proposed by \cite{ribeiro2011utility}, allows to obtain a relevance function that maps each target variabel value into a $[0,1]$ scale of relevance, where 1 represents maximum relevance and 0 represent minimum relevance. The key aspect of this automatic method is the assignment of an higher relevance to the scarcely represented cases which is the most common setting. We will use this automatic method in the ollowing example. A more detail explanation of this method is provided in Section~\ref{sec:methRegres}.

Let us train a model using the original train data. We choose a random forest available through the \texttt{randomForest} package and obtained the predictions on the test set.

\begin{knitrout}\footnotesize
\definecolor{shadecolor}{rgb}{0.969, 0.969, 0.969}\color{fgcolor}\begin{kframe}
\begin{alltt}
\hlkwd{library}\hlstd{(randomForest)}
\hlstd{model} \hlkwb{<-} \hlkwd{randomForest}\hlstd{(Tgt}\hlopt{~}\hlstd{., trainD)}
\hlstd{preds} \hlkwb{<-} \hlkwd{predict}\hlstd{(model, testD)}
\end{alltt}
\end{kframe}
\end{knitrout}

Let us now apply a pre-processing strategy in the original train data and observe the impact on the predictions.

\begin{knitrout}\footnotesize
\definecolor{shadecolor}{rgb}{0.969, 0.969, 0.969}\color{fgcolor}\begin{kframe}
\begin{alltt}
\hlcom{# use the Introduction of Gaussian Noise with the default parameters}
\hlstd{newTrain} \hlkwb{<-} \hlkwd{GaussNoiseRegress}\hlstd{(Tgt}\hlopt{~}\hlstd{., trainD)}
\hlstd{newModel} \hlkwb{<-}\hlkwd{randomForest}\hlstd{(Tgt}\hlopt{~}\hlstd{., newTrain)}
\hlstd{newPreds} \hlkwb{<-} \hlkwd{predict}\hlstd{(newModel, testD)}
\end{alltt}
\end{kframe}
\end{knitrout}

The results obtained by the two random forest models are displayed in Figure~\ref{fig:ImbR_RF1}. In this case, it is clear that the predictions obtained by the model trained with the changed data set are better in the high rare cases. For the higher range of values of $Tgt$ the model trained with the original data displays mostly under-predictions showing a focus in the normal range of values.

\begin{knitrout}\footnotesize
\definecolor{shadecolor}{rgb}{0.969, 0.969, 0.969}\color{fgcolor}\begin{figure}

{\centering \includegraphics[width=0.7\textwidth]{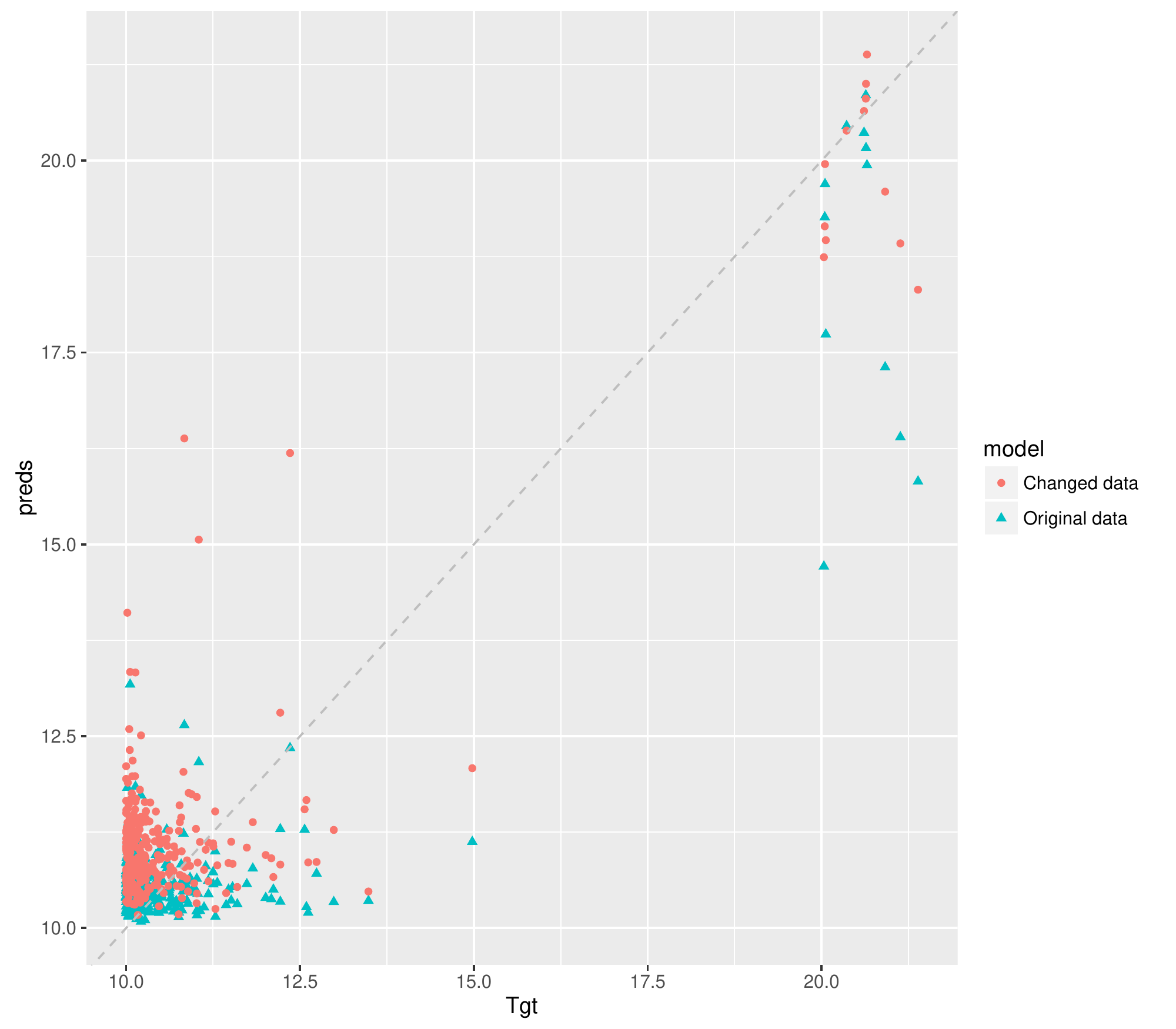} 

}

\caption[Predictions obtained with the original and the new data modified through the Gaussian Noise strategy]{Predictions obtained with the original and the new data modified through the Gaussian Noise strategy.}\label{fig:ImbR_RF1}
\end{figure}

\end{knitrout}

We can also observe the use impact of using the simple random under-sampling pre-processing strategy.

\begin{knitrout}\footnotesize
\definecolor{shadecolor}{rgb}{0.969, 0.969, 0.969}\color{fgcolor}\begin{kframe}
\begin{alltt}
\hlcom{# random under-sampling strategy setting the under-sampling percentage to 0.3}
\hlstd{trainRU} \hlkwb{<-}\hlkwd{RandUnderRegress}\hlstd{(Tgt}\hlopt{~}\hlstd{., trainD,} \hlkwc{C.perc}\hlstd{=}\hlkwd{list}\hlstd{(}\hlnum{0.3}\hlstd{))}
\hlstd{ModelRU} \hlkwb{<-}\hlkwd{randomForest}\hlstd{(Tgt}\hlopt{~}\hlstd{., trainRU)}
\hlstd{PredsRU} \hlkwb{<-} \hlkwd{predict}\hlstd{(ModelRU, testD)}
\end{alltt}
\end{kframe}
\end{knitrout}

\begin{knitrout}\footnotesize
\definecolor{shadecolor}{rgb}{0.969, 0.969, 0.969}\color{fgcolor}\begin{figure}

{\centering \includegraphics[width=0.7\textwidth]{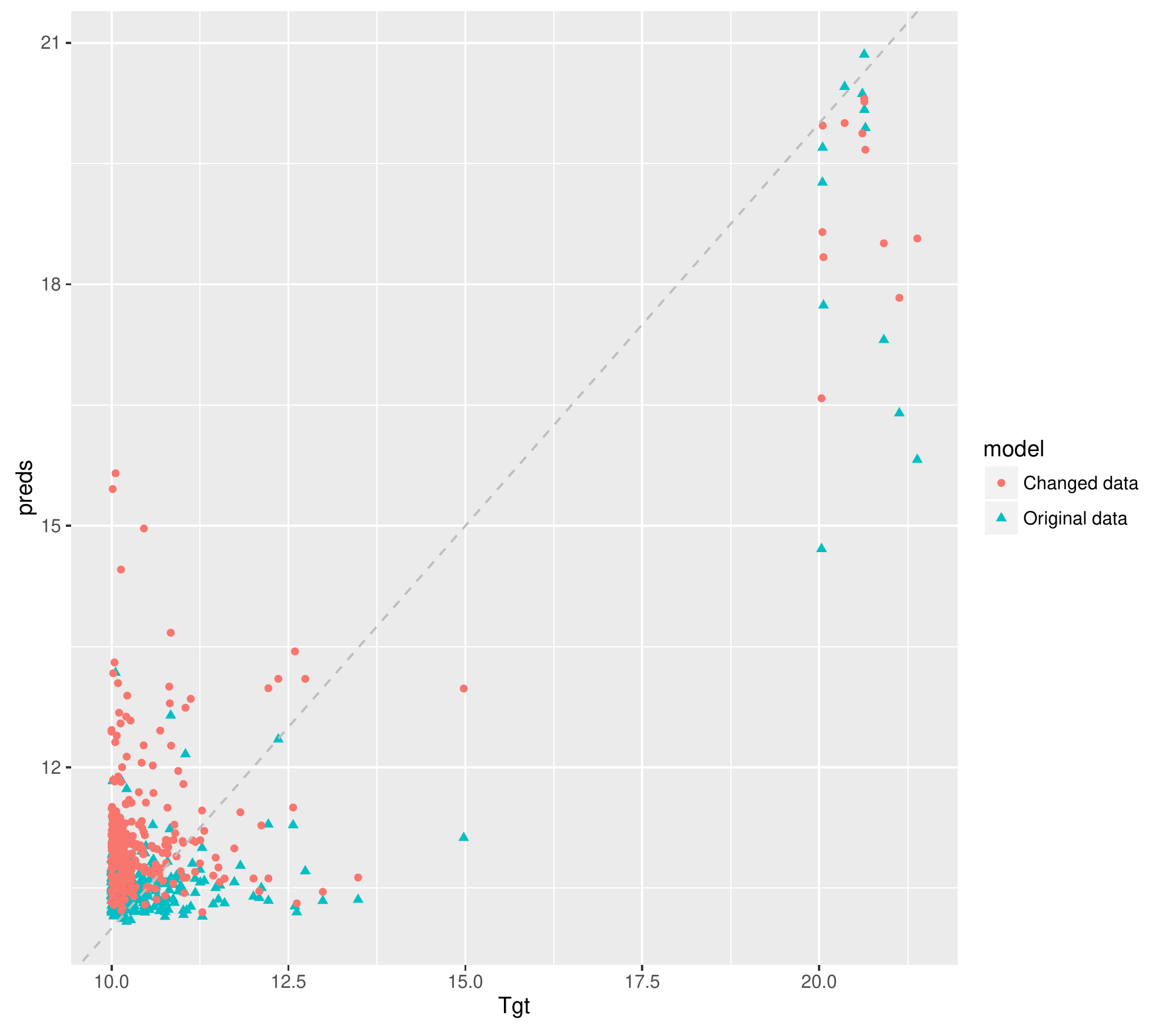} 

}

\caption[Predictions obtained with the original and the new data modified through the random under-sampling strategy]{Predictions obtained with the original and the new data modified through the random under-sampling strategy.}\label{fig:ImbR_RU1}
\end{figure}

\end{knitrout}

Figure~\ref{fig:ImbR_RU1} shows the predictions obtained with the original training set and the training data modified through random under-sampling strategy.

A more complex pre-processing can also be tried. In this case, we apply smoteR algorithm with low percentage of over-sampling. Then, we add more synthetic examples using the Gaussian Noise strategy using only the cases with a relevance value above the 0.8 threshold. Figure~\ref{fig:ImbR_Comb1} shows the predictions obtained with these changes.

\begin{knitrout}\footnotesize
\definecolor{shadecolor}{rgb}{0.969, 0.969, 0.969}\color{fgcolor}\begin{kframe}
\begin{alltt}
\hlstd{train1} \hlkwb{<-} \hlkwd{SmoteRegress}\hlstd{(Tgt}\hlopt{~}\hlstd{., trainD,} \hlkwc{C.perc}\hlstd{=}\hlkwd{list}\hlstd{(}\hlnum{0.9}\hlstd{,} \hlnum{2}\hlstd{))}
\hlstd{train2} \hlkwb{<-} \hlkwd{GaussNoiseRegress}\hlstd{(Tgt}\hlopt{~}\hlstd{., train1,} \hlkwc{thr.rel}\hlstd{=}\hlnum{0.8}\hlstd{,} \hlkwc{C.perc}\hlstd{=}\hlkwd{list}\hlstd{(}\hlnum{0.8}\hlstd{,} \hlnum{2}\hlstd{),} \hlkwc{pert}\hlstd{=}\hlnum{0.01}\hlstd{)}
\hlstd{ModelC} \hlkwb{<-}\hlkwd{randomForest}\hlstd{(Tgt}\hlopt{~}\hlstd{., train2)}
\hlstd{PredsC} \hlkwb{<-} \hlkwd{predict}\hlstd{(ModelC, testD)}
\end{alltt}
\end{kframe}
\end{knitrout}

\begin{knitrout}\footnotesize
\definecolor{shadecolor}{rgb}{0.969, 0.969, 0.969}\color{fgcolor}\begin{figure}

{\centering \includegraphics[width=0.7\textwidth]{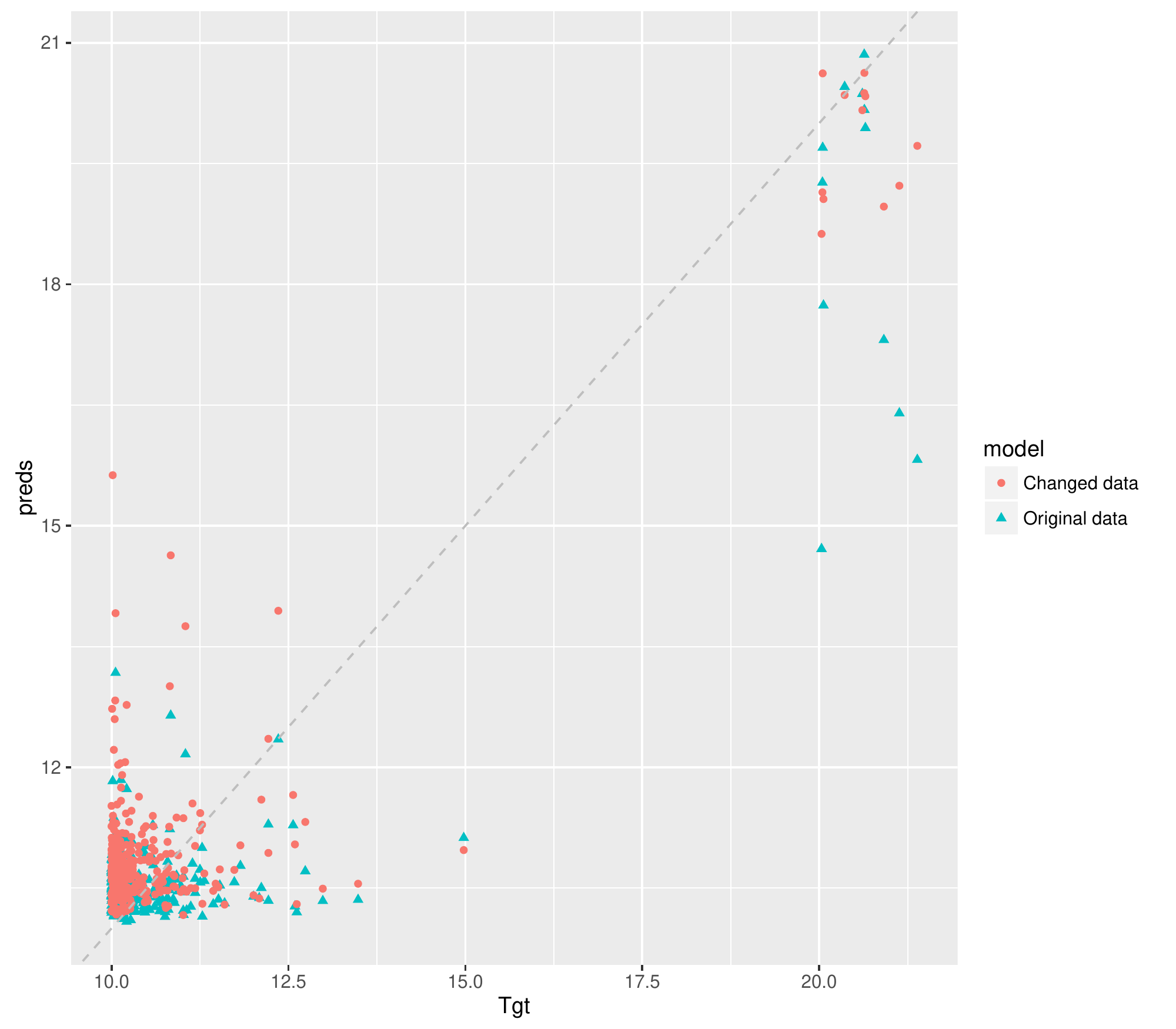} 

}

\caption[Predictions obtained with the original and the new data modified through the combination of strategies]{Predictions obtained with the original and the new data modified through the combination of strategies.}\label{fig:ImbR_Comb1}
\end{figure}

\end{knitrout}

Figure~\ref{fig:ImbR_difs} shows each test set example marked by a point with size and color varying according to the error magnitude for all the models previously obtained. This means that larger blue points represent larger errors and small green points represent a lower magnitude of the error in the example. 

\begin{knitrout}\footnotesize
\definecolor{shadecolor}{rgb}{0.969, 0.969, 0.969}\color{fgcolor}\begin{figure}

{\centering \includegraphics[width=0.9\textwidth]{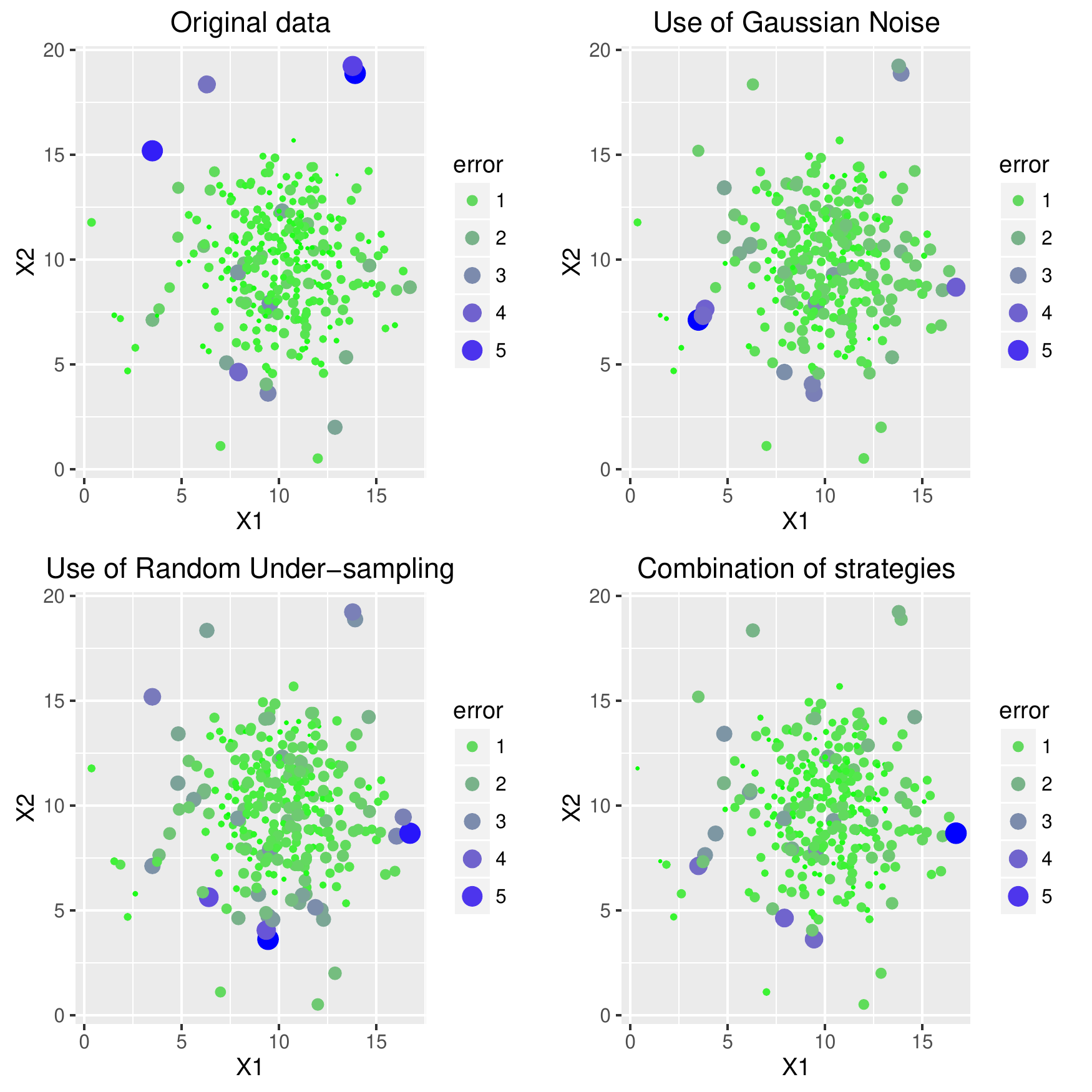} 

}

\caption[Predictions obtained with the original data and the two training sets modified through random under-sampling and Gaussian Noise strategies]{Predictions obtained with the original data and the two training sets modified through random under-sampling and Gaussian Noise strategies.}\label{fig:ImbR_difs}
\end{figure}

\end{knitrout}

% ====================================================================
\section{Methods for Addressing Utility-based Classification Tasks }\label{sec:methClass}

In this section we describe the methods implemented in \pUBL. We provide detailed examples of each function, and discuss how the several parameters can be used and their impact.
The methods explained in this section are the following:
\begin{itemize}
\item \ref{sec:RUClassif}: Random Under-sampling
\item \ref{sec:ROClassif}: Random Over-sampling
\item \ref{sec:ISClassif}: Importance Sampling
\item \ref{sec:Tomek}: Tomek Links
\item \ref{sec:CNN}: Condensed Nearest Neighbors
\item \ref{sec:OSS}: One-sided Selection
\item \ref{sec:ENN}: Edited Nearest Neighbors
\item \ref{sec:NCL}: Neighborhood Cleaning Rule
\item \ref{sec:gnClassif}: Gaussian Noise Introduction
\item \ref{sec:smoteClassif}: Smote Algorithm
\end{itemize}

\subsection{Random Under-sampling}\label{sec:RUClassif}

The random under-sampling strategy is among the simplest strategies for dealing with the class imbalanced problem. To force the learners to focus on the most important and least represented class(es) this technique randomly removes examples from the most represented and less important classes. This process allows to obtain a more balanced data set, although some important data may have been discarded with this technique. Another side effect of this strategy is a big reduction on the number of examples in the data set which facilitates the learners task although some important data may be ignored.

This strategy is implemented in \UBL taking into consideration the possible existence of several minority classes. The user may define through \texttt{C.perc} parameter which are the normal and less important classes and the under-sampling percentages to apply in each one of them. Another possibility is to select ``balance" or ``extreme" for the parameter \texttt{C.perc}. These two options automatically estimate the under-sampling percentages to apply to the classes. The ``balance" option obtains a balanced number of examples in all the existing classes, and the ``extreme" option inverts the existing frequencies, transforming the most frequent classes into the less frequent and vice-versa. The following examples show how these options can be used and their impact.

\begin{knitrout}\footnotesize
\definecolor{shadecolor}{rgb}{0.969, 0.969, 0.969}\color{fgcolor}\begin{kframe}
\begin{alltt}
\hlkwd{library}\hlstd{(UBL)}  \hlcom{# Loading our infra-structure}
\hlkwd{library}\hlstd{(e1071)} \hlcom{# package containing the svm we will use}
\hlkwd{data}\hlstd{(ImbC)}                      \hlcom{# Our synthetic multiclass data set}

\hlkwd{table}\hlstd{(ImbC}\hlopt{$}\hlstd{Class)}
\end{alltt}
\begin{verbatim}
## 
## normal  rare1  rare2 
##    859     10    131
\end{verbatim}
\begin{alltt}
\hlcom{## now, using random under-sampling to create a}
\hlcom{## "balanced problem" automatically}

\hlstd{newData} \hlkwb{<-} \hlkwd{RandUnderClassif}\hlstd{(Class} \hlopt{~} \hlstd{., ImbC)}
\hlkwd{table}\hlstd{(newData}\hlopt{$}\hlstd{Class)}
\end{alltt}
\begin{verbatim}
## 
## normal  rare1  rare2 
##     10     10     10
\end{verbatim}
\end{kframe}
\end{knitrout}

We highlight that, because this method only allows the removal of cases, in order to balance the examples distribution (the function default that was used), it has a strong impact in the total number of examples in the changed data set. This happens because one  of the minority classes has only 10 examples. Figure \ref{fig:Iris_RU1} shows the impact of this strategy in the examples distribution.

\begin{knitrout}\footnotesize
\definecolor{shadecolor}{rgb}{0.969, 0.969, 0.969}\color{fgcolor}\begin{figure}

{\centering \includegraphics[width=0.8\textwidth]{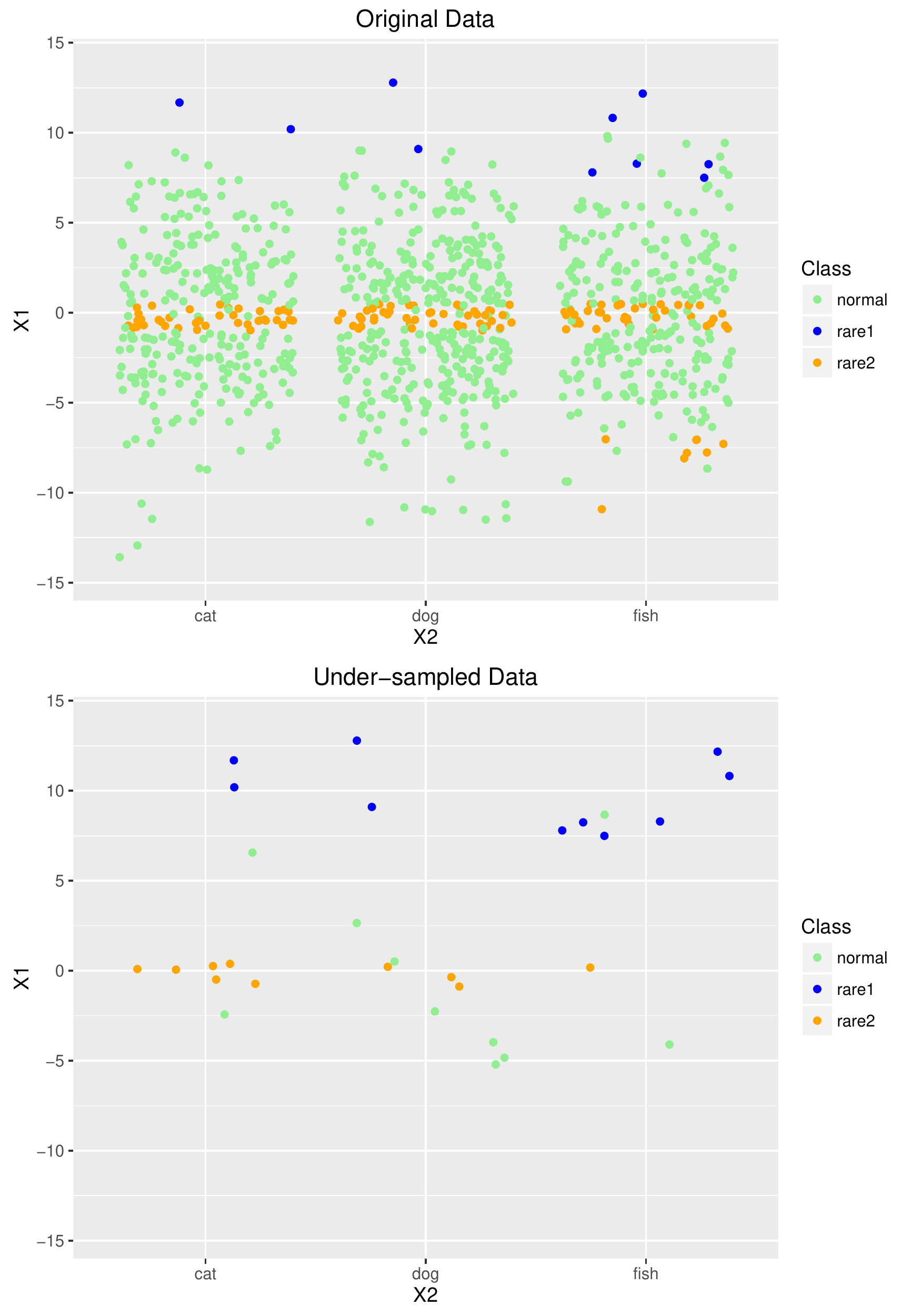} 

}

\caption[The impact of random under-sampling strategy]{The impact of random under-sampling strategy.}\label{fig:Iris_RU1}
\end{figure}

\end{knitrout}

Another example with ImbC data set:

\begin{knitrout}\footnotesize
\definecolor{shadecolor}{rgb}{0.969, 0.969, 0.969}\color{fgcolor}\begin{kframe}
\begin{alltt}
  \hlstd{RUmy} \hlkwb{<-} \hlkwd{RandUnderClassif}\hlstd{(Class}\hlopt{~}\hlstd{., ImbC,} \hlkwd{list}\hlstd{(}\hlkwc{normal}\hlstd{=}\hlnum{0.1}\hlstd{,} \hlkwc{rare2}\hlstd{=}\hlnum{0.9}\hlstd{))}
  \hlstd{RUB} \hlkwb{<-} \hlkwd{RandUnderClassif}\hlstd{(Class}\hlopt{~}\hlstd{., ImbC,} \hlstr{"balance"}\hlstd{)}
  \hlstd{RUE} \hlkwb{<-} \hlkwd{RandUnderClassif}\hlstd{(Class}\hlopt{~}\hlstd{., ImbC,} \hlstr{"extreme"}\hlstd{)}
\end{alltt}
\end{kframe}
\end{knitrout}

\begin{table}[ht]
\centering
\begin{tabular}{rrrr}
  \hline
 & normal & rare1 & rare2 \\ 
  \hline
Original & 859 &  10 & 131 \\ 
  RUmy &  85 &  10 & 117 \\ 
  RUB &  10 &  10 &  10 \\ 
  RUE &   0 &  10 &   0 \\ 
   \hline
\end{tabular}
\caption{Number of examples in each class for different parameters of random under-sampling strategy.} 
\label{tab:RU_tab}
\end{table}

The impact of the strategies on the number of examples in each class of the data set are in Figure\ref{fig:Iris_RU2}.

\begin{knitrout}\footnotesize
\definecolor{shadecolor}{rgb}{0.969, 0.969, 0.969}\color{fgcolor}\begin{figure}

{\centering \includegraphics[width=\maxwidth]{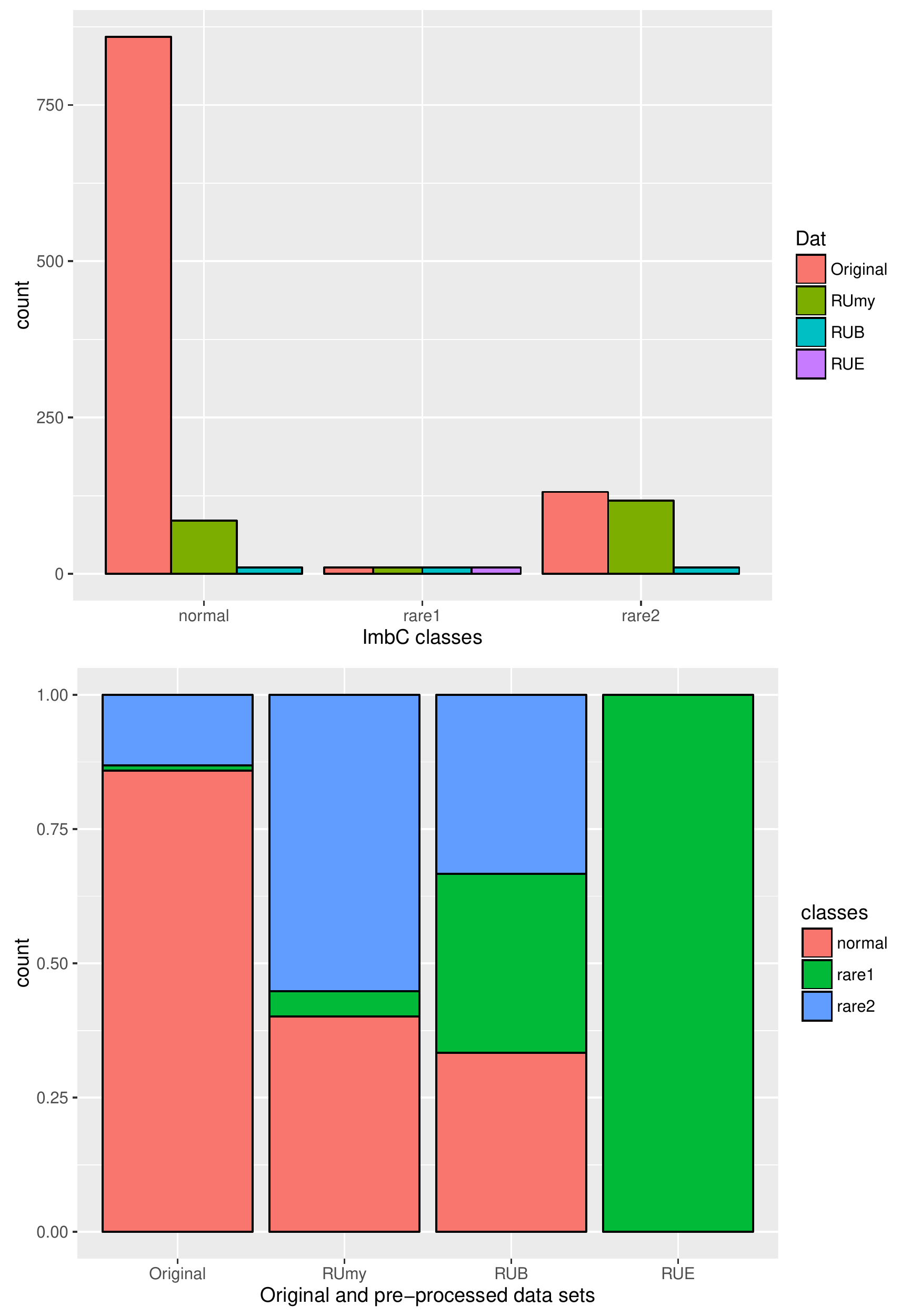} 

}

\caption[Random Under-sampling strategy for different parameters values]{Random Under-sampling strategy for different parameters values.}\label{fig:Iris_RU2}
\end{figure}

\end{knitrout}

\subsection{Random Over-sampling}\label{sec:ROClassif}

The random over-sampling strategy introduces replicas of already existing examples in the data set. The replicas to include are randomly selected among the least populated and more important classes. This allows to obtain a better balanced data set without discarding any examples. However, this method has a strong impact on the number of examples of the new data set which can represent a difficulty to the used learner.

This strategy is implemented in \pUBL taking into consideration the possible existence of several minority classes. The user may define through \texttt{C.perc} parameter which are the most important classes and their respective over-sampling percentages. The parameter \texttt{C.perc} may also be set to ``balance" or ``extreme". These two options automatically estimate the classes and over-sampling percentages to apply. Similarly to the previous strategy the ``balance" option allows to obtain a balanced number of examples in all the existing classes, and the ``extreme" option inverts the existing frequencies, transforming the most frequent classes into the less frequent and vice-versa. The following examples show how these options can be used and their impact:

\begin{knitrout}\footnotesize
\definecolor{shadecolor}{rgb}{0.969, 0.969, 0.969}\color{fgcolor}\begin{kframe}
\begin{alltt}
\hlcom{## now using random over-sampling to create a }
\hlcom{## data with more 500% of examples in the }
\hlcom{## rare1 class}
\hlstd{RO.U1}\hlkwb{<-} \hlkwd{RandOverClassif}\hlstd{(Class} \hlopt{~} \hlstd{., ImbC,}
                        \hlkwc{C.perc}\hlstd{=}\hlkwd{list}\hlstd{(}\hlkwc{rare1}\hlstd{=}\hlnum{5}\hlstd{))}
\hlstd{RO.U2}\hlkwb{<-} \hlkwd{RandOverClassif}\hlstd{(Class} \hlopt{~} \hlstd{., ImbC,}
                        \hlkwc{C.perc}\hlstd{=}\hlkwd{list}\hlstd{(}\hlkwc{rare1}\hlstd{=}\hlnum{4}\hlstd{,} \hlkwc{rare2}\hlstd{=}\hlnum{2.5}\hlstd{))}
\hlstd{RO.B} \hlkwb{<-} \hlkwd{RandOverClassif}\hlstd{(Class} \hlopt{~} \hlstd{., ImbC,} \hlkwc{C.perc}\hlstd{=}\hlstr{"balance"}\hlstd{)}
\hlstd{RO.E} \hlkwb{<-} \hlkwd{RandOverClassif}\hlstd{(Class} \hlopt{~} \hlstd{., ImbC,} \hlkwc{C.perc}\hlstd{=}\hlstr{"extreme"}\hlstd{)}
\end{alltt}
\end{kframe}
\end{knitrout}

\begin{table}[ht]
\centering
\begin{tabular}{rrrr}
  \hline
 & normal & rare1 & rare2 \\ 
  \hline
Original & 859 &  10 & 131 \\ 
  RO.U1 & 859 &  50 & 131 \\ 
  RO.U2 & 859 &  40 & 327 \\ 
  RO.B & 859 & 859 & 859 \\ 
  RO.E & 859 & 73788 & 5633 \\ 
   \hline
\end{tabular}
\caption{Number of examples in each class for different Random over-sampling parameters.} 
\label{tab:RO_tab}
\end{table}

Figure \ref{fig:IrisRO} shows the impact of this strategy in the examples distribution. We have introduced a small perturbation on the examples position to be more clear the replicas that were introduced.

\begin{knitrout}\footnotesize
\definecolor{shadecolor}{rgb}{0.969, 0.969, 0.969}\color{fgcolor}\begin{figure}

{\centering \includegraphics[width=0.8\textwidth]{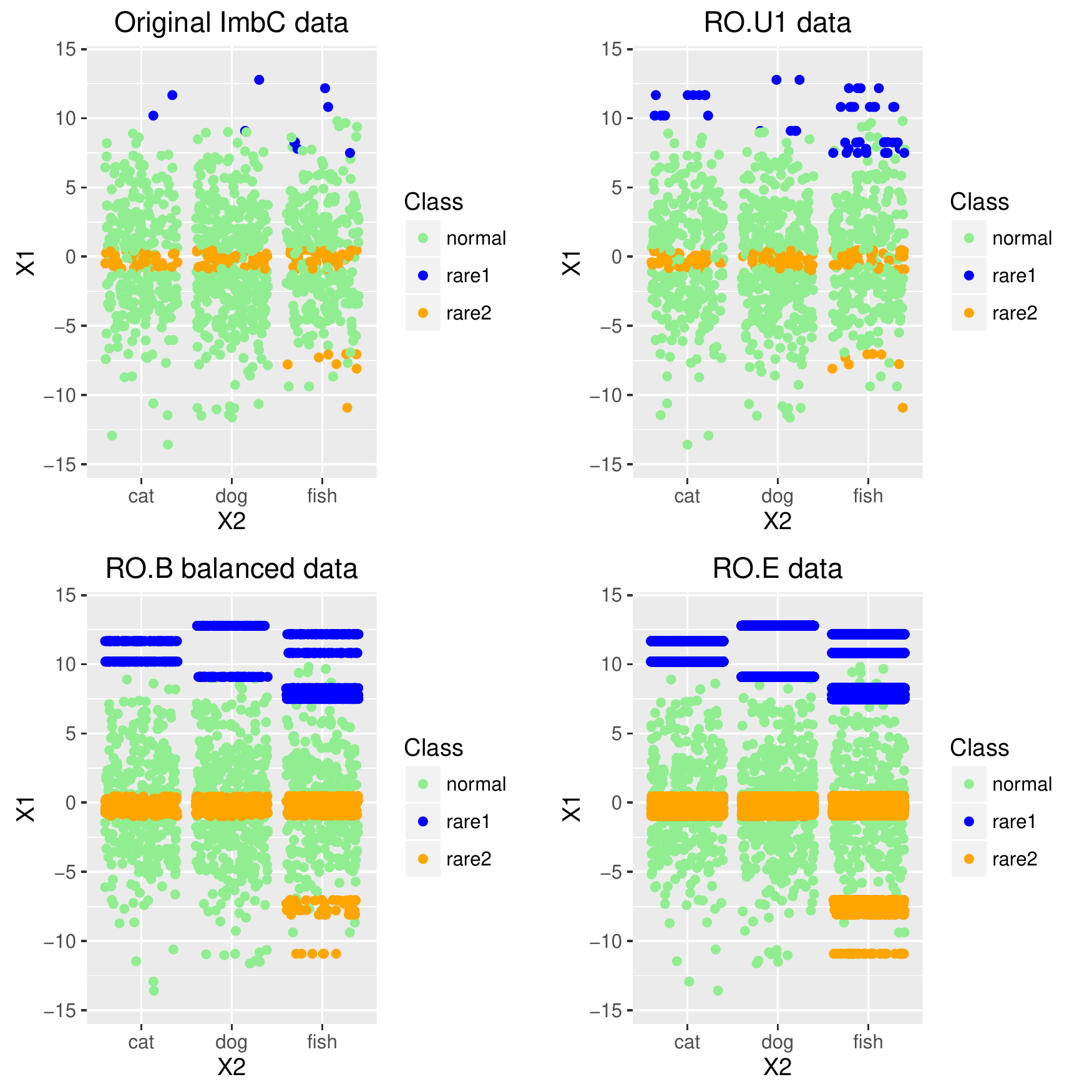} 

}

\caption[The impact of random over-sampling Strategy]{The impact of random over-sampling Strategy.}\label{fig:IrisRO}
\end{figure}

\end{knitrout}

Figure \ref{fig:Iris_RO2} shows the impact of this strategy on the number of examples in the data set.

\begin{knitrout}\footnotesize
\definecolor{shadecolor}{rgb}{0.969, 0.969, 0.969}\color{fgcolor}\begin{figure}

{\centering \includegraphics[width=\maxwidth]{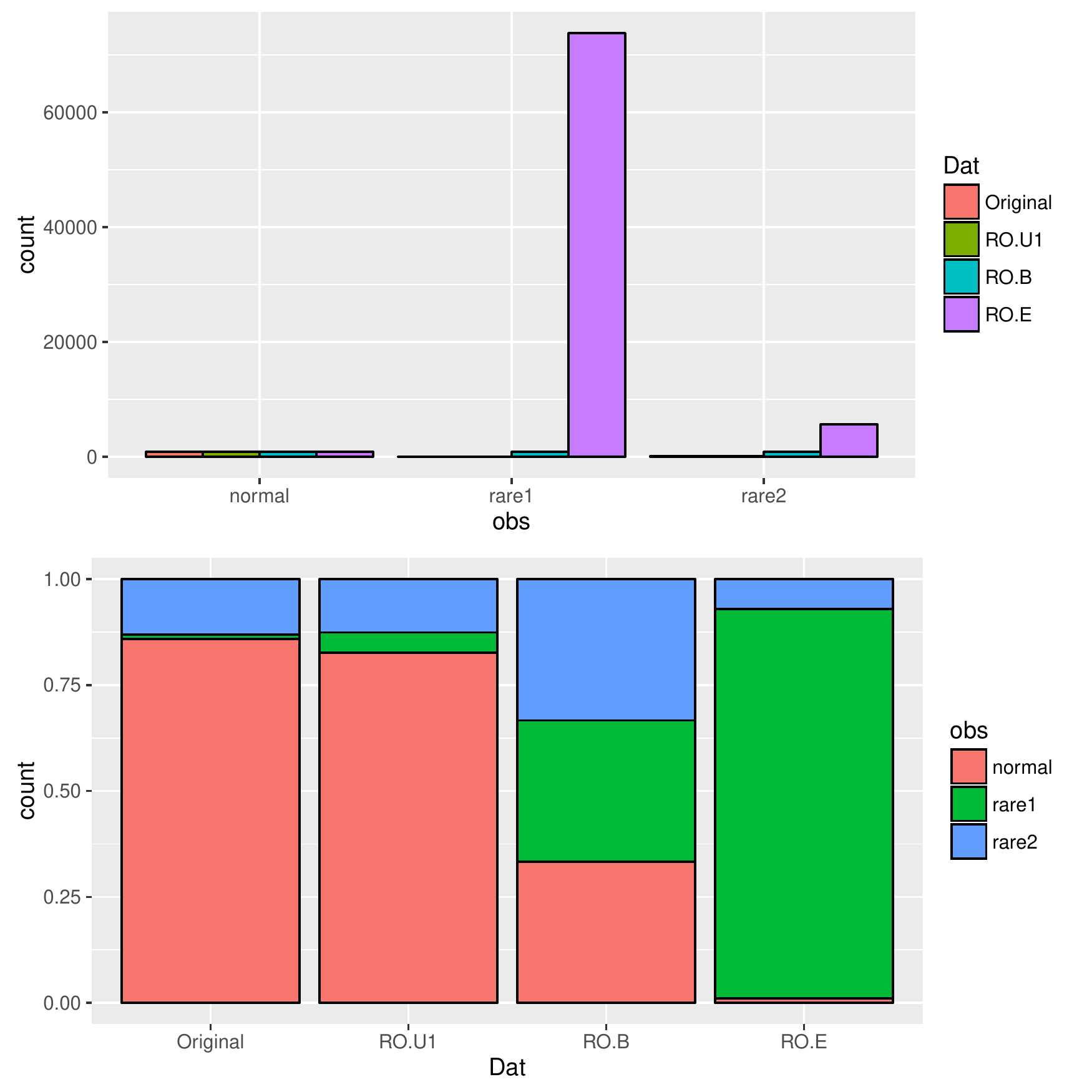} 

}

\caption[Impact of Random over-sampling strategy for different parameters values]{Impact of Random over-sampling strategy for different parameters values.}\label{fig:Iris_RO2}
\end{figure}

\end{knitrout}

\subsection{Importance Sampling}\label{sec:ISClassif}

The main idea of Importance Sampling strategy is to perform random over- or under-sampling in each class according to the importance assigned by the user. This means that for each class the user can specify its relevance. Then, this relevance is used to change each class frequency by selecting randomly examples from each class. Alternatively, the user may consider that each class is equally important or may chose to invert the classes frequencies.

This strategy is available in \UBLp through the function \texttt{ImpSampClassif}. The user may specify using parameter \texttt{C.perc} the classes where over-/under-sampling must be applied, by indicating the corresponding percentages. If all classes are equally important and a perfectly balanced data set should be obtained, the \texttt{C.perc} parameter must be set to ``balance". On the other hand, if the classes frequencies should be inverted, then this parameter should be ``extreme". The following example illustrate the use of this function.

\begin{knitrout}\footnotesize
\definecolor{shadecolor}{rgb}{0.969, 0.969, 0.969}\color{fgcolor}\begin{kframe}
\begin{alltt}
\hlcom{# use the synthetic imbalanced data set ImbC provided with UBL package}
\hlkwd{table}\hlstd{(ImbC}\hlopt{$}\hlstd{Class)}
\end{alltt}
\begin{verbatim}
## 
## normal  rare1  rare2 
##    859     10    131
\end{verbatim}
\begin{alltt}
\hlstd{nds} \hlkwb{<-} \hlkwd{ImpSampClassif}\hlstd{(Class}\hlopt{~}\hlstd{.,ImbC,} \hlkwc{C.perc}\hlstd{=}\hlkwd{list}\hlstd{(}\hlkwc{normal}\hlstd{=}\hlnum{0.4}\hlstd{,} \hlkwc{rare1}\hlstd{=}\hlnum{6}\hlstd{))}
\hlcom{# notice that when a certain class is not specified it remains unaltered}
\hlkwd{table}\hlstd{(nds}\hlopt{$}\hlstd{Class)}
\end{alltt}
\begin{verbatim}
## 
## normal  rare1  rare2 
##    343     60    131
\end{verbatim}
\begin{alltt}
\hlcom{# to obtain a balanced data set}
\hlstd{IS.bal} \hlkwb{<-} \hlkwd{ImpSampClassif}\hlstd{(Class}\hlopt{~}\hlstd{., ImbC)} \hlcom{# or use C.perc="balance"}
\hlkwd{table}\hlstd{(IS.bal}\hlopt{$}\hlstd{Class)}
\end{alltt}
\begin{verbatim}
## 
## normal  rare1  rare2 
##    333    333    333
\end{verbatim}
\begin{alltt}
\hlcom{# to obtain a data set with inverted frequencies }
\hlstd{IS.ext} \hlkwb{<-} \hlkwd{ImpSampClassif}\hlstd{(Class}\hlopt{~}\hlstd{., ImbC,} \hlkwc{C.perc}\hlstd{=}\hlstr{"extreme"}\hlstd{)}
\hlkwd{table}\hlstd{(IS.ext}\hlopt{$}\hlstd{Class)}
\end{alltt}
\begin{verbatim}
## 
## normal  rare1  rare2 
##     11    919     70
\end{verbatim}
\end{kframe}
\end{knitrout}

Figure \ref{fig:ISC} shows the impact on the imbalanced ImbC data set of the changes made in the domain with Importance Sampling.

\begin{knitrout}\footnotesize
\definecolor{shadecolor}{rgb}{0.969, 0.969, 0.969}\color{fgcolor}\begin{figure}

{\centering \includegraphics[width=\maxwidth]{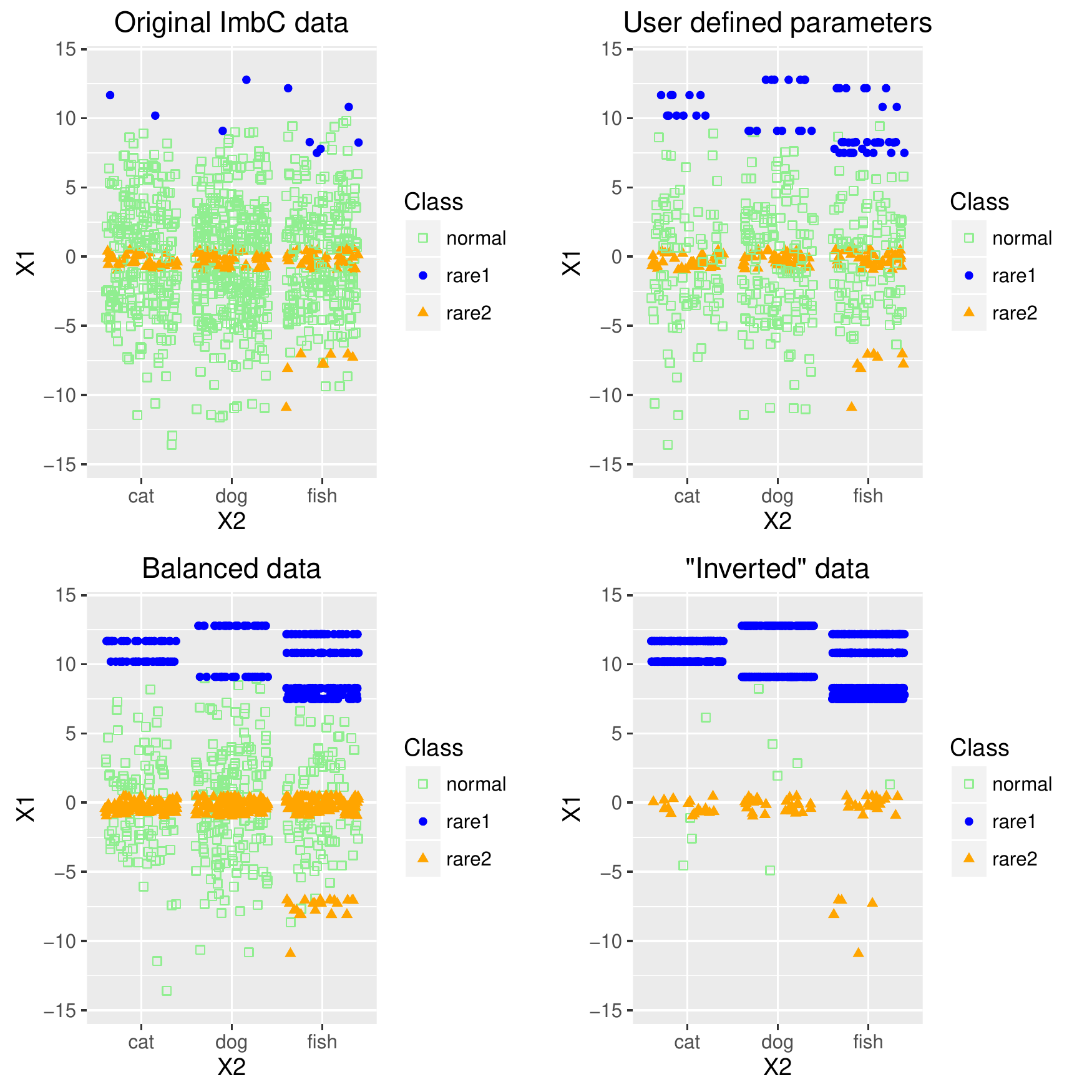} 

}

\caption[Impact of Importance Sampling strategy in ImbC data set]{Impact of Importance Sampling strategy in ImbC data set.}\label{fig:ISC}
\end{figure}

\end{knitrout}

Figure \ref{fig:ISC_2} shows the impact of this strategy on the number of examples in the data set.

\begin{knitrout}\footnotesize
\definecolor{shadecolor}{rgb}{0.969, 0.969, 0.969}\color{fgcolor}\begin{figure}

{\centering \includegraphics[width=\maxwidth]{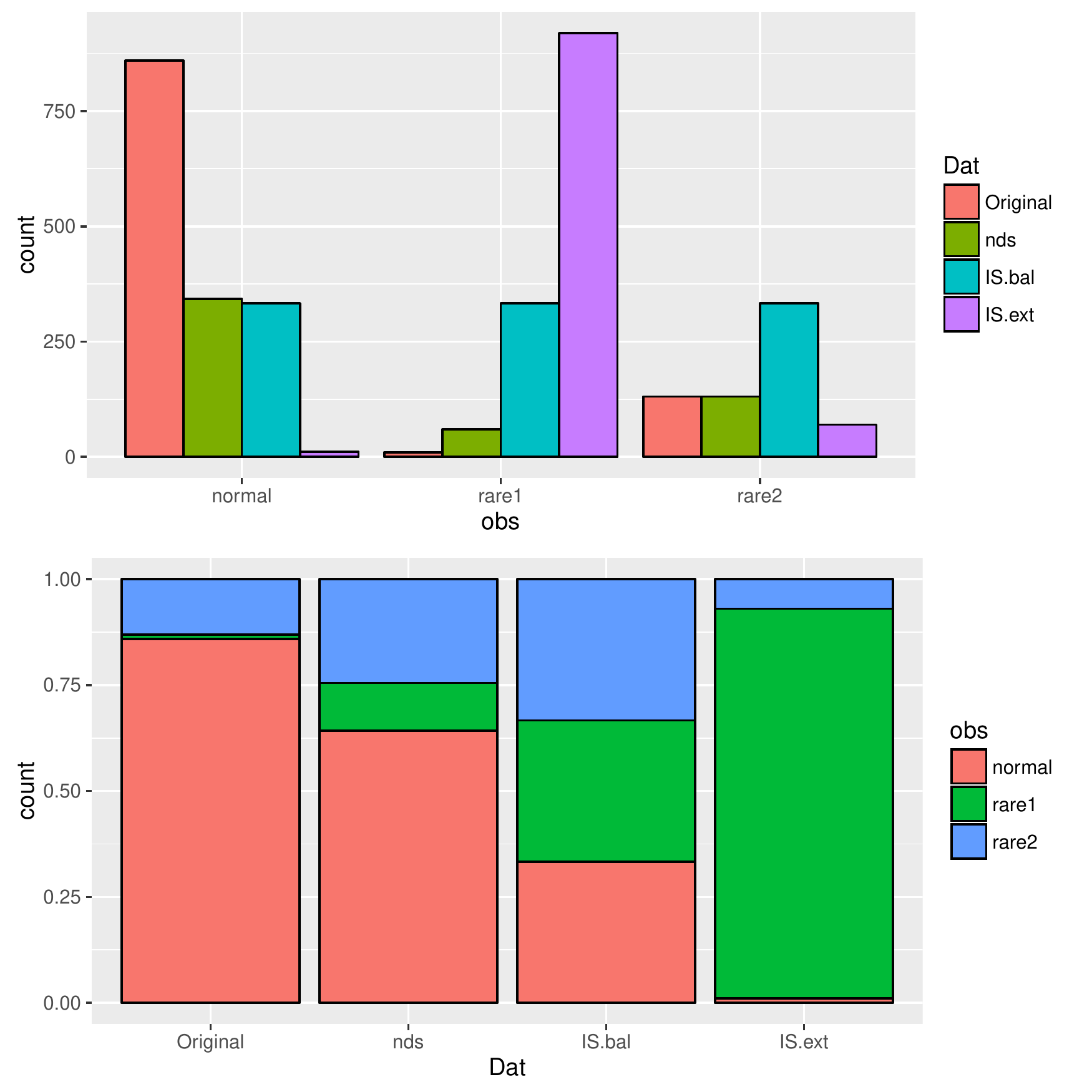} 

}

\caption[Impact of Importance Sampling strategy]{Impact of Importance Sampling strategy.}\label{fig:ISC_2}
\end{figure}

\end{knitrout}

We must highlight that random under- and over-sampling also allow to balance and invert the classes frequencies. Importance Sampling strategy, although also allowing this type of impact, acts differently because it combines both under- and over-sampling strategies. This means that a balanced data set can be obtained through random under-sampling, random over-sampling or importance sampling strategy. However, the resulting data sets will be different. If we use random under-sampling the final size of the data set is reduced, while if we use the random over-sampling approach the changed data set is significantly larger than the original one. If we select the importance sampling, the combination of the strategies allows to roughly maintain the data set size.

\subsection{Tomek Links}\label{sec:Tomek}

Tomek Links \cite{tomek1976two} can be defined as follows: two examples form a Tomek Link if and only if they belong to different classes and are each other nearest neighbors. This is a property existing between a pair of examples $(S_i, S_j)$ having different class labels and for which 

$\nexists S_k : dist(S_i,S_k) < dist(S_i,S_j) \vee dist(S_j, S_k)<dist(S_i,S_j)$

\noindent Having determined the examples which form Tomek Links, these connections may be explained because either the examples are both borderline examples or one of the examples may be considered as noise.
Therefore, there are two possibilities of using Tomek links to accomplish under-sampling:
\begin{itemize}
  \item remove the two examples forming a Tomek link, or
  \item only remove the example from the most populated class which forms a Tomek link.
\end{itemize}

These two options correspond to using Tomek link as cleaning technique (by removing both borderline examples) or as an under-sampling method for balancing the classes (by removing the majority class example).

In \pUBL we have adapted this technique for being able to deal with multiclass imbalanced problems. 
For working with more than two classes some issues were considered: 
\begin{itemize}
\item allow the user to select which classes should be under-sampled (if not defined, the default is to under-sample all the existing classes);
\item if the user selects a given number of classes what to do to break the link, i.e., how to decide which example(s) to remove (if any). 
\end{itemize}
So, in \UBL the user may chose from which classes he is interested in removing examples through the \texttt{Cl} parameter. Moreover, the user can also decide if both examples are removed or if just one is discarded using the \texttt{rem} parameter. If this can be easily understood in two class problems, the impact of these parameters may not be so clear for multiclass imbalanced tasks. 
In fact,the options set for \texttt{Cl} and \texttt{rem} parameters may ``disagree". In those cases, the preference is given to the \texttt{Cl} options once the user choose that specific set of classes to under-sample and not the other ones (even if the defined classes are not the larger ones). This means that, when making a decision on how many and which examples will be removed the first criteria used will be the \texttt{Cl} definition.

For a better clarification of the behavior stated we now provide some possible scenarios for multiclass problems and the corresponding expected behavior:

\begin{itemize}
\item \texttt{Cl} is set to one class which is neither the most nor the least frequent, and \texttt{rem} is set to ``maj". The expected behavior is the following:
- if a Tomek link exists connecting the largest class and another class(not included in \texttt{Cl}): no example is removed;
- if a Tomek link exists connecting the larger class and the class defined in \texttt{Cl}: the example from the \texttt{Cl} class is removed (because the user expressly indicates that only examples from class \texttt{Cl} should be removed);

\item \texttt{Cl} includes two classes and \texttt{rem} is set to ``both". This function will do the following:
- if a Tomek link exists between an example with class in \texttt{Cl} and another example with class not in \texttt{Cl}, then, only the example with class in \texttt{Cl} is removed;
- if the Tomek link exists between two examples with classes in \texttt{Cl}, then, both are removed.

\item \texttt{Cl} includes two classes and \texttt{rem} is set to ``maj". The behavior of this function is the following:
-if a Tomek link exists connecting two classes included in \texttt{Cl}, then only the example belonging to the more populated class is removed;
-if a Tomek link exists connecting an example from a class included in \texttt{Cl} and another example whose class is not in \texttt{Cl} and is the largest class, then, no example is removed.

\end{itemize}

We must also highlight that this strategy strongly depends on the distance metric considered for the nearest neighbors computation. We provide in \pUBL several different distance measures which are able to deal with numeric and/or nominal features, such as Manhattan distance, Euclidean Distance, HEOM or HVDM. For more details on the available distance functions check Section \ref{sec:distFunc}. The user may set the desired distance metric through the \texttt{dist} parameter.

The implementation provided in this package returns a list containing: the new data set modified through the Tomek links strategy and the indexes of the examples removed. Under certain situations, this strategy is not able to remove any example of the data set. In this case, a warning is issued to advert the user that no example was removed.

The following examples with mbC data set show how Tomek links can be applied.

\begin{knitrout}\footnotesize
\definecolor{shadecolor}{rgb}{0.969, 0.969, 0.969}\color{fgcolor}\begin{kframe}
\begin{alltt}
\hlcom{# using the default parameters except for the distance function}
\hlcom{# ImbC has nominal and numeric features and this requires the use of }
\hlcom{# specific distance functions}
  \hlstd{ds} \hlkwb{<-} \hlkwd{TomekClassif}\hlstd{(Class}\hlopt{~}\hlstd{., ImbC,} \hlkwc{dist}\hlstd{=}\hlstr{"HEOM"}\hlstd{)}
\hlcom{# using HEOM distance metric, and selecting only one class to under-sample}
  \hlstd{dsHEOM} \hlkwb{<-} \hlkwd{TomekClassif}\hlstd{(Class}\hlopt{~}\hlstd{., ImbC,} \hlkwc{dist}\hlstd{=}\hlstr{"HEOM"}\hlstd{,}
                         \hlkwc{Cl}\hlstd{=}\hlstr{"normal"}\hlstd{)}

\hlcom{# check the new dsHEOM data set}
  \hlkwd{summary}\hlstd{(dsHEOM[[}\hlnum{1}\hlstd{]])}
\end{alltt}
\begin{verbatim}
##        X1              X2         Class    
##  Min.   :-13.5843   cat :294   normal:843  
##  1st Qu.: -2.7206   dog :394   rare1 : 10  
##  Median : -0.1603   fish:296   rare2 :131  
##  Mean   : -0.1141                          
##  3rd Qu.:  2.5029                          
##  Max.   : 12.7836
\end{verbatim}
\begin{alltt}
\hlcom{# check the indexes of the examples removed:}
  \hlstd{dsHEOM[[}\hlnum{2}\hlstd{]]}
\end{alltt}
\begin{verbatim}
##  [1] 359 830 103 172 182 748 314 322 341 981 943 544 703 681 920 996
\end{verbatim}
\begin{alltt}
\hlcom{# using HVDM distance, enable the removal of examples from all classes, and}
\hlcom{# select to break the link by only removing the example from the majority class}
  \hlstd{dsHVDMM} \hlkwb{<-} \hlkwd{TomekClassif}\hlstd{(Class}\hlopt{~}\hlstd{., ImbC,} \hlkwc{dist}\hlstd{=}\hlstr{"HVDM"}\hlstd{,} \hlkwc{Cl}\hlstd{=}\hlstr{"all"}\hlstd{,} \hlkwc{rem}\hlstd{=}\hlstr{"maj"}\hlstd{)}
\hlcom{# similar but breaking the Tomek link by removing both examples that for it}
  \hlstd{dsHVDMB} \hlkwb{<-} \hlkwd{TomekClassif}\hlstd{(Class}\hlopt{~}\hlstd{., ImbC,} \hlkwc{dist}\hlstd{=}\hlstr{"HVDM"}\hlstd{,} \hlkwc{Cl}\hlstd{=}\hlstr{"all"}\hlstd{,} \hlkwc{rem}\hlstd{=}\hlstr{"both"}\hlstd{)}
\end{alltt}
\end{kframe}
\end{knitrout}

\begin{table}[ht]
\centering
\begin{tabular}{rrrr}
  \hline
 & normal & rare1 & rare2 \\ 
  \hline
Original & 859 &  10 & 131 \\ 
  ds & 843 &   9 & 116 \\ 
  dsHEOM & 843 &  10 & 131 \\ 
  dsHVDMM & 833 &  10 & 131 \\ 
  dsHVDMB & 833 &   8 & 107 \\ 
   \hline
\end{tabular}
\caption{Number of examples in each class for different Tomek Links parameters.} 
\label{tab:TL_table}
\end{table}

Figure \ref{fig:TL_difPar2} shows the impact in ImbC examples of the last experiences.
\begin{knitrout}\footnotesize
\definecolor{shadecolor}{rgb}{0.969, 0.969, 0.969}\color{fgcolor}\begin{figure}

{\centering \includegraphics[width=\maxwidth]{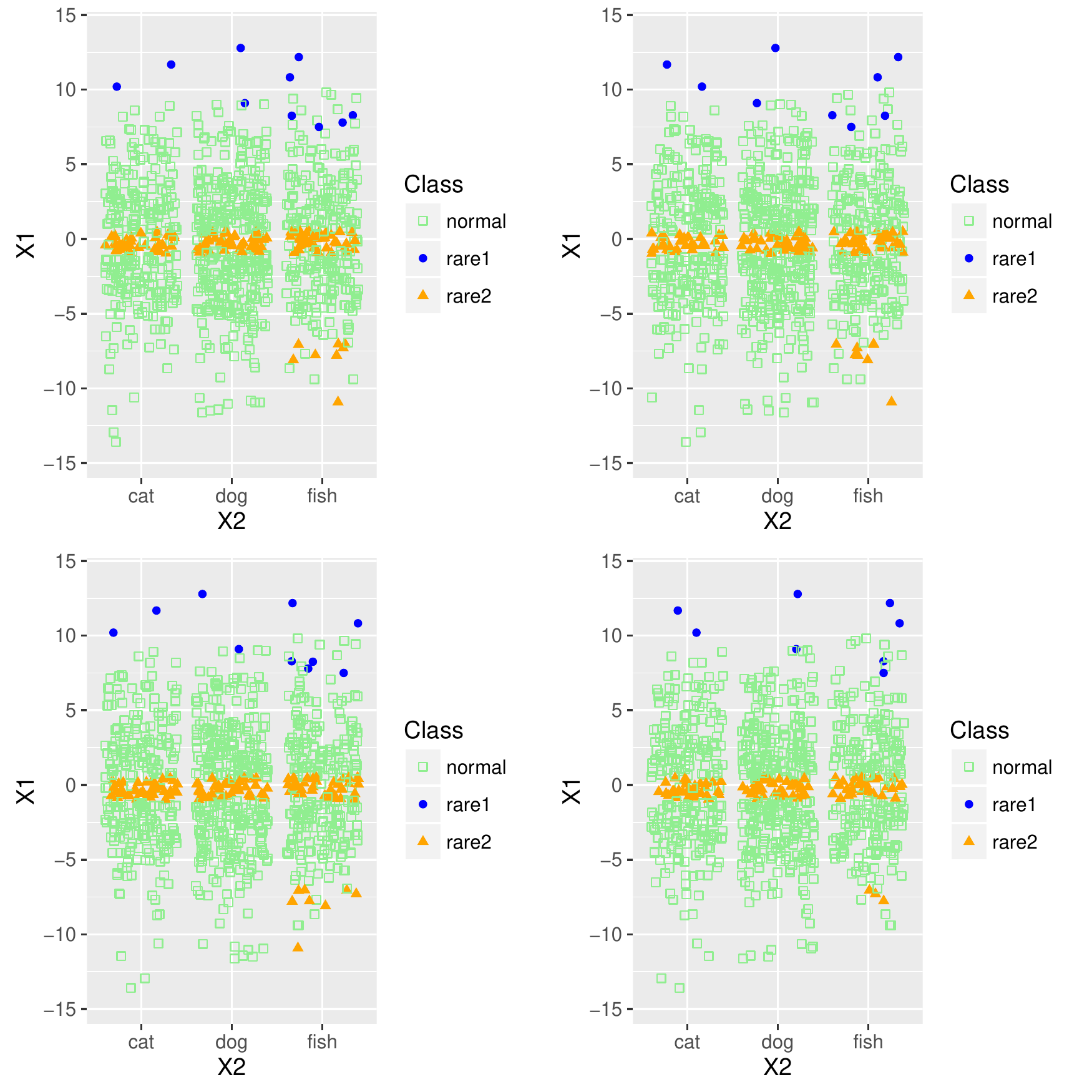} 

}

\caption[Impact of Tomek links strategy in ImbC synthetic data set]{Impact of Tomek links strategy in ImbC synthetic data set. (top left: ImbC data; top right: ds data; bottom left:dsHVDMM data; and bottom right: dsHVDMB data)}\label{fig:TL_difPar2}
\end{figure}

\end{knitrout}

Let us now consider the iris data set, changed with the goal of having an imbalanced distribution of the classes. 
\begin{knitrout}\footnotesize
\definecolor{shadecolor}{rgb}{0.969, 0.969, 0.969}\color{fgcolor}\begin{kframe}
\begin{alltt}
\hlkwd{data}\hlstd{(iris)}
\hlstd{dat} \hlkwb{<-} \hlstd{iris[}\hlopt{-}\hlkwd{c}\hlstd{(}\hlnum{61}\hlopt{:}\hlnum{85}\hlstd{,} \hlnum{116}\hlopt{:}\hlnum{150}\hlstd{),}\hlkwd{c}\hlstd{(}\hlnum{1}\hlstd{,}\hlnum{2}\hlstd{,}\hlnum{5}\hlstd{)]}
\hlkwd{summary}\hlstd{(dat)}
\end{alltt}
\begin{verbatim}
##   Sepal.Length    Sepal.Width          Species  
##  Min.   :4.300   Min.   :2.300   setosa    :50  
##  1st Qu.:5.000   1st Qu.:2.900   versicolor:25  
##  Median :5.350   Median :3.150   virginica :15  
##  Mean   :5.497   Mean   :3.170                  
##  3rd Qu.:5.800   3rd Qu.:3.475                  
##  Max.   :7.600   Max.   :4.400
\end{verbatim}
\end{kframe}
\end{knitrout}

Let us observe the impact of applying Tomek links strategy in this data.

\begin{knitrout}\footnotesize
\definecolor{shadecolor}{rgb}{0.969, 0.969, 0.969}\color{fgcolor}\begin{kframe}
\begin{alltt}
\hlcom{# using the default in all parameters}
  \hlstd{ir} \hlkwb{<-} \hlkwd{TomekClassif}\hlstd{(Species}\hlopt{~}\hlstd{., dat)}
\hlcom{# using chebyshev distance metric, and selecting only two classes to under-sample}
  \hlstd{irCheb} \hlkwb{<-} \hlkwd{TomekClassif}\hlstd{(Species}\hlopt{~}\hlstd{., dat,} \hlkwc{dist}\hlstd{=}\hlstr{"Chebyshev"}\hlstd{,}
                         \hlkwc{Cl}\hlstd{=}\hlkwd{c}\hlstd{(}\hlstr{"virginica"}\hlstd{,} \hlstr{"setosa"}\hlstd{))}
\hlcom{# using Manhattan distance, enable the removal of examples from all classes, and}
\hlcom{# select to break the link by only removing the example from the majority class}
  \hlstd{irManM} \hlkwb{<-} \hlkwd{TomekClassif}\hlstd{(Species}\hlopt{~}\hlstd{., dat,} \hlkwc{dist}\hlstd{=}\hlstr{"Manhattan"}\hlstd{,} \hlkwc{Cl}\hlstd{=}\hlstr{"all"}\hlstd{,} \hlkwc{rem}\hlstd{=}\hlstr{"maj"}\hlstd{)}
  \hlstd{irManB} \hlkwb{<-} \hlkwd{TomekClassif}\hlstd{(Species}\hlopt{~}\hlstd{., dat,} \hlkwc{dist}\hlstd{=}\hlstr{"Manhattan"}\hlstd{,} \hlkwc{Cl}\hlstd{=}\hlstr{"all"}\hlstd{,} \hlkwc{rem}\hlstd{=}\hlstr{"both"}\hlstd{)}
\end{alltt}
\end{kframe}
\end{knitrout}

\begin{table}[ht]
\centering
\begin{tabular}{rrrr}
  \hline
 & setosa & versicolor & virginica \\ 
  \hline
Original &  50 &  25 &  15 \\ 
  ir &  50 &  19 &   9 \\ 
  irCheb &  50 &  25 &  10 \\ 
  irManM &  50 &  19 &  15 \\ 
  irManB &  50 &  19 &   9 \\ 
   \hline
\end{tabular}
\caption{Number of examples in each class for different Tomek Links parameters.} 
\label{tab:irTL_table}
\end{table}

Figure \ref{fig:TL_difPar} shows the impact of the previously described Tomek links strategies in the iris subset considered.

\begin{knitrout}\footnotesize
\definecolor{shadecolor}{rgb}{0.969, 0.969, 0.969}\color{fgcolor}\begin{figure}

{\centering \includegraphics[width=\maxwidth]{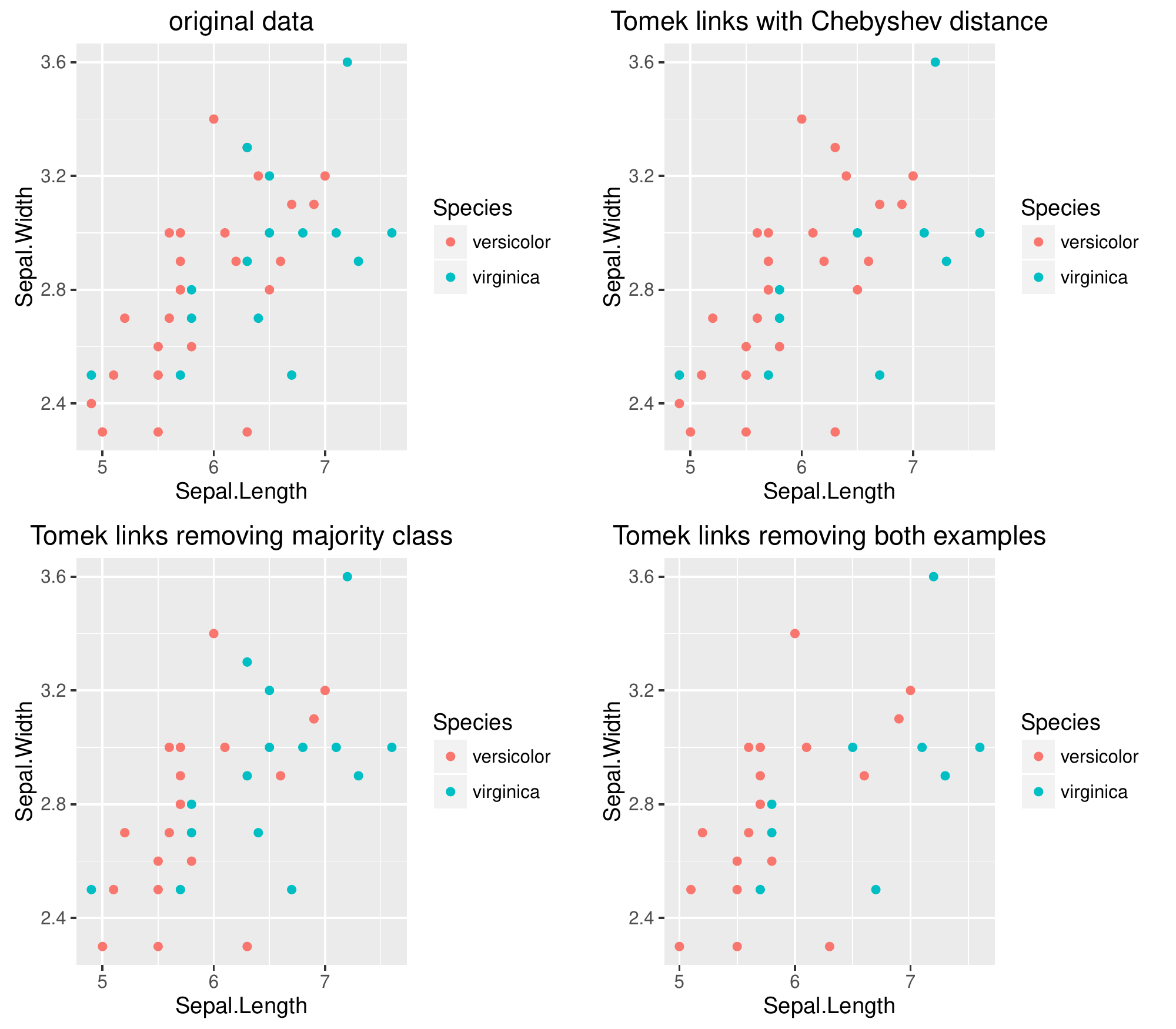} 

}

\caption[Impact of Tomek links strategy in classes virginica and versicolor of the subset of iris data (top left]{Impact of Tomek links strategy in classes virginica and versicolor of the subset of iris data (top left: original iris subset; top right: irCheb data; bottom left: irManM data and bottom right: irManB data).}\label{fig:TL_difPar}
\end{figure}

\end{knitrout}

\subsection{Condensed Nearest Neighbors}\label{sec:CNN}

The Condensed nearest neighbors rule (CNN) was presented by \cite{cnn}. The goal of this strategy is to perform under-sampling by building a subset of examples which is consistent with the original data. A subset is consistent with another if the elements in the subset classify correctly all the original examples using a 1-NN.

To build a consistent subset we have adapted the proposal of \cite{KM97} to multiclass problems. The user starts by defining which are the most relevant classes in the data set using the \texttt{Cl} parameter. If the user prefers, an automatic option that corresponds to setting \texttt{Cl} to ``smaller", evaluates the distribution of the classes and determines which classes are candidates for being the smaller and most important. By default, this parameter is set to ``smaller" which means that the most relevant classes are automatically estimated from the data and correspond to those classes containing less than $\frac{\text{total number of examples}}{\text{number of classes}}$ examples. For instance, if a data set has 5 classes and a total number of examples of 100, the classes with less than 20 $(\frac{100}{5})$ examples will be considered the most important. The examples of the most relevant classes are then joined with one randomly selected example from each of the other classes. A 1-NN is computed with the distance metric provided by the user through the \texttt{dist} parameter. Then, all the examples from the original data set which were mislabeled in this procedure are also added to the reduced data set. This allows to obtain a smaller data set by removing examples from the larger and less important classes which are farther from the decision border.

This strategy is available through the \texttt{CNNClassif} function. This function returns a list containing: the modified data set, the classes that were considered important, and finally the unimportant classes.

We can now see some examples of this approach on the sythetic ImbC data and in the subset of iris data previously defined.

\begin{knitrout}\footnotesize
\definecolor{shadecolor}{rgb}{0.969, 0.969, 0.969}\color{fgcolor}\begin{kframe}
\begin{alltt}
\hlcom{# select a distance that is appropriate for dealing with}
\hlcom{# both nominal and numeric features}
\hlcom{# the default considers the two minority classes as the most important ones}
\hlstd{IHEOM} \hlkwb{<-} \hlkwd{CNNClassif}\hlstd{(Class}\hlopt{~}\hlstd{., ImbC,} \hlkwc{dist}\hlstd{=}\hlstr{"HEOM"}\hlstd{)}

\hlcom{# considering only rare1 class is important}
\hlstd{IHEOM1} \hlkwb{<-} \hlkwd{CNNClassif}\hlstd{(Class}\hlopt{~}\hlstd{., ImbC,} \hlkwc{dist}\hlstd{=}\hlstr{"HEOM"}\hlstd{,} \hlkwc{Cl}\hlstd{=}\hlstr{"rare1"}\hlstd{)}

\hlcom{# considering only rare2 class as important}
\hlstd{IHEOM2} \hlkwb{<-} \hlkwd{CNNClassif}\hlstd{(Class}\hlopt{~}\hlstd{., ImbC,} \hlkwc{dist}\hlstd{=}\hlstr{"HEOM"}\hlstd{,} \hlkwc{Cl}\hlstd{=}\hlstr{"rare2"}\hlstd{)}

\hlcom{# use HVDM distance and the default of conisdering }
\hlcom{# both minority classes as the most important}
\hlstd{IHVDM} \hlkwb{<-} \hlkwd{CNNClassif}\hlstd{(Class}\hlopt{~}\hlstd{., ImbC,} \hlkwc{dist}\hlstd{=}\hlstr{"HVDM"}\hlstd{)}

\hlcom{# now we select rare1 as the important class}
\hlstd{IHVDM1} \hlkwb{<-} \hlkwd{CNNClassif}\hlstd{(Class}\hlopt{~}\hlstd{., ImbC,} \hlkwc{dist}\hlstd{=}\hlstr{"HVDM"}\hlstd{,} \hlkwc{Cl}\hlstd{=}\hlstr{"rare1"}\hlstd{)}

\hlcom{# this selects the class rare2 as the most important one}
\hlstd{IHVDM2} \hlkwb{<-} \hlkwd{CNNClassif}\hlstd{(Class}\hlopt{~}\hlstd{., ImbC,} \hlkwc{dist}\hlstd{=}\hlstr{"HVDM"}\hlstd{,} \hlkwc{Cl}\hlstd{=}\hlstr{"rare2"}\hlstd{)}
\end{alltt}
\end{kframe}
\end{knitrout}

\begin{table}[ht]
\centering
\begin{tabular}{rrrr}
  \hline
 & normal & rare1 & rare2 \\ 
  \hline
Original & 859 &  10 & 131 \\ 
  IHEOM & 777 &  10 & 131 \\ 
  IHEOM1 & 640 &  10 &  87 \\ 
  IHEOM2 & 698 &   5 & 131 \\ 
  IHVDM & 637 &  10 & 131 \\ 
  IHVDM1 & 535 &  10 &   1 \\ 
  IHVDM2 & 685 &   1 & 131 \\ 
   \hline
\end{tabular}
\caption{Number of examples in each class of IMbC pre-processed data sets for different CNN parameters.} 
\label{tab:ImbC_CNN_table}
\end{table}

Table~\ref{tab:ImbC_CNN_table} shows the number of examples that remain in each class of the pre-processed data sets. In this case, it is evident that both the distance function selected and the classes provided to be considered important have a significant impact on this strategy. 
Figures~\ref{fig:ImbC_CNN_HEOM} and \ref{fig:ImbC_CNN_HVDM} show the impact of the previously described strategies on ImbC data using HEOM and HVDM distances.

\begin{knitrout}\footnotesize
\definecolor{shadecolor}{rgb}{0.969, 0.969, 0.969}\color{fgcolor}\begin{figure}

{\centering \includegraphics[width=\maxwidth]{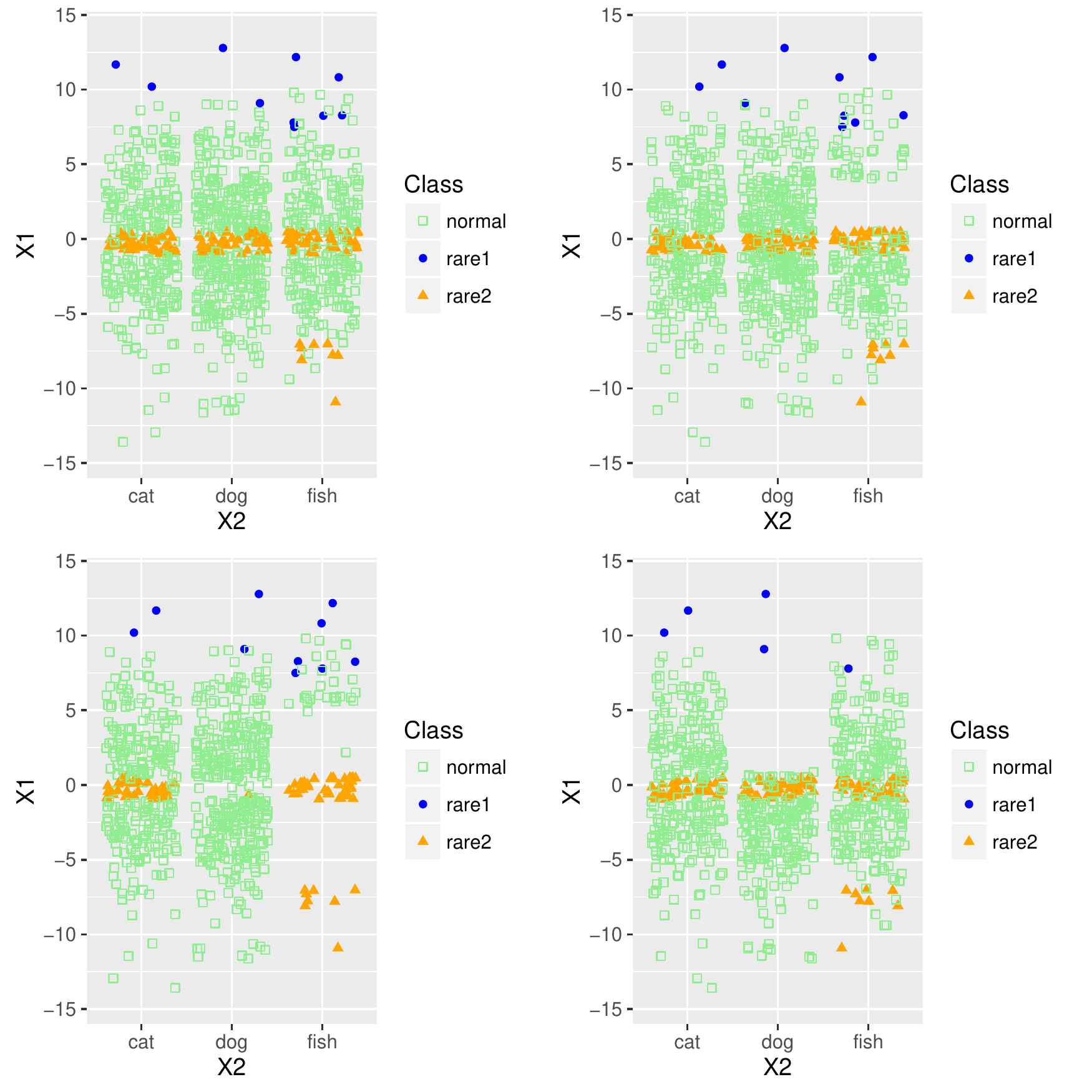} 

}

\caption[Impact on ImbC data of CNN method with HEOM distance for different values of parameter Cl]{Impact on ImbC data of CNN method with HEOM distance for different values of parameter Cl.}\label{fig:ImbC_CNN_HEOM}
\end{figure}

\end{knitrout}

\begin{knitrout}\footnotesize
\definecolor{shadecolor}{rgb}{0.969, 0.969, 0.969}\color{fgcolor}\begin{figure}

{\centering \includegraphics[width=\maxwidth]{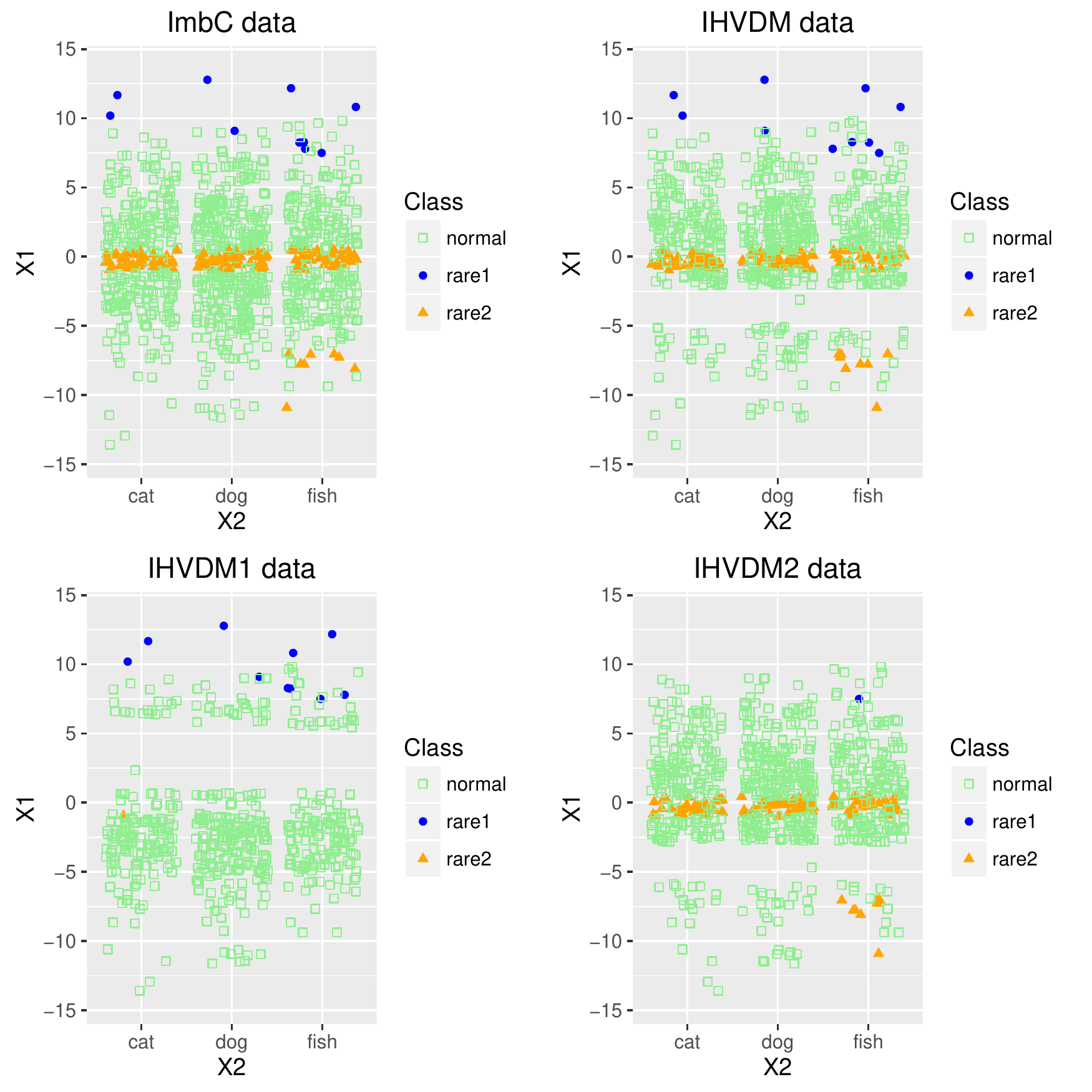} 

}

\caption[Impact on ImbC data of CNN method with HVDM distance for different values of parameter Cl]{Impact on ImbC data of CNN method with HVDM distance for different values of parameter Cl.}\label{fig:ImbC_CNN_HVDM}
\end{figure}

\end{knitrout}

Let us now consider the subset of iris data. In this case, the two features are numeric which allows the use of different distance functions.

\begin{knitrout}\footnotesize
\definecolor{shadecolor}{rgb}{0.969, 0.969, 0.969}\color{fgcolor}\begin{kframe}
\begin{alltt}
\hlcom{# just to remember the considered subset of iris data set}
  \hlkwd{summary}\hlstd{(dat)}
\end{alltt}
\begin{verbatim}
##   Sepal.Length    Sepal.Width          Species  
##  Min.   :4.300   Min.   :2.300   setosa    :50  
##  1st Qu.:5.000   1st Qu.:2.900   versicolor:25  
##  Median :5.350   Median :3.150   virginica :15  
##  Mean   :5.497   Mean   :3.170                  
##  3rd Qu.:5.800   3rd Qu.:3.475                  
##  Max.   :7.600   Max.   :4.400
\end{verbatim}
\begin{alltt}
\hlcom{# use of the default distance: Euclidean }
  \hlstd{myCNN} \hlkwb{<-} \hlkwd{CNNClassif}\hlstd{(Species}\hlopt{~}\hlstd{., dat,} \hlkwc{Cl}\hlstd{=}\hlkwd{c}\hlstd{(}\hlstr{"setosa"}\hlstd{,} \hlstr{"virginica"}\hlstd{))}
  \hlstd{CNN1} \hlkwb{<-} \hlkwd{CNNClassif}\hlstd{(Species}\hlopt{~}\hlstd{., dat,} \hlkwc{Cl}\hlstd{=}\hlstr{"smaller"}\hlstd{)}
\hlcom{# try other distance functions}
  \hlstd{CNN2} \hlkwb{<-} \hlkwd{CNNClassif}\hlstd{(Species}\hlopt{~}\hlstd{., dat,} \hlkwc{dist}\hlstd{=}\hlstr{"Chebyshev"}\hlstd{,} \hlkwc{Cl}\hlstd{=}\hlstr{"versicolor"}\hlstd{)}
  \hlstd{CNN3} \hlkwb{<-} \hlkwd{CNNClassif}\hlstd{(Species}\hlopt{~}\hlstd{., dat,} \hlkwc{dist}\hlstd{=}\hlstr{"HVDM"}\hlstd{,} \hlkwc{Cl}\hlstd{=}\hlstr{"virginica"}\hlstd{)}
  \hlstd{CNN4} \hlkwb{<-} \hlkwd{CNNClassif}\hlstd{(Species}\hlopt{~}\hlstd{., dat,} \hlkwc{dist}\hlstd{=}\hlstr{"p-norm"}\hlstd{,} \hlkwc{p}\hlstd{=}\hlnum{3}\hlstd{,} \hlkwc{Cl}\hlstd{=}\hlstr{"setosa"}\hlstd{)}

\hlcom{# check the new data set obtained in CNN1}
\hlkwd{summary}\hlstd{(CNN1[[}\hlnum{1}\hlstd{]]}\hlopt{$}\hlstd{Species)}
\end{alltt}
\begin{verbatim}
##     setosa versicolor  virginica 
##         17         25         15
\end{verbatim}
\begin{alltt}
\hlcom{# check the classes which were considered important}
\hlstd{CNN1[[}\hlnum{2}\hlstd{]]}
\end{alltt}
\begin{verbatim}
## [1] "versicolor" "virginica"
\end{verbatim}
\begin{alltt}
\hlcom{# check the classes which were considered unimportant}
\hlstd{CNN1[[}\hlnum{3}\hlstd{]]}
\end{alltt}
\begin{verbatim}
## [1] "setosa"
\end{verbatim}
\end{kframe}
\end{knitrout}

\begin{table}[ht]
\centering
\begin{tabular}{rrrr}
  \hline
 & setosa & versicolor & virginica \\ 
  \hline
Original &  50 &  25 &  15 \\ 
  myCNN &  50 &  24 &  15 \\ 
  CNN1 &  17 &  25 &  15 \\ 
  CNN2 &  36 &  25 &  15 \\ 
  CNN3 &  23 &  20 &  15 \\ 
   \hline
\end{tabular}
\caption{Number of examples in each class for different CNN parameters.} 
\label{tab:CNN_table}
\end{table}

It is clear from these examples that this method entails a significant reduction on the number of examples left in the modified data set. Moreover, since there is a random selection of points belonging to the less important class(es) the obtained data set may differ for different runs. Figure \ref{fig:CNN_plot} provides a visual illustration of the impact of this method in the previously considered subset of iris data.

\begin{knitrout}\footnotesize
\definecolor{shadecolor}{rgb}{0.969, 0.969, 0.969}\color{fgcolor}\begin{figure}

{\centering \includegraphics[width=\maxwidth]{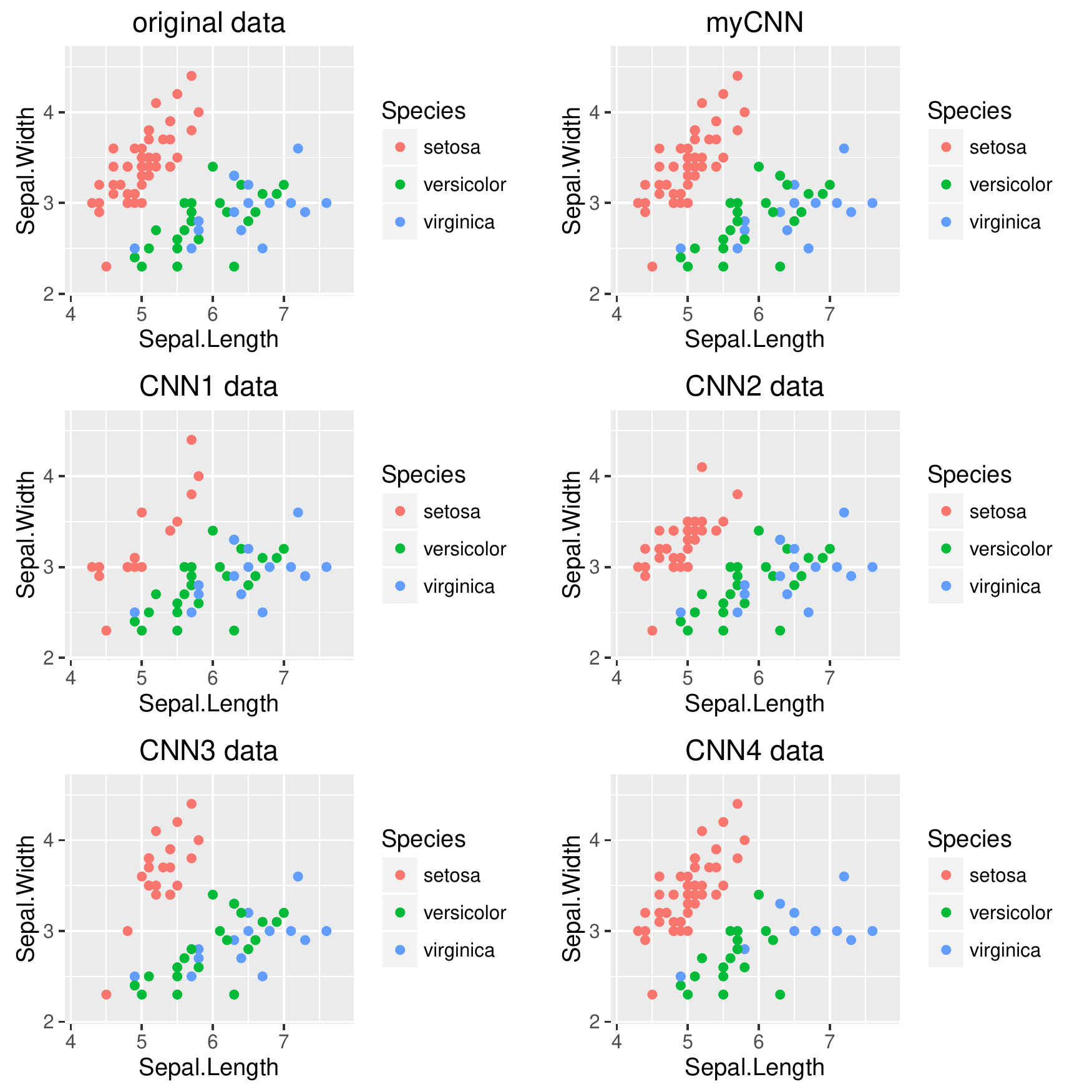} 

}

\caption[Impact of CNN method for different values of parameter Cl and different distance functions on the subset of iris data (top left]{Impact of CNN method for different values of parameter Cl and different distance functions on the subset of iris data (top left: original iris subset; top right:myCNN data; middle left: CNN1 data; middle right: CNN2 data; bottom left: CNN3 data; and bottom right: CNN4 data).}\label{fig:CNN_plot}
\end{figure}

\end{knitrout}

\subsection{One-sided Selection}\label{sec:OSS}

\cite{KM97} proposed a new method for modifying a given data set by applying the Tomek links under-sampling strategy and afterwards the CNN technique. \cite{batista2004study} also tested the reverse order for applying the techniques: first apply CNN method and then Tomek links. The main motivation for this was to apply Tomek links to an already reduced data set because Tomek links technique is a more computationally demanding task.

In \UBL we have gathered under the same function, \texttt{OSSClassif}, both techniques. To distinguish between the two methods, we included a parameter \texttt{start} which defaults to ``CNN". The user may therefore select the order in which we want to apply the two techniques: CNN and Tomek links. In this implementation, when Tomek links are applied, they always imply the removal of both examples forming the Tomek link. 

We have adapted both methods for dealing with multiclass imbalanced problems. To do so, we have included the parameter \texttt{Cl} which allows the user to specify the most important classes. Similarly to the behavior of CNN strategy, the user may define for the \texttt{Cl} parameter the value ``smaller". In this case, the most important classes are automatically determined using the same method presented in CNN strategy. When the relevant classes are chosen with this automatic method, the less frequent classes (which are considered the most relevant ones) are those which have a frequency below $\frac{number of examples}{number of classes}$. This means that all the classes with a frequency below the mean frequency of the data set classes is considered a minority class. The \texttt{OSSClassif} function also allows to specify which distance metric should be used in the neighbors computation. For more details on the available distance functions see Section \ref{sec:distFunc}. We must also mention that this strategy may potentially produce warnings due to the use of Tomek links strategy. As previously mentioned when Tomek links approach was presented, this method may not change the provided data set. In this case a warning is issued to advert the user. This warning may also occur when using OSS strategy if the Tomek links method produce it.

Let us observe how this method can be used with ImbC data.

\begin{knitrout}\footnotesize
\definecolor{shadecolor}{rgb}{0.969, 0.969, 0.969}\color{fgcolor}\begin{kframe}
\begin{alltt}
\hlcom{# OSS method with HEOM distance}
\hlstd{HEOM1} \hlkwb{<-} \hlkwd{OSSClassif}\hlstd{(Class}\hlopt{~}\hlstd{., ImbC,} \hlkwc{dist}\hlstd{=}\hlstr{"HEOM"}\hlstd{)}
\hlstd{HEOM2} \hlkwb{<-} \hlkwd{OSSClassif}\hlstd{(Class}\hlopt{~}\hlstd{., ImbC,} \hlkwc{dist}\hlstd{=}\hlstr{"HEOM"}\hlstd{,} \hlkwc{start}\hlstd{=}\hlstr{"Tomek"}\hlstd{,} \hlkwc{Cl}\hlstd{=}\hlstr{"rare1"}\hlstd{)}

\hlcom{# OSS method with HVDM distance}
\hlstd{HVDM1} \hlkwb{<-} \hlkwd{OSSClassif}\hlstd{(Class}\hlopt{~}\hlstd{., ImbC,} \hlkwc{dist}\hlstd{=}\hlstr{"HVDM"}\hlstd{,} \hlkwc{Cl}\hlstd{=}\hlstr{"rare1"}\hlstd{)}
\hlstd{HVDM2} \hlkwb{<-} \hlkwd{OSSClassif}\hlstd{(Class}\hlopt{~}\hlstd{., ImbC,} \hlkwc{dist}\hlstd{=}\hlstr{"HVDM"}\hlstd{,} \hlkwc{start}\hlstd{=}\hlstr{"Tomek"}\hlstd{,} \hlkwc{Cl}\hlstd{=}\hlstr{"rare2"}\hlstd{)}
\end{alltt}
\end{kframe}
\end{knitrout}

Table~\ref{tab:ImbC_oss_table} shows the impact of OSS strategy with different parameters on the number of examples in each class of the pre-processed data sets and Figure~\ref{fig:ImbC_oss} shows the examples distribution on the pre-processed data sets.

\begin{table}[ht]
\centering
\begin{tabular}{rrrr}
  \hline
 & normal & rare1 & rare2 \\ 
  \hline
Original & 859 &  10 & 131 \\ 
  HEOM1 & 777 &  10 & 131 \\ 
  HEOM2 & 635 &  10 &  73 \\ 
  HVDM1 & 328 &  10 & 123 \\ 
  HVDM2 & 659 &   3 & 131 \\ 
   \hline
\end{tabular}
\caption{Number of examples in each class for different OSS parameters.} 
\label{tab:ImbC_oss_table}
\end{table}

\begin{knitrout}\footnotesize
\definecolor{shadecolor}{rgb}{0.969, 0.969, 0.969}\color{fgcolor}\begin{figure}

{\centering \includegraphics[width=\maxwidth]{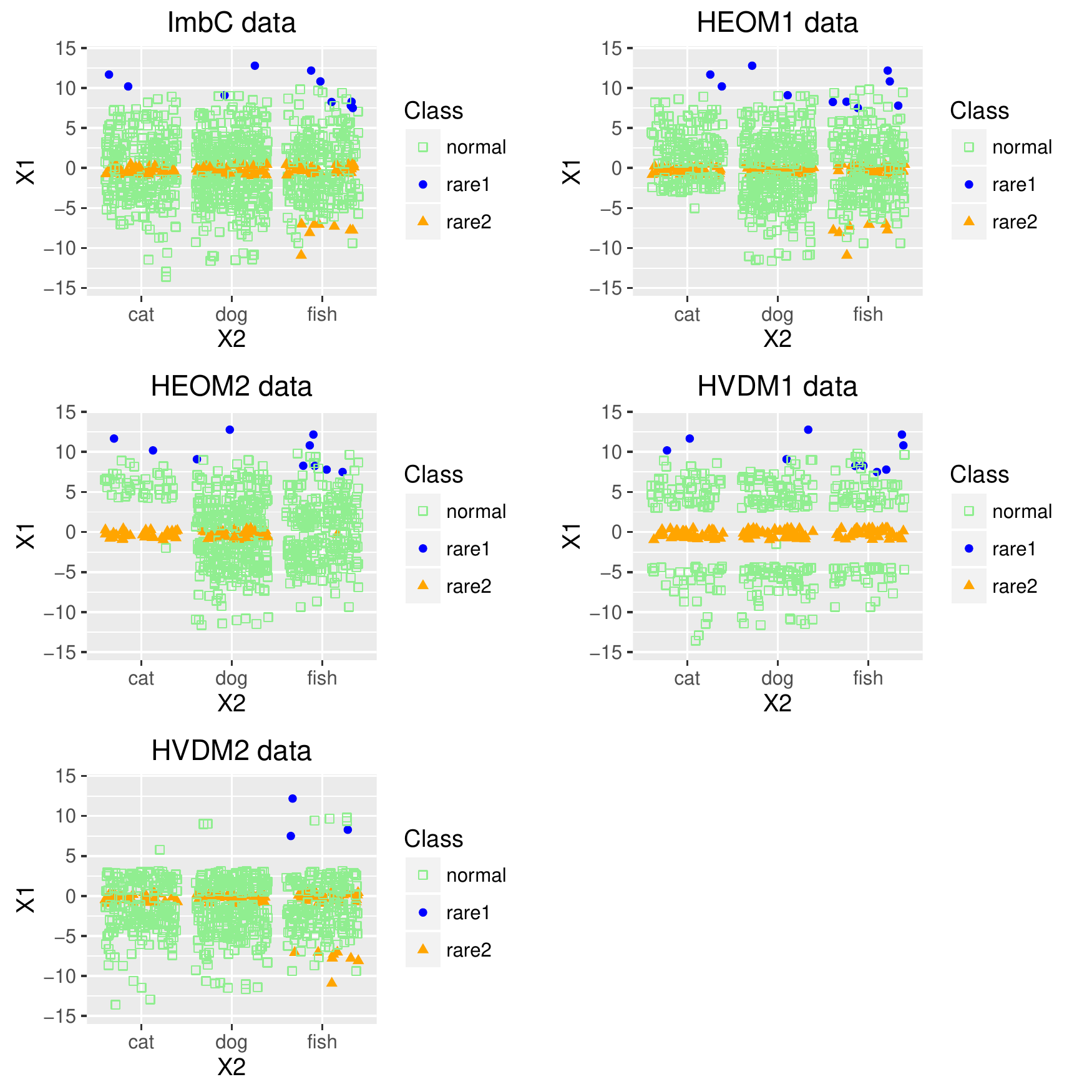} 

}

\caption[Impact on ImbC data of OSS method with different parameters]{Impact on ImbC data of OSS method with different parameters.}\label{fig:ImbC_oss}
\end{figure}

\end{knitrout}

The use of this method with data sets with all numeric features allows the use of several other distance functions. Let us briefly observe the impact of this method in the previously defined iris subset.

\begin{knitrout}\footnotesize
\definecolor{shadecolor}{rgb}{0.969, 0.969, 0.969}\color{fgcolor}\begin{kframe}
\begin{alltt}
\hlkwd{set.seed}\hlstd{(}\hlnum{1234}\hlstd{)}

\hlcom{# using all the defaults}
\hlstd{ir1} \hlkwb{<-} \hlkwd{OSSClassif}\hlstd{(Species}\hlopt{~}\hlstd{., dat)}
\end{alltt}

{\ttfamily\noindent\color{warningcolor}{\#\# Warning: TomekClassif found no examples to remove!}}\begin{alltt}
\hlcom{# using distance functions only suitable for numeric features}
\hlstd{ir2} \hlkwb{<-} \hlkwd{OSSClassif}\hlstd{(Species}\hlopt{~}\hlstd{., dat,} \hlkwc{dist}\hlstd{=}\hlstr{"p-norm"}\hlstd{,} \hlkwc{p}\hlstd{=}\hlnum{3}\hlstd{,}
                  \hlkwc{Cl}\hlstd{=}\hlstr{"virginica"}\hlstd{)}
\hlstd{ir3} \hlkwb{<-} \hlkwd{OSSClassif}\hlstd{(Species}\hlopt{~}\hlstd{., dat,} \hlkwc{dist}\hlstd{=}\hlstr{"Chebyshev"}\hlstd{,}
                  \hlkwc{Cl}\hlstd{=}\hlkwd{c}\hlstd{(}\hlstr{"versicolor"}\hlstd{,} \hlstr{"virginica"}\hlstd{),} \hlkwc{start}\hlstd{=}\hlstr{"Tomek"}\hlstd{)}
\end{alltt}

{\ttfamily\noindent\color{warningcolor}{\#\# Warning: TomekClassif found no examples to remove!}}\begin{alltt}
\hlkwd{summary}\hlstd{(ir1}\hlopt{$}\hlstd{Species)}
\end{alltt}
\begin{verbatim}
##     setosa versicolor  virginica 
##         23         25         15
\end{verbatim}
\begin{alltt}
\hlkwd{summary}\hlstd{(ir2}\hlopt{$}\hlstd{Species)}
\end{alltt}
\begin{verbatim}
##     setosa versicolor  virginica 
##         13         15         15
\end{verbatim}
\begin{alltt}
\hlkwd{summary}\hlstd{(ir3}\hlopt{$}\hlstd{Species)}
\end{alltt}
\begin{verbatim}
##     setosa versicolor  virginica 
##         22         25         15
\end{verbatim}
\end{kframe}
\end{knitrout}

The results obtained with the variants of OSS method on iris subset can be visualized in Figure \ref{fig:OSS_plot}.

\begin{knitrout}\footnotesize
\definecolor{shadecolor}{rgb}{0.969, 0.969, 0.969}\color{fgcolor}\begin{figure}

{\centering \includegraphics[width=\maxwidth]{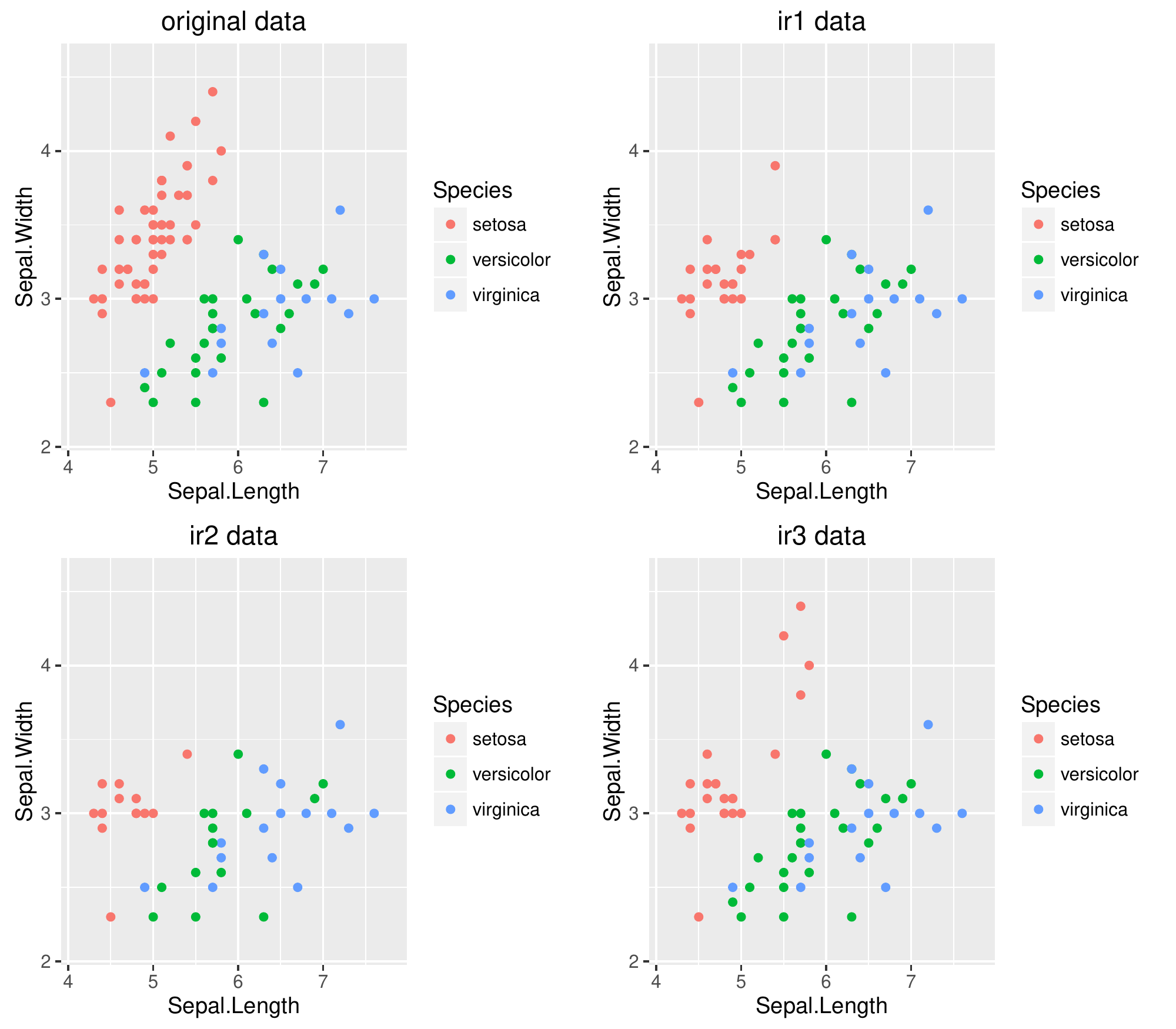} 

}

\caption[OSS technique with different parameters applied to the imbalanced iris subset]{OSS technique with different parameters applied to the imbalanced iris subset.}\label{fig:OSS_plot}
\end{figure}

\end{knitrout}

\subsection{Edited Nearest Neighbors}\label{sec:ENN}

The Edited Nearest Neighbor (ENN) algorithm was proposed by \cite{wilson1972asymptotic}. This method falls within the under-sampling approaches and has been used to address imbalanced classification problems. The original ENN algorithm uses a 3-NN classifier to remove the examples whose class is different from the class of at least two of its neighbors. 

We have implemented this approach for being able to tackle multiclass problems, allowing the user to specify through the \texttt{Cl} parameter a subset of classes which should be under-sampled. Moreover, in our implementation, the user may also define the number of nearest neighbors that should be considered by the algorithm. This means that an example is removed if its class label is different from the class label of at least half of its k-nearest neighbors and if it belongs to the subset of classes candidates for removal. The ENN algorithm is available in \UBL through the function \texttt{ENNClassif}. The number of neighbors to consider is set through the parameter \texttt{k} and the subset of classes that are candidates for being under-sampled are defined through the \texttt{Cl} parameter. The default of \texttt{Cl} is ``all", meaning that all classes are candidates for having examples removed. The user can also specify which distance metric he wants to use in the nearest neighbors computation. The function \texttt{ENNClassif} returns a list containing the new under-sampled data set and the indexes of the examples removed.

It is possible that ENN finds no examples to remove, which means that, for the parameters selected, there are no examples satisfying the necessary conditions to be removed. In this case, a warning is issued with the goal of adverting the user that the strategy is not modifying the data set provided.

We can use this strategy in ImbC data as follows:

\begin{knitrout}\footnotesize
\definecolor{shadecolor}{rgb}{0.969, 0.969, 0.969}\color{fgcolor}\begin{kframe}
\begin{alltt}
\hlcom{# use of default parameters except for the distance function}
\hlstd{ENN1} \hlkwb{<-} \hlkwd{ENNClassif}\hlstd{(Class}\hlopt{~}\hlstd{., ImbC,} \hlkwc{dist}\hlstd{=}\hlstr{"HVDM"}\hlstd{)}

\hlstd{ENN2} \hlkwb{<-} \hlkwd{ENNClassif}\hlstd{(Class}\hlopt{~}\hlstd{., ImbC,} \hlkwc{dist}\hlstd{=}\hlstr{"HVDM"}\hlstd{,} \hlkwc{Cl}\hlstd{=}\hlstr{"rare1"}\hlstd{)}

\hlstd{ENN3} \hlkwb{<-} \hlkwd{ENNClassif}\hlstd{(Class}\hlopt{~}\hlstd{., ImbC,} \hlkwc{dist}\hlstd{=}\hlstr{"HVDM"}\hlstd{,} \hlkwc{Cl}\hlstd{=}\hlstr{"rare2"}\hlstd{)}

\hlcom{# now using the HEOM distance}
\hlstd{ENN4} \hlkwb{<-} \hlkwd{ENNClassif}\hlstd{(Class}\hlopt{~}\hlstd{., ImbC,} \hlkwc{dist}\hlstd{=}\hlstr{"HEOM"}\hlstd{)}

\hlcom{# vary the number of neighbors considered by this method}
\hlstd{ENN5} \hlkwb{<-} \hlkwd{ENNClassif}\hlstd{(Class}\hlopt{~}\hlstd{., ImbC,} \hlkwc{k}\hlstd{=}\hlnum{5}\hlstd{,} \hlkwc{dist}\hlstd{=}\hlstr{"HEOM"}\hlstd{)}

\hlstd{ENN6} \hlkwb{<-} \hlkwd{ENNClassif}\hlstd{(Class}\hlopt{~}\hlstd{., ImbC,} \hlkwc{k}\hlstd{=}\hlnum{1}\hlstd{,} \hlkwc{dist}\hlstd{=}\hlstr{"HEOM"}\hlstd{)}
\end{alltt}
\end{kframe}
\end{knitrout}

Table~\ref{tab:ImbC_ENN_table} shows the impact of ENN strategy on ImbC synthetic data set with different parameters and Figure~\ref{fig:ImbC_ENN_plot} illustrates the examples distribution in the original and changed data sets.

\begin{table}[ht]
\centering
\begin{tabular}{rrrr}
  \hline
 & normal & rare1 & rare2 \\ 
  \hline
Original & 859 &  10 & 131 \\ 
  ENN1 & 826 &   4 & 112 \\ 
  ENN2 & 859 &   4 & 131 \\ 
  ENN3 & 859 &  10 & 112 \\ 
  ENN4 & 829 &   1 & 114 \\ 
  ENN5 & 823 &   1 & 119 \\ 
  ENN6 & 834 &   5 & 106 \\ 
   \hline
\end{tabular}
\caption{Number of examples in each class for different parameters of ENN strategyapplied in ImbC data.} 
\label{tab:ImbC_ENN_table}
\end{table}

\begin{knitrout}\footnotesize
\definecolor{shadecolor}{rgb}{0.969, 0.969, 0.969}\color{fgcolor}\begin{figure}

{\centering \includegraphics[width=\maxwidth]{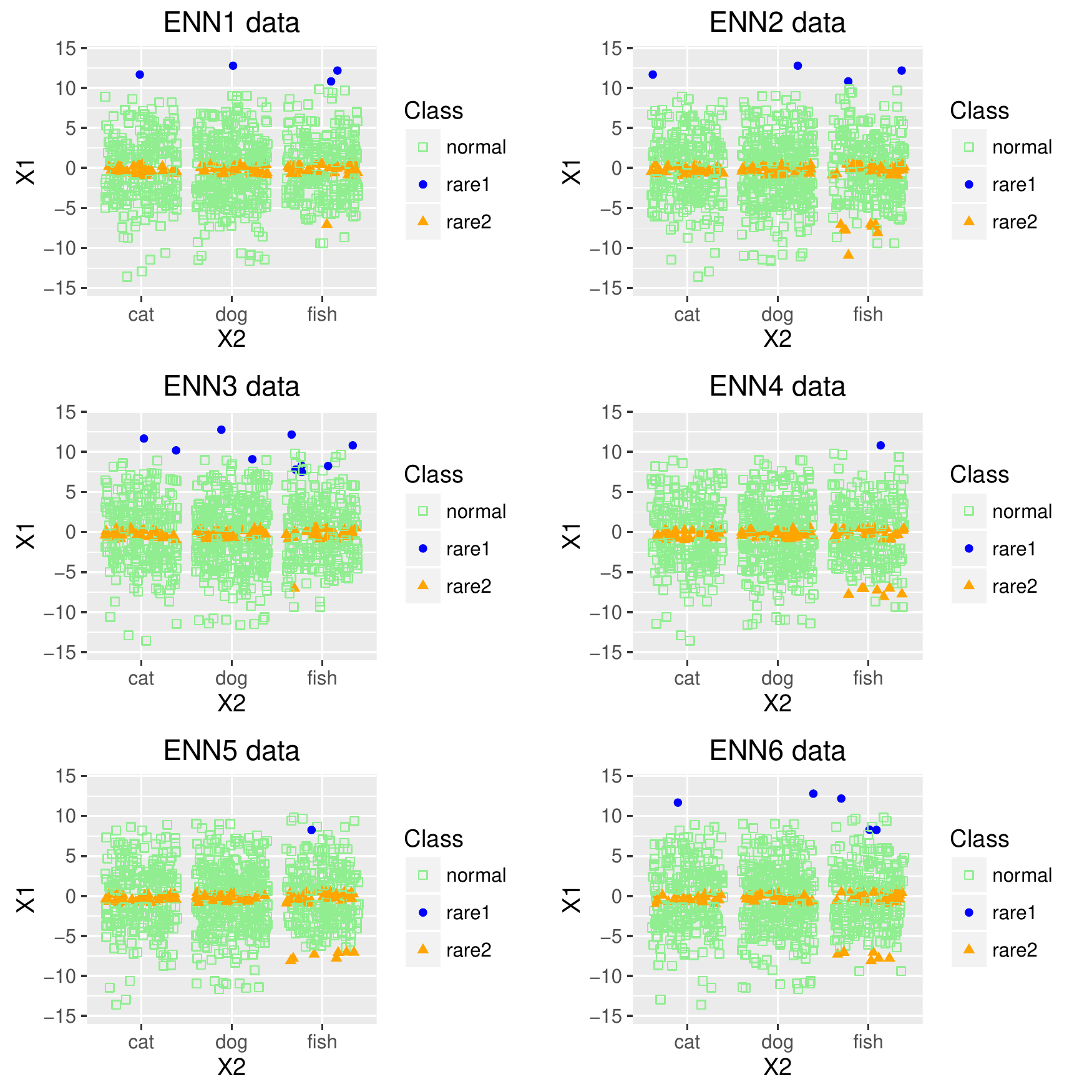} 

}

\caption[Impact on ImbC data of ENN method with different parameters]{Impact on ImbC data of ENN method with different parameters.}\label{fig:ImbC_ENN_plot}
\end{figure}

\end{knitrout}

We can also use this strategy on the imbalanced iris subset previously defined. In this case, given that the data contains only numeric features it is possible to use different distance functions.

\begin{knitrout}\footnotesize
\definecolor{shadecolor}{rgb}{0.969, 0.969, 0.969}\color{fgcolor}\begin{kframe}
\begin{alltt}
  \hlkwd{set.seed}\hlstd{(}\hlnum{123}\hlstd{)}
  \hlstd{Man5} \hlkwb{<-} \hlkwd{ENNClassif}\hlstd{(Species}\hlopt{~}\hlstd{., dat,} \hlkwc{k}\hlstd{=}\hlnum{5}\hlstd{,} \hlkwc{dist}\hlstd{=}\hlstr{"Manhattan"}\hlstd{,} \hlkwc{Cl}\hlstd{=}\hlstr{"all"}\hlstd{)}
  \hlstd{Default} \hlkwb{<-} \hlkwd{ENNClassif}\hlstd{(Species}\hlopt{~}\hlstd{., dat)}
  \hlstd{ChebSub7} \hlkwb{<-} \hlkwd{ENNClassif}\hlstd{(Species}\hlopt{~}\hlstd{., dat,} \hlkwc{k}\hlstd{=}\hlnum{7}\hlstd{,} \hlkwc{dist}\hlstd{=}\hlstr{"Chebyshev"}\hlstd{,}
                         \hlkwc{Cl}\hlstd{=}\hlkwd{c}\hlstd{(}\hlstr{"virginica"}\hlstd{,} \hlstr{"setosa"}\hlstd{))}
  \hlstd{ChebAll7} \hlkwb{<-} \hlkwd{ENNClassif}\hlstd{(Species}\hlopt{~}\hlstd{., dat,} \hlkwc{k}\hlstd{=}\hlnum{7}\hlstd{,} \hlkwc{dist}\hlstd{=}\hlstr{"Chebyshev"}\hlstd{)}
  \hlstd{HVDM3} \hlkwb{<-} \hlkwd{ENNClassif}\hlstd{(Species}\hlopt{~}\hlstd{., dat,} \hlkwc{k}\hlstd{=}\hlnum{3}\hlstd{,} \hlkwc{dist}\hlstd{=}\hlstr{"HVDM"}\hlstd{)}
\end{alltt}
\end{kframe}
\end{knitrout}

In Table \ref{tab:iris_ENN_table} we can observe the examples distributions for some parameters settings in ENN strategy and in Figure \ref{fig:ir_ENN_plot} we can visualize that distribution.

\begin{table}[ht]
\centering
\begin{tabular}{rrrr}
  \hline
 & setosa & versicolor & virginica \\ 
  \hline
Original &  50 &  25 &  15 \\ 
  Man5 &  49 &  14 &   3 \\ 
  Default &  49 &  16 &   2 \\ 
  ChebSub7 &  49 &  25 &   2 \\ 
  ChebAll7 &  49 &  19 &   2 \\ 
  HVDM3 &  49 &  18 &   3 \\ 
   \hline
\end{tabular}
\caption{Number of examples in each class for different parameters of ENN strategy.} 
\label{tab:iris_ENN_table}
\end{table}

\begin{knitrout}\footnotesize
\definecolor{shadecolor}{rgb}{0.969, 0.969, 0.969}\color{fgcolor}\begin{figure}

{\centering \includegraphics[width=\maxwidth,height=0.5\textheight]{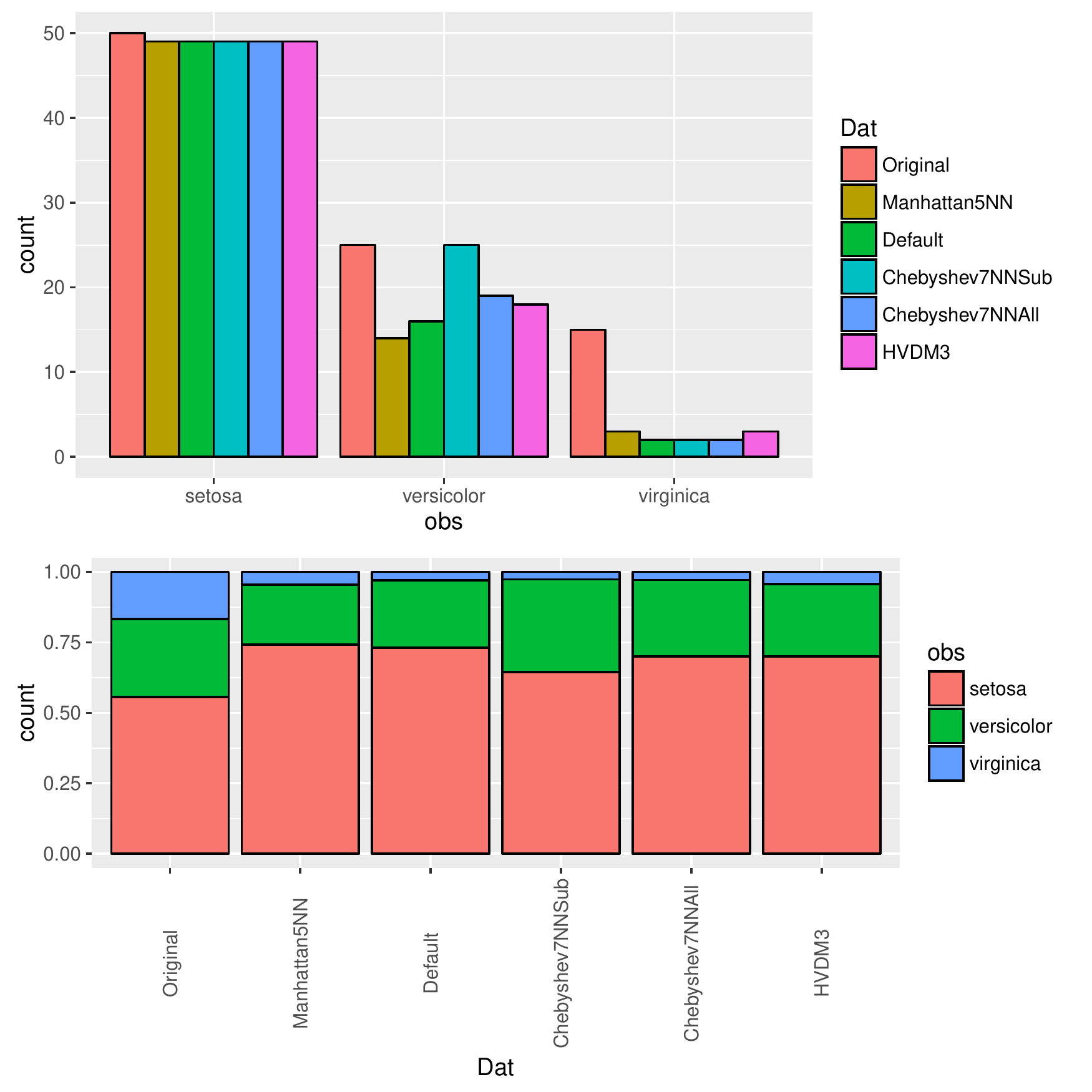} 

}

\caption[Impact in the subset of iris data of several parameters for ENN strategy]{Impact in the subset of iris data of several parameters for ENN strategy.}\label{fig:ir_ENN_plot}
\end{figure}

\end{knitrout}

This strategy has an unexpected behavior at first sight. In fact, the ENN method has further reduced the already minority classes. This can be explained by the goal of the ENN method which, being a cleaning technique, discards examples which may introduce errors no mater to which class they belong. As we know, in the iris data set the classes versicolor and virginica are the ones which are more difficult to classify. Therefore, the applied ENN strategy will try to remove examples exactly from those two classes.

Sometimes, ENN method is not capable of removing any example. When this happens, the original data set remains unchanged and an warning is issued. This warning is made only to advert the user that no examples were removed. On the other hand, with some data sets, this algorithm may completely remove one or more classes. This behavior may jeopardize the use of standard learning algorithms because they are provided with data set with only one class in the target variable. To overcome this issue, when a class is completely removed with the ENN strategy we randomly chose one example of that class to add to the under-sampled data set.

\subsection{Neighborhood Cleaning Rule}\label{sec:NCL}

The Neighborhood Cleaning Rule (NCL) algorithm was proposed in \cite{laurikkala2001improving}. This approach starts by splitting the data set $D$ in two: a subset $C$ with the examples belonging to the most important (an usually less frequent) class(es) and another subset $O$ containing the examples from the less important class(es). A new set $A_1$ of examples is formed with the noisy examples belonging to the subset $O$ which are identified using the ENN method.
Then, another set $A_2$ of examples is built as follows. For each class $C_i$ in $O$, the k nearest neighbors of each example in $C_i$ are scanned. The example is included in $A_2$ if all the scanned k nearest neighbors have a class label not contained in $C$ and if the example belongs to a class which has a cardinal of at least $\frac{1}{2}$ of the cardinal of smaller class in $C$. This last constraint forces the algorithm to keep the examples of classes with to few examples.
Finally, the examples in $A_1$ and $A_2$ are removed from the original data set.

Since this strategy internally uses the ENN approach we highlight that it is possible that warnings are issued. As mentioned before, the user is always adverted if ENN does not alter the data set. This can also happen with NCL if internally the ENN does not remove any example.

The NCL approach is available in \UBL through the \texttt{NCLClassif} function. In addition to providing a formula describing the prediction problem (\texttt{form}) and a data set (\texttt{dat}) the user may set the parameters corresponding to the number of neighbors considered (\texttt{k}), the distance function used (\texttt{dist}) and the classes that should be under-sampled (\texttt{Cl}). This last parameter may be set to \texttt{smaller}. In this case, the smaller classes are automatically estimated, and assumed to be the most important ones. All the other least important classes are candidates for the under-sampling of NCL method to be applied. We now provide some examples of application of the NCL method.

We will begin using the ImbC data set. As this data contains numeric and nominal features it is necessary to use suitable distance functions, such as "HEOM" or "HVDM".

\begin{knitrout}\footnotesize
\definecolor{shadecolor}{rgb}{0.969, 0.969, 0.969}\color{fgcolor}\begin{kframe}
\begin{alltt}
\hlstd{IHEOM1} \hlkwb{<-} \hlkwd{NCLClassif}\hlstd{(Class}\hlopt{~}\hlstd{., ImbC,} \hlkwc{k}\hlstd{=}\hlnum{10}\hlstd{,} \hlkwc{dist}\hlstd{=}\hlstr{"HEOM"}\hlstd{,} \hlkwc{Cl}\hlstd{=}\hlstr{"smaller"}\hlstd{)}
\hlstd{IHEOM2} \hlkwb{<-} \hlkwd{NCLClassif}\hlstd{(Class}\hlopt{~}\hlstd{., ImbC,} \hlkwc{k}\hlstd{=}\hlnum{1}\hlstd{,} \hlkwc{dist}\hlstd{=}\hlstr{"HEOM"}\hlstd{)}
\hlstd{IHEOM3} \hlkwb{<-} \hlkwd{NCLClassif}\hlstd{(Class}\hlopt{~}\hlstd{., ImbC,} \hlkwc{k}\hlstd{=}\hlnum{1}\hlstd{,} \hlkwc{dist}\hlstd{=}\hlstr{"HEOM"}\hlstd{,} \hlkwc{Cl}\hlstd{=}\hlstr{"rare1"}\hlstd{)}

\hlstd{IHVDM1}\hlkwb{<-} \hlkwd{NCLClassif}\hlstd{(Class}\hlopt{~}\hlstd{., ImbC,} \hlkwc{k}\hlstd{=}\hlnum{10}\hlstd{,} \hlkwc{dist}\hlstd{=}\hlstr{"HVDM"}\hlstd{,} \hlkwc{Cl}\hlstd{=}\hlstr{"smaller"}\hlstd{)}
\hlstd{IHVDM2}\hlkwb{<-} \hlkwd{NCLClassif}\hlstd{(Class}\hlopt{~}\hlstd{., ImbC,} \hlkwc{k}\hlstd{=}\hlnum{5}\hlstd{,} \hlkwc{dist}\hlstd{=}\hlstr{"HVDM"}\hlstd{,} \hlkwc{Cl}\hlstd{=}\hlstr{"rare1"}\hlstd{)}
\hlstd{IHVDM3}\hlkwb{<-} \hlkwd{NCLClassif}\hlstd{(Class}\hlopt{~}\hlstd{., ImbC,} \hlkwc{k}\hlstd{=}\hlnum{1}\hlstd{,} \hlkwc{dist}\hlstd{=}\hlstr{"HVDM"}\hlstd{,} \hlkwc{Cl}\hlstd{=}\hlstr{"rare2"}\hlstd{)}
\end{alltt}
\end{kframe}
\end{knitrout}

Table~\ref{tab:ImbC_NCL_table} summarizes the impact produced in the number of examples in the classes on the new data sets chenaged through NCL strategy and Figure~\ref{fig:ImbC_NCL_plot} show the examples distributions in these data sets.

\begin{table}[ht]
\centering
\begin{tabular}{rrrr}
  \hline
 & normal & rare1 & rare2 \\ 
  \hline
Original & 859 &  10 & 131 \\ 
  IHEOM1 & 849 &  10 & 131 \\ 
  IHEOM2 & 835 &  10 & 131 \\ 
  IHEOM3 & 831 &  10 & 106 \\ 
  IHVDM1 & 839 &  10 & 131 \\ 
  IHVDM2 & 818 &  10 & 116 \\ 
  IHVDM3 & 828 &   5 & 131 \\ 
   \hline
\end{tabular}
\caption{Number of examples in each class for different parameters of NCL strategyon ImbC data set.} 
\label{tab:ImbC_NCL_table}
\end{table}

\begin{knitrout}\footnotesize
\definecolor{shadecolor}{rgb}{0.969, 0.969, 0.969}\color{fgcolor}\begin{figure}

{\centering \includegraphics[width=\maxwidth]{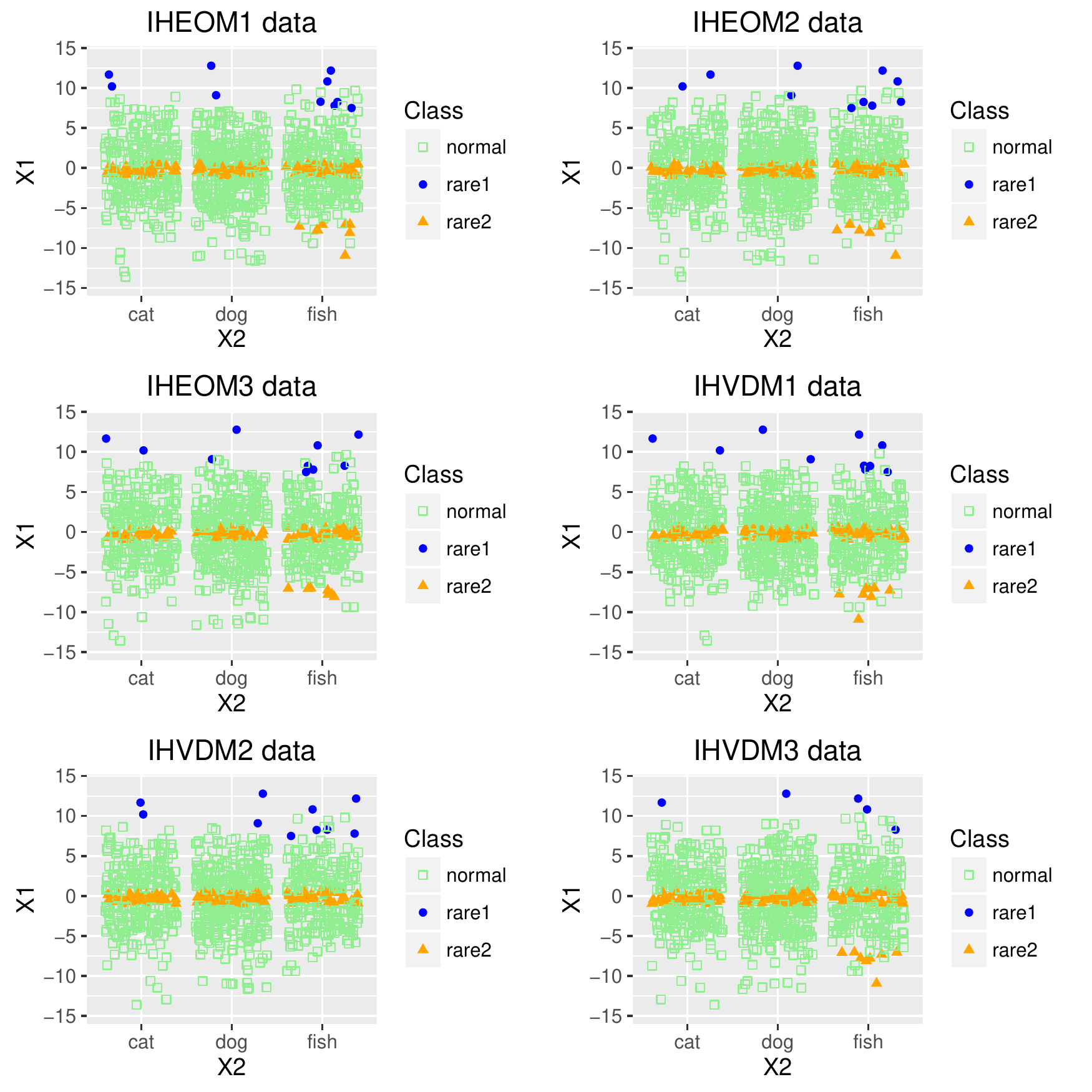} 

}

\caption[Impact on ImbC data of NCL method with different parameters]{Impact on ImbC data of NCL method with different parameters.}\label{fig:ImbC_NCL_plot}
\end{figure}

\end{knitrout}

Let us now observe how this technique can be used in the imbalanced iris subset previously defined.

\begin{knitrout}\footnotesize
\definecolor{shadecolor}{rgb}{0.969, 0.969, 0.969}\color{fgcolor}\begin{kframe}
\begin{alltt}
\hlkwd{set.seed}\hlstd{(}\hlnum{1234}\hlstd{)}
\hlstd{ir.M1} \hlkwb{<-} \hlkwd{NCLClassif}\hlstd{(Species}\hlopt{~}\hlstd{., dat,} \hlkwc{k}\hlstd{=}\hlnum{3}\hlstd{,} \hlkwc{dist}\hlstd{=}\hlstr{"p-norm"}\hlstd{,} \hlkwc{p}\hlstd{=}\hlnum{1}\hlstd{,} \hlkwc{Cl}\hlstd{=}\hlstr{"smaller"}\hlstd{)}
\hlstd{ir.M2}\hlkwb{<-} \hlkwd{NCLClassif}\hlstd{(Species}\hlopt{~}\hlstd{., dat,} \hlkwc{k}\hlstd{=}\hlnum{1}\hlstd{,} \hlkwc{dist}\hlstd{=}\hlstr{"p-norm"}\hlstd{,} \hlkwc{p}\hlstd{=}\hlnum{1}\hlstd{,} \hlkwc{Cl}\hlstd{=}\hlstr{"setosa"}\hlstd{)}
\hlstd{ir.Def} \hlkwb{<-} \hlkwd{NCLClassif}\hlstd{(Species}\hlopt{~}\hlstd{., dat)}
\hlstd{ir.Ch} \hlkwb{<-} \hlkwd{NCLClassif}\hlstd{(Species}\hlopt{~}\hlstd{., dat,} \hlkwc{k}\hlstd{=}\hlnum{7}\hlstd{,} \hlkwc{dist}\hlstd{=}\hlstr{"Chebyshev"}\hlstd{,} \hlkwc{Cl}\hlstd{=}\hlstr{"virginica"}\hlstd{)}
\end{alltt}

{\ttfamily\noindent\color{warningcolor}{\#\# Warning: ENNClassif found no examples to remove!}}\begin{alltt}
\hlstd{ir.Eu} \hlkwb{<-} \hlkwd{NCLClassif}\hlstd{(Species}\hlopt{~}\hlstd{., dat,} \hlkwc{k}\hlstd{=}\hlnum{3}\hlstd{,} \hlkwc{dist}\hlstd{=}\hlstr{"Euclidean"}\hlstd{,}
                    \hlkwc{Cl}\hlstd{=}\hlkwd{c}\hlstd{(}\hlstr{"setosa"}\hlstd{,} \hlstr{"virginica"}\hlstd{))}
\end{alltt}
\end{kframe}
\end{knitrout}

Table \ref{tab:iris_NCL_table} provides the number of examples in each class for different parameters of NCL method and in Figure \ref{fig:NCL_plot} the changes produced by the use of this method may be visualized.

\begin{table}[ht]
\centering
\begin{tabular}{rrrr}
  \hline
 & setosa & versicolor & virginica \\ 
  \hline
Original &  50 &  25 &  15 \\ 
  ir.M1 &  50 &  25 &  15 \\ 
  ir.M2 &  50 &  16 &   3 \\ 
  ir.Def &  50 &  25 &  15 \\ 
  ir.Ch &  47 &  21 &  15 \\ 
  ir.Eu &  50 &  19 &  15 \\ 
   \hline
\end{tabular}
\caption{Number of examples in each class for different parameters of NCL strategy.} 
\label{tab:iris_NCL_table}
\end{table}

\begin{knitrout}\footnotesize
\definecolor{shadecolor}{rgb}{0.969, 0.969, 0.969}\color{fgcolor}\begin{figure}

{\centering \includegraphics[width=\maxwidth]{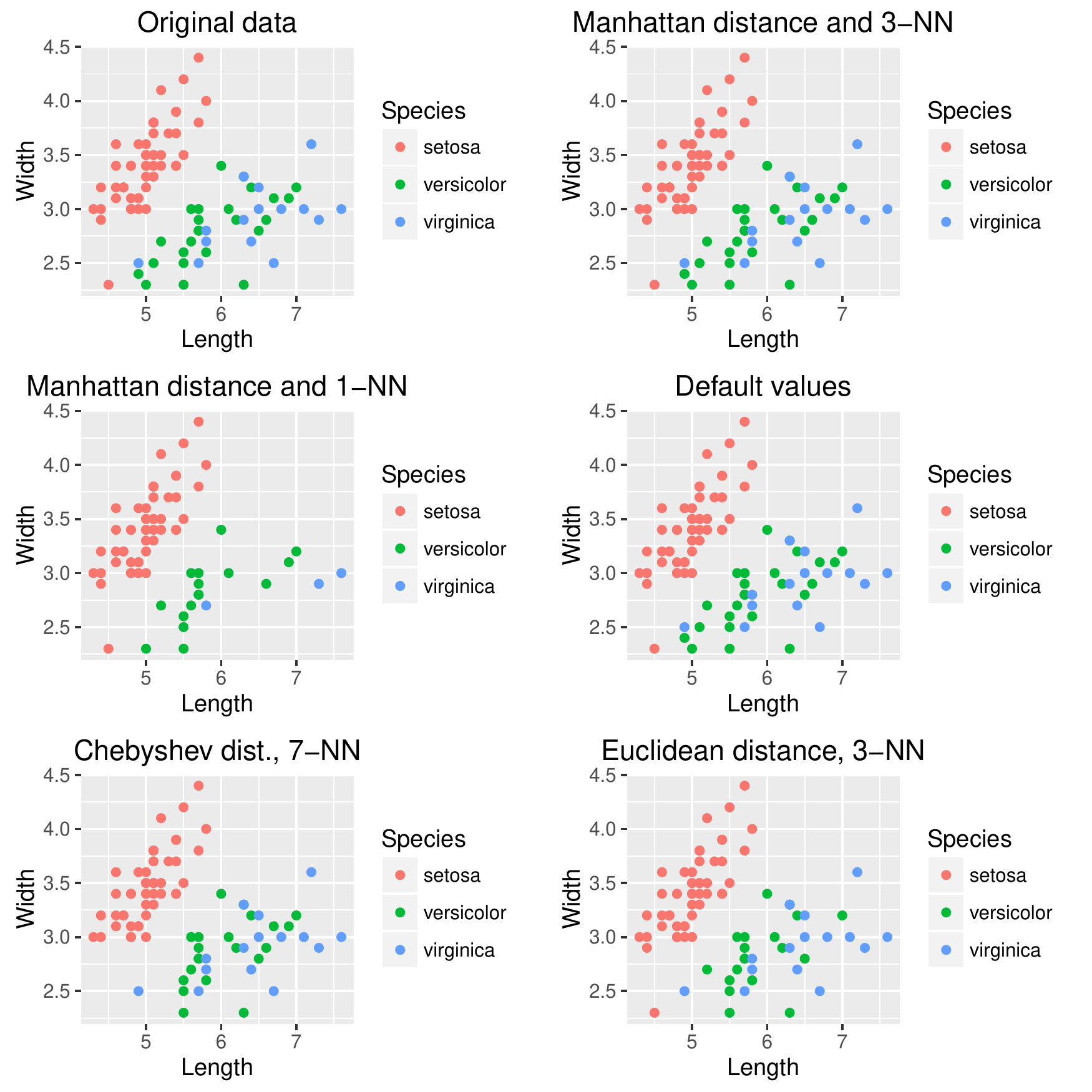} 

}

\caption[NCL techniques applied to a multiclass imbalanced problem]{NCL techniques applied to a multiclass imbalanced problem.}\label{fig:NCL_plot}
\end{figure}

\end{knitrout}

\subsection{Generation of synthetic examples by the introduction of Gaussian Noise}\label{sec:gnClassif}

The use of Gaussian Noise to introduce a small perturbation in the data set examples was proposed by \cite{lee1999regularization} and then extended in \cite{lee2000noisy}. The proposed method consisted of producing replicas of the examples of the minority class by introducing normally distributed noise. In this approach, the majority class remained unchanged while the minority class was increased. The noise introduced depends on a fraction of the standard deviation of each numeric feature.

We have adapted this technique to multiclass imbalanced problems. Moreover, we have also included the possibility of combining this over-sampling procedure with the random under-sampling technique described in Section \ref{sec:RUClassif}. 

Regarding the over-sampling method, a new example from an important class is obtained by perturbing each numeric feature according to a random value following a normally distributed percentage of its standard deviation (with the standard deviation evaluated on the examples of that class). This means that, for a given value of \texttt{pert} defined by the user, each feature value ($i$) of the new example ($new_i$) is built as follows: $new_i=ex_i+rnorm(0,sd(i)\times pert) $, where $ex_i$ represents the original example value for feature $i$, and $sd(i)$ represents the evaluated standard deviation for feature $i$ in the class under consideration. For nominal features, the new example selects a label with a probability directly proportional to the frequency of the existing labels(with the frequency evaluated on the examples of that class).

The user may express which are the most relevant and the less important classes of the data set through the parameter \texttt{C.perc}. With this parameter the user also indicates the percentages of under and over-sampling to apply in each class. If a class is not referred in this parameter it will remain unchanged. Moreover, this parameter can also be set to ``balance" or ``extreme", cases where the under and over-sampling percentages are automatically estimated to achieve a balanced data set or a data set with the frequencies of the classes inverted. The perturbation applied to the numeric features is set using the \texttt{pert} parameter. Finally, the user may also specify if, when performing the random under-sampling strategy, it is allowed to perform sampling with repetition or not.

We now present an example of the impact of applying this technique for different values of the parameters in IMbC data.

\begin{knitrout}\footnotesize
\definecolor{shadecolor}{rgb}{0.969, 0.969, 0.969}\color{fgcolor}\begin{kframe}
\begin{alltt}
\hlcom{# using the default parameters that balance the problem classes}
\hlstd{GN1} \hlkwb{<-}\hlkwd{GaussNoiseClassif}\hlstd{(Class}\hlopt{~}\hlstd{., ImbC)}

\hlcom{# increase the neighborhood radius for generating the new synthetic examples}
\hlcom{# the default is pert = 0.1}
\hlstd{GN2} \hlkwb{<-} \hlkwd{GaussNoiseClassif}\hlstd{(Class}\hlopt{~}\hlstd{., ImbC,} \hlkwc{pert} \hlstd{=} \hlnum{0.5}\hlstd{)}
\hlcom{# select the percentages to apply in each class}
\hlstd{GN3} \hlkwb{<-} \hlkwd{GaussNoiseClassif}\hlstd{(Class}\hlopt{~}\hlstd{., ImbC,}
                         \hlkwc{C.perc} \hlstd{=} \hlkwd{list}\hlstd{(}\hlstr{"normal"} \hlstd{=} \hlnum{0.5}\hlstd{,} \hlstr{"rare1"} \hlstd{=} \hlnum{10}\hlstd{,} \hlstr{"rare2"} \hlstd{=} \hlnum{3}\hlstd{))}

\hlcom{#select the re-sampling percentages and the perturbation radius }
\hlstd{GN4} \hlkwb{<-} \hlkwd{GaussNoiseClassif}\hlstd{(Class}\hlopt{~}\hlstd{., ImbC,}
                         \hlkwc{C.perc} \hlstd{=} \hlkwd{list}\hlstd{(}\hlstr{"normal"} \hlstd{=} \hlnum{0.3}\hlstd{,} \hlstr{"rare1"} \hlstd{=} \hlnum{5}\hlstd{,} \hlstr{"rare2"} \hlstd{=} \hlnum{2}\hlstd{),}
                         \hlkwc{pert} \hlstd{=} \hlnum{0.05}\hlstd{)}

\hlcom{# use the option the inverts the classes frequencies}
\hlstd{GN5} \hlkwb{<-} \hlkwd{GaussNoiseClassif}\hlstd{(Class}\hlopt{~}\hlstd{., ImbC,} \hlkwc{C.perc}\hlstd{=}\hlstr{"extreme"}\hlstd{)}
\end{alltt}
\end{kframe}
\end{knitrout}

Table \ref{tab:iris_GN_table} presents the impact on the number of examples for the considered parameters of this strategy. In Figure \ref{fig:ir_GN_plot} we can observe the number of examples on the changed data sets for the parameters considered and Figure \ref{fig:ir_GN_plot2} presents the distribution of examples for those parameters.

\begin{table}[ht]
\centering
\begin{tabular}{rrrr}
  \hline
 & normal & rare1 & rare2 \\ 
  \hline
Original & 859 &  10 & 131 \\ 
  GN1 & 333 & 332 & 332 \\ 
  GN2 & 333 & 332 & 332 \\ 
  GN3 & 429 & 100 & 393 \\ 
  GN4 & 257 &  50 & 262 \\ 
  GN5 &  11 & 919 &  70 \\ 
   \hline
\end{tabular}
\caption{Number of examples in each class for different parameters of Gaussian Noise strategy applied in ImbC data.} 
\label{tab:iris_GN_table}
\end{table}

\begin{knitrout}\footnotesize
\definecolor{shadecolor}{rgb}{0.969, 0.969, 0.969}\color{fgcolor}\begin{figure}

{\centering \includegraphics[width=\maxwidth,height=0.5\textheight]{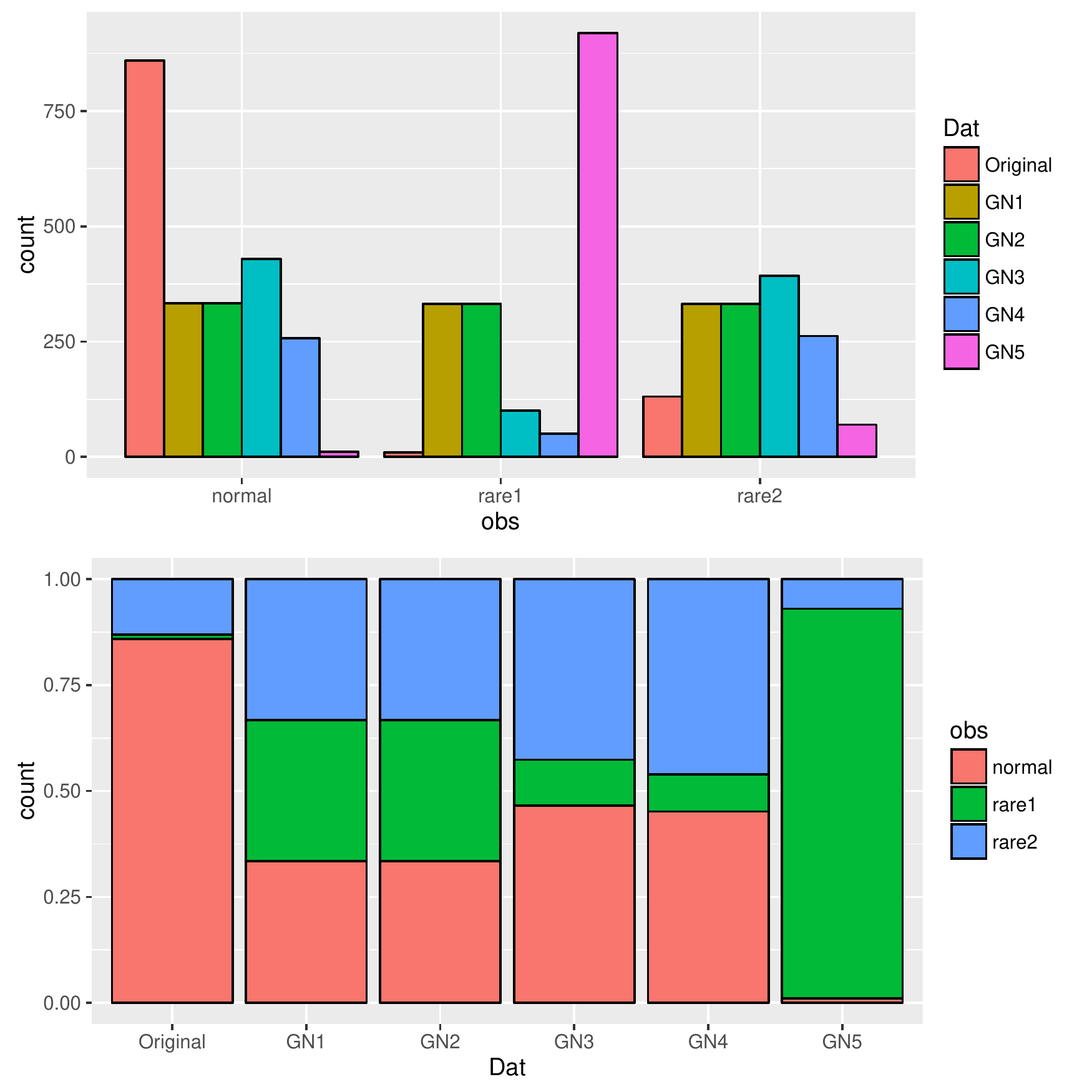} 

}

\caption[Impact in the Original ImbC data set of several parameters in Gaussian noise strategy]{Impact in the Original ImbC data set of several parameters in Gaussian noise strategy. }\label{fig:ir_GN_plot}
\end{figure}

\end{knitrout}

\begin{knitrout}\footnotesize
\definecolor{shadecolor}{rgb}{0.969, 0.969, 0.969}\color{fgcolor}\begin{figure}

{\centering \includegraphics[width=\maxwidth]{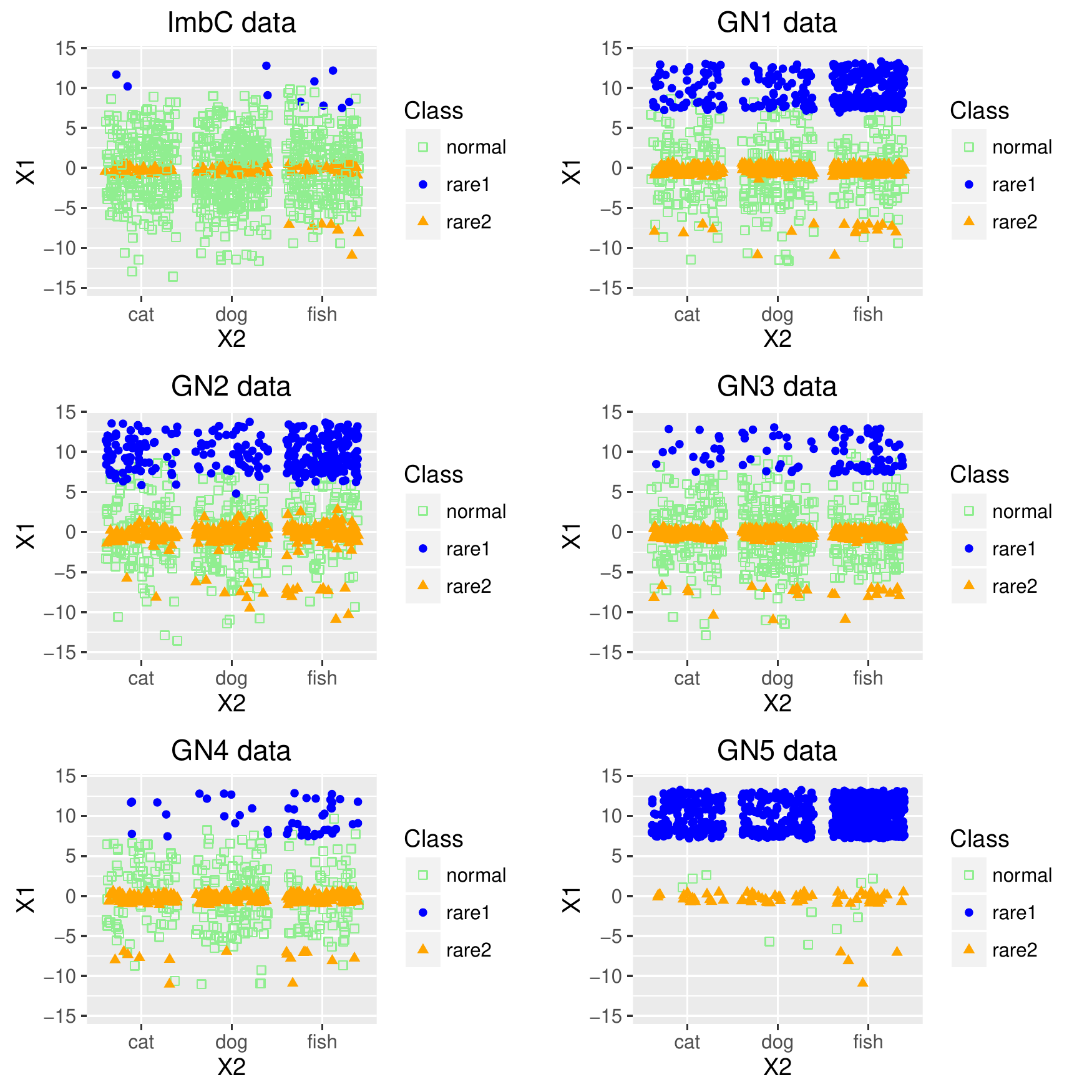} 

}

\caption[Impact on the examples distribution of ImbC data for different parameters in Gaussian Noise strategy]{Impact on the examples distribution of ImbC data for different parameters in Gaussian Noise strategy.}\label{fig:ir_GN_plot2}
\end{figure}

\end{knitrout}

\subsection{The Smote Algorithm}\label{sec:smoteClassif}

The well known Smote algorithm was proposed by \cite{CBOK02}. This algorithm presents a new strategy to address the problem of imbalanced domains through the generation of synthetic examples. The new synthetic cases are generated by interpolation of two cases from the minority (positive) class. To obtain a new example from the minority class, the algorithm uses a seed example from that class, and randomly selects one of its k nearest neighbors. Then, having the two examples, a new synthetic case is obtained by interpolating the examples features. This procedure is illustrated in Figure \ref{fig:smote_illust}.

\begin{knitrout}\footnotesize
\definecolor{shadecolor}{rgb}{0.969, 0.969, 0.969}\color{fgcolor}\begin{figure}

{\centering \includegraphics[width=0.6\textwidth,height=0.3\textheight]{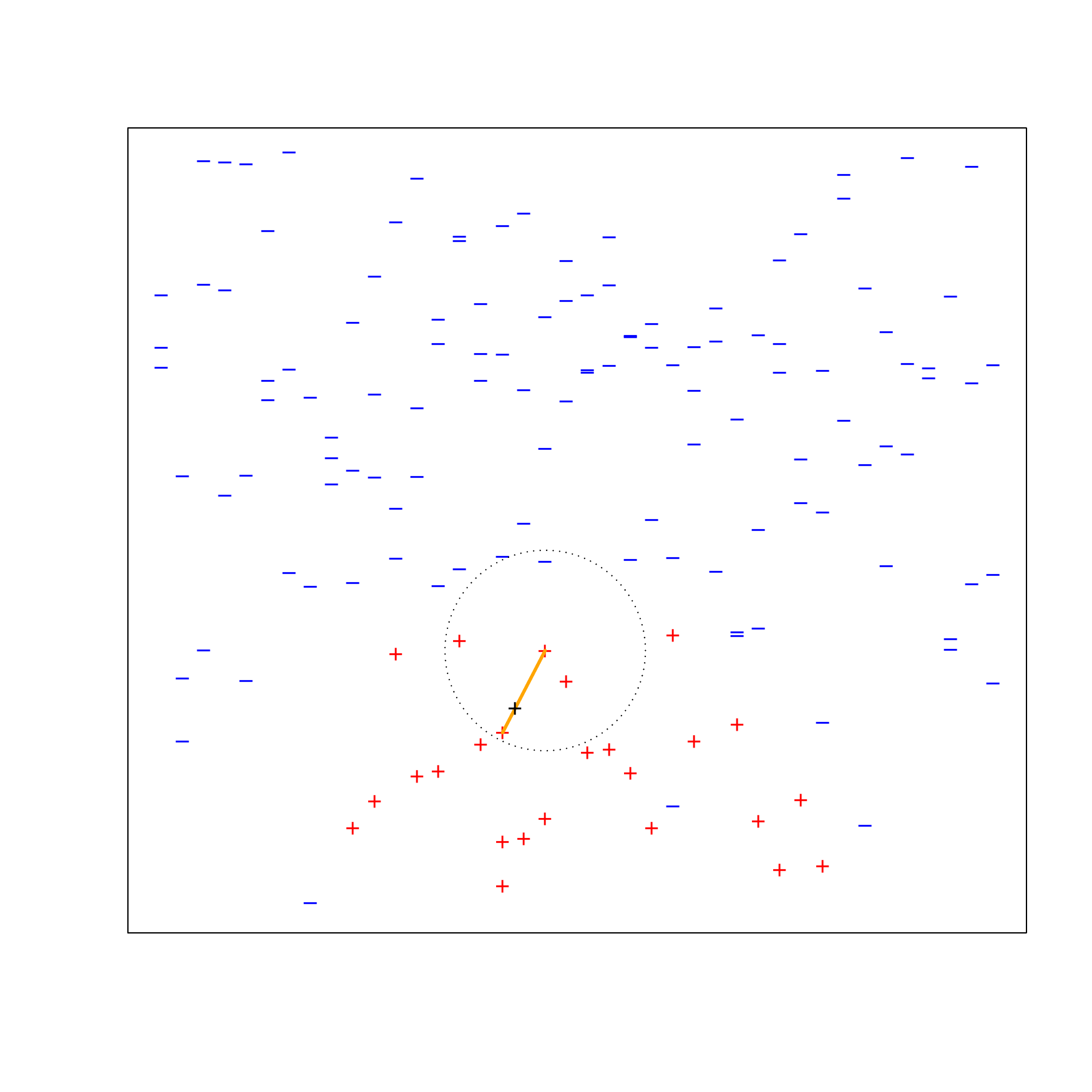} 

}

\caption[Generation of synthetic examples through Smote algorithm]{Generation of synthetic examples through Smote algorithm.\label{smote_illust}}\label{fig:smote_illust}
\end{figure}

\end{knitrout}

This over-sampling strategy was also combined with random under-sampling of the majority class in \cite{CBOK02}.

Our implementation of this method is available through the \texttt{SmoteClassif} function and is able to deal with multiclass tasks. The user can specify which are the most important and the less relevant classes using the \texttt{C.perc} parameter. Using the same parameter the user also expresses the percentages of over and under-sampling that should be applied to each class. When the data set includes nominal features, the interpolation of two examples for these features is solved by randomly selecting among the two values of the seed examples. Two automatic methods are provided for both the estimation of the relevant classes and the percentages of over and under-sampling to apply. These methods are available through the \texttt{C.perc} parameter which can be set to ``balance" or ``extreme". In both cases, it is ensured that the new obtained data set includes roughly the same number of examples as the original data set. When ``balance" or ``extreme" are chosen, both the minority/majority classes and the percentages of over/under-sampling are automatically estimated. The ``balance" option provides a balanced data set and the ``extreme" option provides a data set with the classes frequencies inverted, i.e., the most frequent classes in the original data set are the less frequent on the new data set and vice-versa.

Finally, the user may also express if the under-sampling process may include repetition of examples or not (using the \texttt{repl} parameter), may choose the number of nearest neighbors to use (parameter \texttt{k}) and can select the distance metric to be used in the nearest neighbors evaluation (parameter \texttt{dist}). 

The following example shows how this strategy can be used to modify the synthetic ImbC data set.

\begin{knitrout}\footnotesize
\definecolor{shadecolor}{rgb}{0.969, 0.969, 0.969}\color{fgcolor}\begin{kframe}
\begin{alltt}
\hlstd{IC1} \hlkwb{<-} \hlkwd{SmoteClassif}\hlstd{(Class}\hlopt{~}\hlstd{., ImbC,} \hlkwc{dist} \hlstd{=} \hlstr{"HEOM"}\hlstd{)}

\hlstd{IC2} \hlkwb{<-} \hlkwd{SmoteClassif}\hlstd{(Class}\hlopt{~}\hlstd{., ImbC,} \hlkwc{k} \hlstd{=} \hlnum{1}\hlstd{,} \hlkwc{dist}\hlstd{=}\hlstr{"HEOM"}\hlstd{)}

\hlstd{IC3} \hlkwb{<-} \hlkwd{SmoteClassif}\hlstd{(Class}\hlopt{~}\hlstd{., ImbC,}
                    \hlkwc{C.perc} \hlstd{=} \hlkwd{list}\hlstd{(}\hlstr{"normal"} \hlstd{=} \hlnum{0.4}\hlstd{,} \hlstr{"rare1"} \hlstd{=} \hlnum{8}\hlstd{,} \hlstr{"rare2"} \hlstd{=} \hlnum{6}\hlstd{),}
                    \hlkwc{dist} \hlstd{=} \hlstr{"HEOM"}\hlstd{)}

\hlstd{IC4} \hlkwb{<-} \hlkwd{SmoteClassif}\hlstd{(Class}\hlopt{~}\hlstd{., ImbC,} \hlkwc{dist} \hlstd{=} \hlstr{"HVDM"}\hlstd{)}

\hlstd{IC5} \hlkwb{<-} \hlkwd{SmoteClassif}\hlstd{(Class}\hlopt{~}\hlstd{., ImbC,} \hlkwc{k} \hlstd{=} \hlnum{1}\hlstd{,} \hlkwc{dist} \hlstd{=} \hlstr{"HVDM"}\hlstd{)}

\hlcom{# class rare2 is not refered in the C.perc parameter. This means that}
\hlcom{# this class will remain unchanged}
\hlstd{IC6} \hlkwb{<-} \hlkwd{SmoteClassif}\hlstd{(Class}\hlopt{~}\hlstd{., ImbC,} \hlkwc{dist} \hlstd{=} \hlstr{"HVDM"}\hlstd{,}
                    \hlkwc{C.perc} \hlstd{=} \hlkwd{list}\hlstd{(}\hlstr{"normal"}\hlstd{=}\hlnum{0.2}\hlstd{,} \hlstr{"rare1"}\hlstd{=}\hlnum{10}\hlstd{))}

\hlstd{IC7} \hlkwb{<-} \hlkwd{SmoteClassif}\hlstd{(Class}\hlopt{~}\hlstd{., ImbC,} \hlkwc{dist} \hlstd{=} \hlstr{"HVDM"}\hlstd{,} \hlkwc{C.perc}\hlstd{=}\hlstr{"extreme"}\hlstd{)}
\end{alltt}
\end{kframe}
\end{knitrout}

Table \ref{tab:iris_smote_table} show the impact on the number of examples in each class for several parameters of smote technique.

\begin{table}[ht]
\centering
\begin{tabular}{rrrr}
  \hline
 & normal & rare1 & rare2 \\ 
  \hline
Original & 859 &  10 & 131 \\ 
  IC1 & 333 & 332 & 332 \\ 
  IC2 & 333 & 332 & 332 \\ 
  IC3 & 343 &  80 & 786 \\ 
  IC4 & 333 & 332 & 332 \\ 
  IC5 & 333 & 332 & 332 \\ 
  IC6 & 171 & 100 & 131 \\ 
  IC7 &  11 & 919 &  70 \\ 
   \hline
\end{tabular}
\caption{Number of examples in each class for different parameters of smote strategy.} 
\label{tab:iris_smote_table}
\end{table}

Figures \ref{fig:smote_plot_hist} and \ref{fig:smote_plot} present the impact of applying smote strategy on an imbalanced data set.

\begin{knitrout}\footnotesize
\definecolor{shadecolor}{rgb}{0.969, 0.969, 0.969}\color{fgcolor}\begin{figure}

{\centering \includegraphics[width=\maxwidth,height=0.5\textheight]{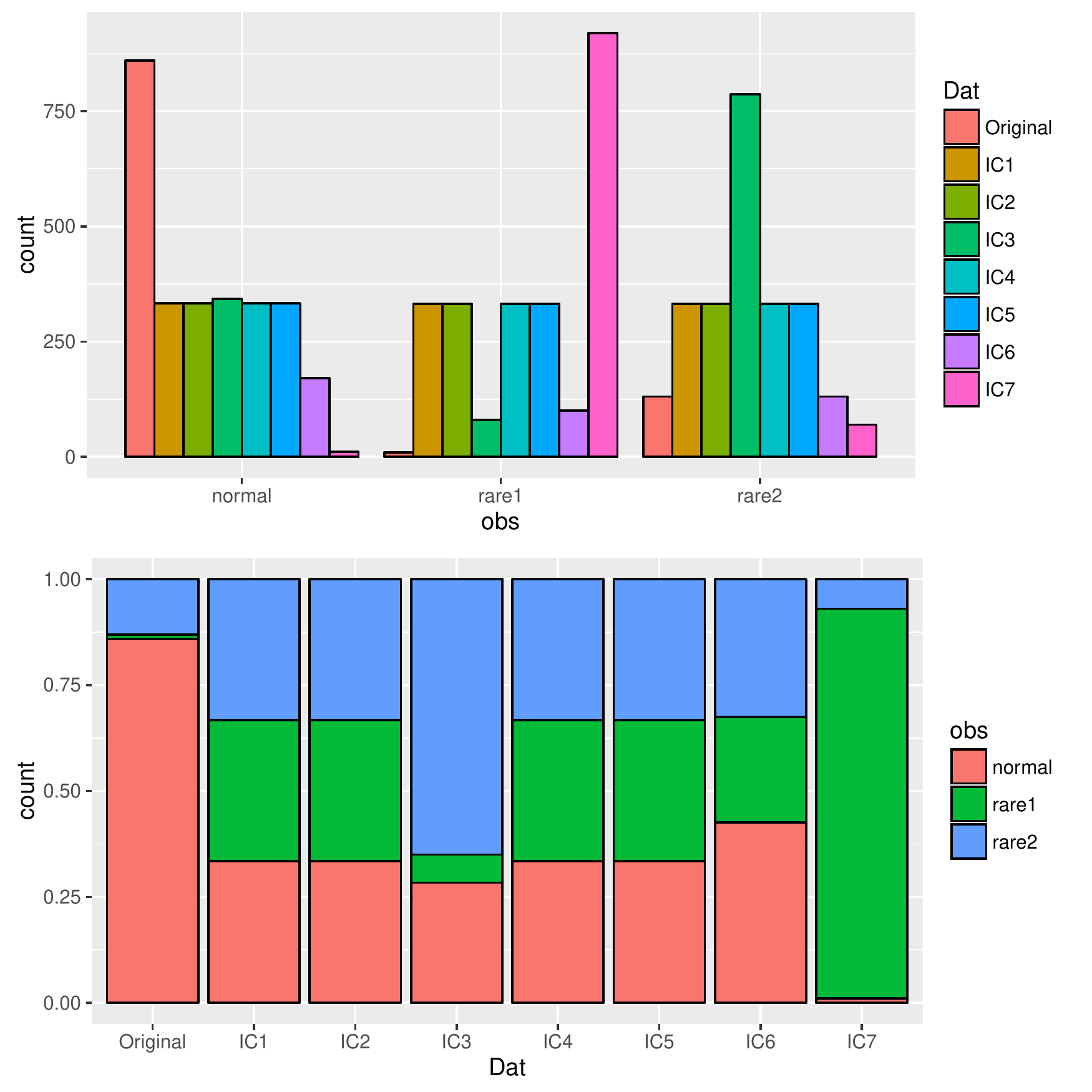} 

}

\caption[Impact in the Original data set of several parameters in smote strategy]{Impact in the Original data set of several parameters in smote strategy. }\label{fig:smote_plot_hist}
\end{figure}

\end{knitrout}

\begin{knitrout}\footnotesize
\definecolor{shadecolor}{rgb}{0.969, 0.969, 0.969}\color{fgcolor}\begin{figure}

{\centering \includegraphics[width=\maxwidth]{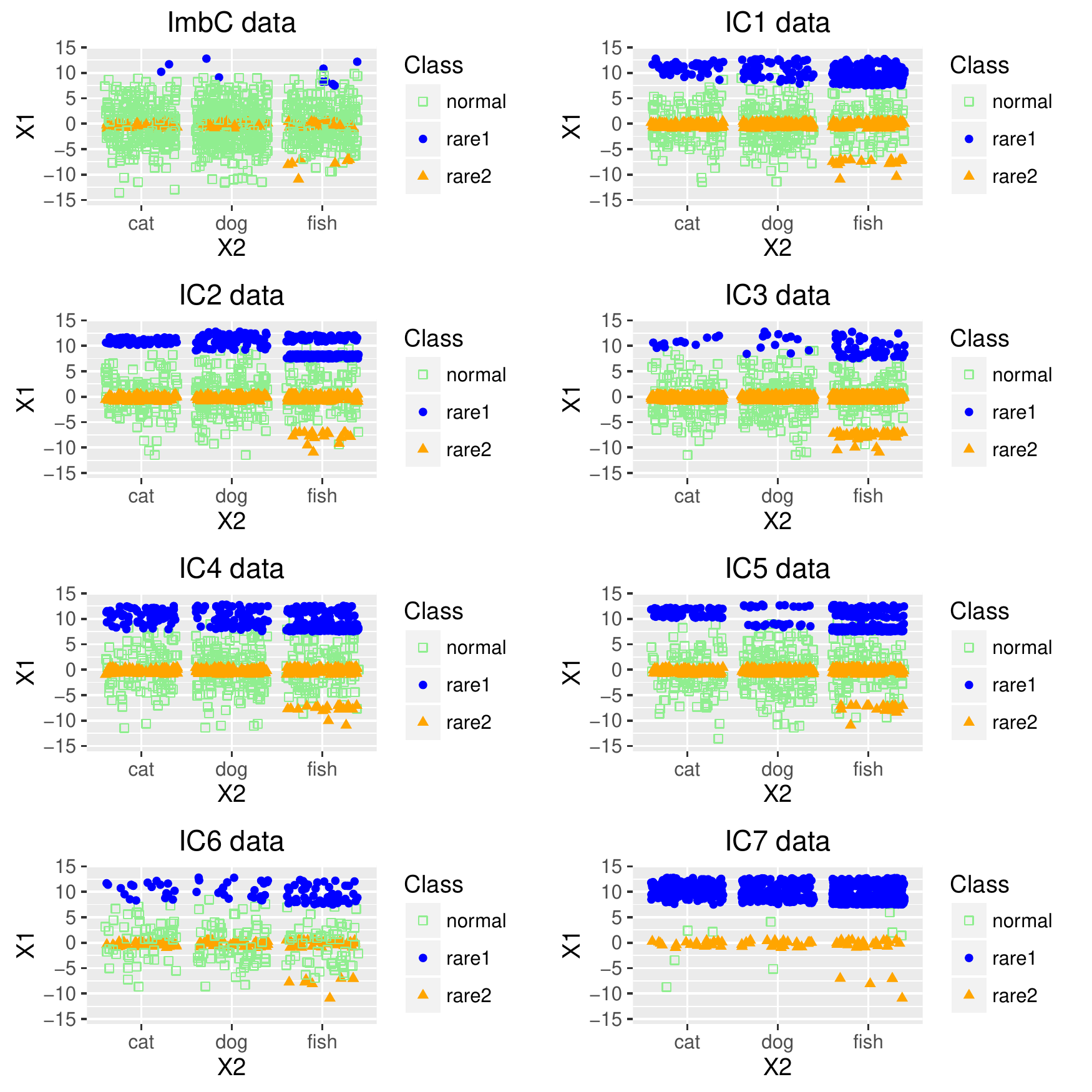} 

}

\caption[Smote strategy applied to ImbC data with different parameters]{Smote strategy applied to ImbC data with different parameters.}\label{fig:smote_plot}
\end{figure}

\end{knitrout}

% ====================================================================
\section{Methods for Addressing Utility-based Regression Tasks}\label{sec:methRegres}

Utility-based problems also occur for regression tasks. However, for these problems we have a continuous target variable and therefore are no classes defined. Instead, the user may consider some ranges of the target variable domain more important (which are usually less represented) while other regions of that variable are less important. As proposed by \cite{torgo2007utility, ribeiro2011utility}, utility-based regression problems depend on the definition of a continuous relevance function ($\phi()$) which expresses the importance of the target variable values across its domain. This function $\phi()$, varies between 0 and 1, where 0 represents points in the target variable domain which are not relevant and 1 identifies the most important values. Usually, the user is also asked to provide a relevance threshold (a numeric value in $[0,1]$) which helps to clearly distinguish between the important and unimportant values.

\cite{ribeiro2011utility} proposed a framework for defining the relevance function of a given continuous target variable. This framework has an automatic method that allows to obtain the relevance function from the target variable distribution. The assumption made to achieve this goal regards the most usual setting, where the extreme rare values are the most important to the user. This framework also allows the user to manually specify which are the relevant and irrelevant values using a matrix. The R package \texttt{uba} \cite{uba}, available in \url{http://www.dcc.fc.up.pt/~rpribeiro/uba/}, includes several other functionalities for dealing with utility-based regression. We use in \UBLp the functions regarding the relevance function.

Considering a target variable with domain $[0,10]$, a possible relevance function could be the one represented in Figure \ref{fig:relev_ex}. For this particular regression task, the relevance function selected and the chosen relevance threshold of 0.5 characterize the most important ranges of the target variable and the bumps of relevance. In this case, we have established two bumps which include the most important values (also named ``rare" cases) of the target variable ($[0, 1.5]$ and $[4.5, 7]$ represented in green in Figure~\ref{fig:relev_ex}). On the other hand, the target values falling in the intervals $]1.5, 4.5[$ and $]7,10]$ (represented in red in Figure~\ref{fig:relev_ex}) are the less relevant and ``normal" cases.

\begin{knitrout}\footnotesize
\definecolor{shadecolor}{rgb}{0.969, 0.969, 0.969}\color{fgcolor}\begin{figure}

{\centering \includegraphics[width=0.8\textwidth]{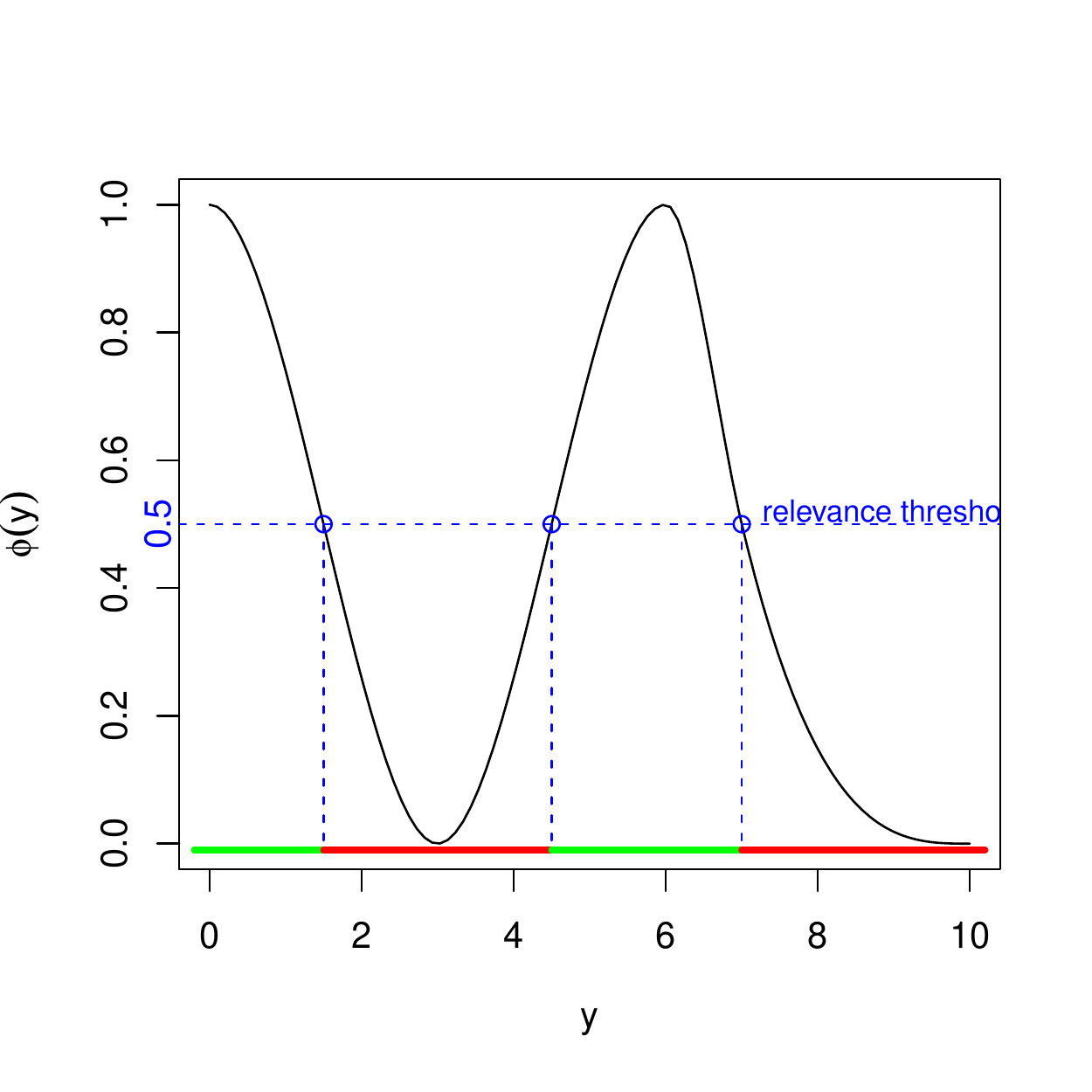} 

}

\caption[Example of a relevance function]{Example of a relevance function.}\label{fig:relev_ex}
\end{figure}

\end{knitrout}

The user has the responsibility of defining a relevance function suitable for the regression task he is considering. We provide the mechanism proposed by \cite{ribeiro2011utility} and also implemented in \texttt{uba} package to assist the user in this task. This method is called \texttt{range}, and depends on the introduction by the user of reference points for the $y$, and corresponding $\phi()$ and $\phi'()$ values. With the method \texttt{range} the relevance function may be manually defined with a 3-column matrix containing the interpolating points as follows:

\begin{knitrout}\footnotesize
\definecolor{shadecolor}{rgb}{0.969, 0.969, 0.969}\color{fgcolor}\begin{kframe}
\begin{alltt}
\hlcom{# relevance function represented in the previous example}

\hlcom{## method: range}
\hlcom{# the user should provide a matrix with y, phi(y), phi'(y)}

\hlstd{rel} \hlkwb{<-} \hlkwd{matrix}\hlstd{(}\hlnum{0}\hlstd{,} \hlkwc{ncol} \hlstd{=} \hlnum{3}\hlstd{,} \hlkwc{nrow} \hlstd{=} \hlnum{0}\hlstd{)}

\hlcom{# for the target value of zero the relevance function should be one and}
\hlcom{# the derivative at that point should be zero}
\hlstd{rel} \hlkwb{<-} \hlkwd{rbind}\hlstd{(rel,} \hlkwd{c}\hlstd{(}\hlnum{0}\hlstd{,}\hlnum{1}\hlstd{,}\hlnum{0}\hlstd{))}

\hlcom{# for the value three the relevance assigned is zero and the derivative is zero}
\hlstd{rel} \hlkwb{<-} \hlkwd{rbind}\hlstd{(rel,} \hlkwd{c}\hlstd{(}\hlnum{3}\hlstd{,}\hlnum{0}\hlstd{,}\hlnum{0}\hlstd{))}
\hlstd{rel} \hlkwb{<-} \hlkwd{rbind}\hlstd{(rel,} \hlkwd{c}\hlstd{(}\hlnum{6}\hlstd{,}\hlnum{1}\hlstd{,}\hlnum{0}\hlstd{))}
\hlstd{rel} \hlkwb{<-} \hlkwd{rbind}\hlstd{(rel,} \hlkwd{c}\hlstd{(}\hlnum{7}\hlstd{,}\hlnum{0.5}\hlstd{,}\hlnum{1}\hlstd{))}
\hlstd{rel} \hlkwb{<-} \hlkwd{rbind}\hlstd{(rel,} \hlkwd{c}\hlstd{(}\hlnum{10}\hlstd{,}\hlnum{0}\hlstd{,}\hlnum{0}\hlstd{))}
\hlcom{# after defining the relevance function the user may obtain the }
\hlcom{# phi values as follows:}

\hlcom{# use method "range" when defining a matrix}
\hlstd{phiF.args} \hlkwb{<-} \hlkwd{phi.control}\hlstd{(y,}\hlkwc{method} \hlstd{=} \hlstr{"range"}\hlstd{,} \hlkwc{control.pts} \hlstd{= rel)}

\hlcom{# obtain the relevance values for the target variable y}
\hlstd{y.phi} \hlkwb{<-} \hlkwd{phi}\hlstd{(y,}\hlkwc{control.parms} \hlstd{= phiF.args)}
\end{alltt}
\end{kframe}
\end{knitrout}

In order to facilitate the user task, we also provide an automatic mechanism proposed by \cite{ribeiro2011utility} and also implemented in \texttt{uba} package, for defining the relevance function. This automatic method, called \texttt{extremes} is based on the boxplot of the target variable values and assigns a larger importance to the least represented values. In this case, the user does not need to provide interpolating points because this method assumes that the least represented ranges of the target variable are the most important. We now provide an example of how to use this automatic method.

\begin{knitrout}\footnotesize
\definecolor{shadecolor}{rgb}{0.969, 0.969, 0.969}\color{fgcolor}\begin{kframe}
\begin{alltt}
\hlcom{## method: extremes}

\hlcom{## for considering only the high extremes}
\hlstd{phiF.args} \hlkwb{<-} \hlkwd{phi.control}\hlstd{(y,}\hlkwc{method} \hlstd{=} \hlstr{"extremes"}\hlstd{,}\hlkwc{extr.type} \hlstd{=} \hlstr{"high"}\hlstd{)}
\hlstd{y.phi} \hlkwb{<-} \hlkwd{phi}\hlstd{(y,}\hlkwc{control.parms} \hlstd{= phiF.args)}

\hlcom{## for considering only the low extremes}
\hlstd{phiF.args} \hlkwb{<-} \hlkwd{phi.control}\hlstd{(y,}\hlkwc{method} \hlstd{=} \hlstr{"extremes"}\hlstd{,}\hlkwc{extr.type} \hlstd{=} \hlstr{"low"}\hlstd{)}
\hlstd{y.phi} \hlkwb{<-} \hlkwd{phi}\hlstd{(y,}\hlkwc{control.parms} \hlstd{= phiF.args)}

\hlcom{## for considering both extreme types (low and high)}
\hlstd{phiF.args} \hlkwb{<-} \hlkwd{phi.control}\hlstd{(y,}\hlkwc{method} \hlstd{=} \hlstr{"extremes"}\hlstd{,}\hlkwc{extr.type} \hlstd{=} \hlstr{"both"}\hlstd{)}
\hlstd{y.phi} \hlkwb{<-} \hlkwd{phi}\hlstd{(y,}\hlkwc{control.parms} \hlstd{= phiF.args)}
\end{alltt}
\end{kframe}
\end{knitrout}

All the implemented methods for utility-based regression tasks depend on the definition of a relevance function, and the majority of them also rely on a user-defined relevance threshold.

Let us now observe the impact of using the automatic method for defining a relevance function with the synthetic ImbR data provided with \UBLp.

\begin{knitrout}\footnotesize
\definecolor{shadecolor}{rgb}{0.969, 0.969, 0.969}\color{fgcolor}\begin{kframe}
\begin{alltt}
\hlcom{# define that the automatic method will be used and }
\hlcom{# specify that we are only interested in the high extreme values}
\hlstd{phiF.args} \hlkwb{<-} \hlkwd{phi.control}\hlstd{(ImbR}\hlopt{$}\hlstd{Tgt,} \hlkwc{method} \hlstd{=} \hlstr{"extremes"}\hlstd{,} \hlkwc{extr.type} \hlstd{=} \hlstr{"high"}\hlstd{)}
\hlstd{y.phi} \hlkwb{<-} \hlkwd{phi}\hlstd{(}\hlkwd{sort}\hlstd{(ImbR}\hlopt{$}\hlstd{Tgt),}\hlkwc{control.parms} \hlstd{= phiF.args)}
\end{alltt}
\end{kframe}
\end{knitrout}

However, the user has also the possibility to define its own relevance function as follows:

\begin{knitrout}\footnotesize
\definecolor{shadecolor}{rgb}{0.969, 0.969, 0.969}\color{fgcolor}\begin{kframe}
\begin{alltt}
\hlcom{# specify the y, phi(y) and phi'(y) in each row of the matrix}
\hlstd{rel} \hlkwb{<-} \hlkwd{matrix}\hlstd{(}\hlkwd{c}\hlstd{(}\hlnum{10}\hlstd{,} \hlnum{1}\hlstd{,} \hlnum{0}\hlstd{,} \hlnum{11}\hlstd{,} \hlnum{0}\hlstd{,} \hlnum{0}\hlstd{,} \hlnum{18}\hlstd{,} \hlnum{0.5}\hlstd{,} \hlnum{1}\hlstd{,} \hlnum{19}\hlstd{,} \hlnum{0.8}\hlstd{,} \hlnum{0}\hlstd{,} \hlnum{21}\hlstd{,} \hlnum{1}\hlstd{,} \hlnum{0}\hlstd{),}
              \hlkwc{ncol} \hlstd{=} \hlnum{3}\hlstd{,} \hlkwc{nrow} \hlstd{=} \hlnum{5}\hlstd{,} \hlkwc{byrow} \hlstd{=} \hlnum{TRUE}\hlstd{)}
\hlstd{phiF.argsR} \hlkwb{<-} \hlkwd{phi.control}\hlstd{(ImbR}\hlopt{$}\hlstd{Tgt,} \hlkwc{method} \hlstd{=} \hlstr{"range"}\hlstd{,} \hlkwc{control.pts} \hlstd{= rel)}
\hlstd{y.phiR} \hlkwb{<-} \hlkwd{phi}\hlstd{(}\hlkwd{sort}\hlstd{(ImbR}\hlopt{$}\hlstd{Tgt),} \hlkwc{control.parms} \hlstd{= phiF.argsR)}
\end{alltt}
\end{kframe}
\end{knitrout}

Figures~\ref{fig:Rel1} and \ref{fig:Rel2} show the two relevance functions previously obtained for ImbC data (the first one is built with the automatic and the second uses the matrix with interpolating points provided by the user). The automatic method "extremes" takes into account the examples distribution while the "range" method uses the information provided by the user regardless of the domain distribution. In the relevance function specified with "range" method the lower and higher values are both considered extremely relevant.

\begin{knitrout}\footnotesize
\definecolor{shadecolor}{rgb}{0.969, 0.969, 0.969}\color{fgcolor}\begin{figure}

{\centering \includegraphics[width=0.6\textwidth]{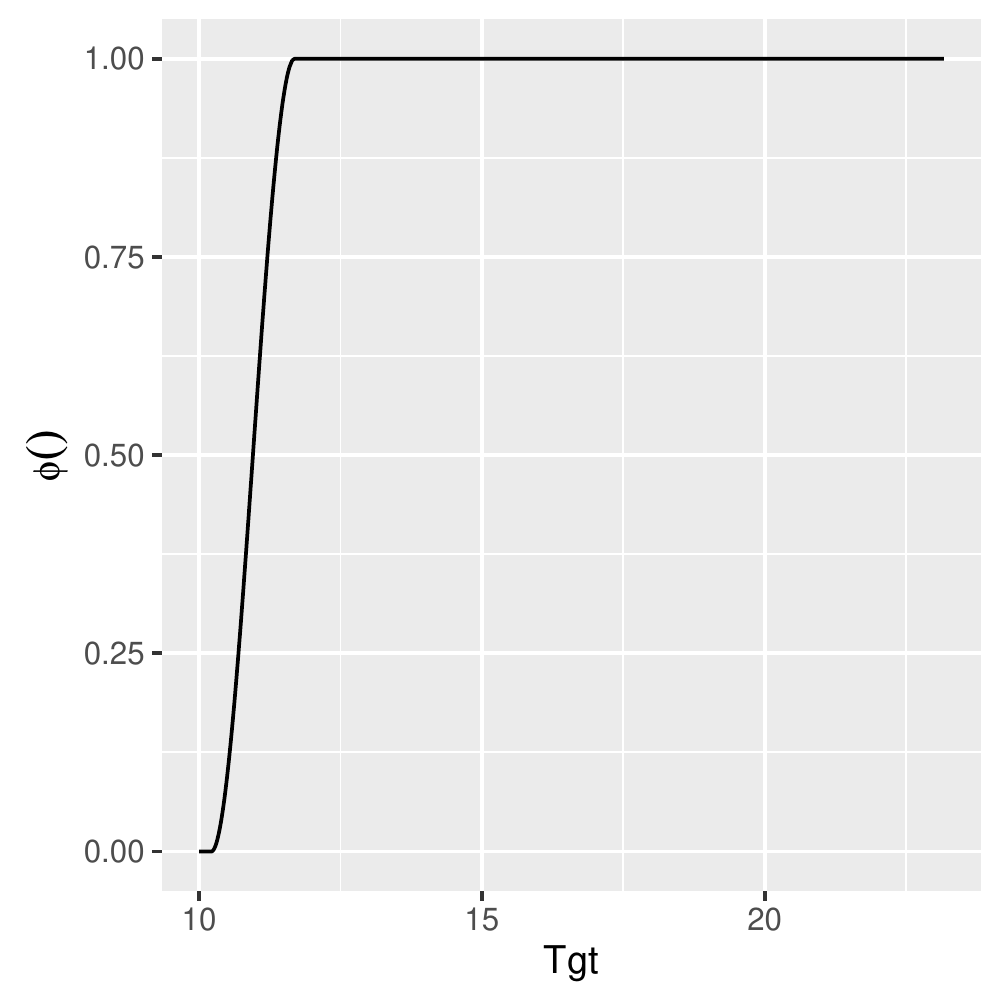} 

}

\caption[Relevance function automatically obtained]{Relevance function automatically obtained.}\label{fig:Rel1}
\end{figure}

\end{knitrout}

\begin{knitrout}\footnotesize
\definecolor{shadecolor}{rgb}{0.969, 0.969, 0.969}\color{fgcolor}\begin{figure}

{\centering \includegraphics[width=0.6\textwidth]{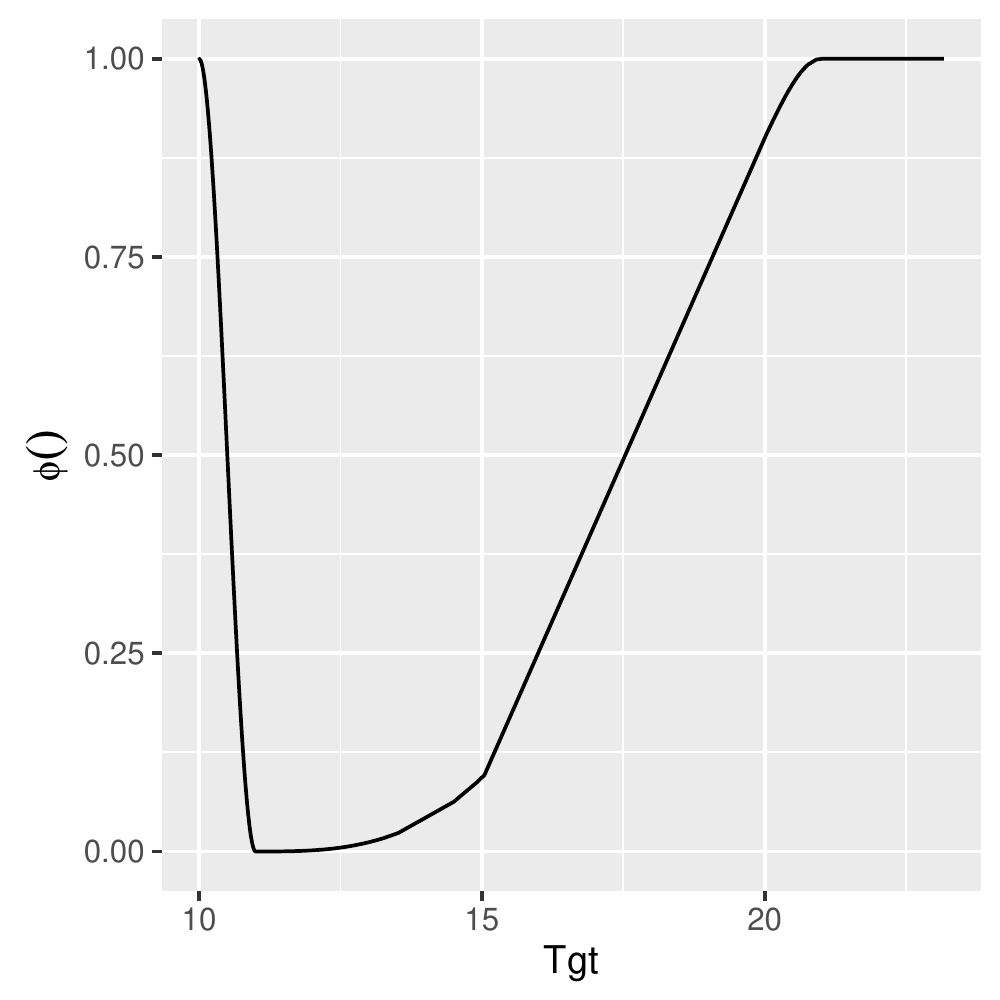} 

}

\caption[Relevance function obtained through a matrix with interpolating points provided by the user]{Relevance function obtained through a matrix with interpolating points provided by the user.}\label{fig:Rel2}
\end{figure}

\end{knitrout}

In the next sections we describe the following methods for tackling utility-based regression tasks:
\begin{itemize}
\item \ref{sec:RURegress} Random Under-sampling
\item \ref{sec:RORegress} Random Over-sampling
\item \ref{sec:gnRegress} Gaussian Noise Introduction
\item \ref{sec:smoteR} SmoteR Algorithm
\item \ref{sec:ISRegress} Importance Sampling

\end{itemize}

\subsection{Random Under-sampling}\label{sec:RURegress}

Random under-sampling strategy for regression problems was first proposed by \cite{torgo2013smote}. This strategy is similar to the strategy presented for classification. It depends on the definition of both a relevance function and a relevance threshold. In this proposal, all the target values below the relevance threshold are considered normal and uninteresting and thus are regarded as candidates to be under-sampled. The user is also asked to set another parameter that establishes the proportion between normal (unimportant) and rare (important) cases that the new under-sampled data set should contain.

In the implementation of this strategy in \pUBL, we ask for the user to define the relevance function (manually through the method ``range" or using the automatic method, called ``extremes", previously described). This means that the user may define as many relevance bumps as wanted. Parameter \texttt{rel} is used to indicate the relevance function. For using the automatic method the parameter \texttt{rel} should be set to ``auto" (the default). If the user wants to apply the range method, then, as previously explained, a 3-column matrix should be provided. It is also necessary for the user to define a relevance threshold through the \texttt{thr.rel} parameter. Having this set, all the target variable values with relevance below the relevance threshold are candidates to be under-sampled. Finally, the user can also express using the \texttt{C.perc} parameter which under-sampling percentage should be applied in each bump with uninteresting values, or alternatively this parameter may be set to ``balance" or ``extreme". If ``balance" is chosen the under-sampling percentage is automatically estimated in order to balance the normal/important and rare/unimportant cases. On the other hand, the ``extreme" option will invert the existing frequencies. The following example uses the regression data set provided with \UBLp, ImbR, to show how these parameters can be set and their impact on the changed data.

\begin{knitrout}\footnotesize
\definecolor{shadecolor}{rgb}{0.969, 0.969, 0.969}\color{fgcolor}\begin{kframe}
\begin{alltt}
\hlkwd{data}\hlstd{(ImbR)} \hlcom{# load the synthetic data set provided with UBL}

\hlcom{# Using the automatic method for defining the relevance function}
\hlcom{# This is the default behaviour, therefore, we can simply }
\hlcom{# not mention the "rel" parameter}

\hlcom{# default of C.perc parameter balances the examples in the bumps}
\hlstd{IRU1} \hlkwb{<-} \hlkwd{RandUnderRegress}\hlstd{(Tgt}\hlopt{~}\hlstd{., ImbR)}

\hlstd{IRU2} \hlkwb{<-} \hlkwd{RandUnderRegress}\hlstd{(Tgt}\hlopt{~}\hlstd{., ImbR,} \hlkwc{C.perc} \hlstd{=} \hlstr{"extreme"}\hlstd{)}

\hlcom{# the automatic method for the relevance function generates only }
\hlcom{# one bump with uninteresting values, thus we only need to set }
\hlcom{# one under-sampling percentage}
\hlstd{IRU3} \hlkwb{<-} \hlkwd{RandUnderRegress}\hlstd{(Tgt}\hlopt{~}\hlstd{., ImbR,} \hlkwc{C.perc} \hlstd{=} \hlkwd{list}\hlstd{(}\hlnum{0.5}\hlstd{))}
\end{alltt}
\end{kframe}
\end{knitrout}

Figure~\ref{fig:ImbR_RU_ex1} shows the impact of the applied strategies on the density of target variable of ImbR data.
\begin{knitrout}\footnotesize
\definecolor{shadecolor}{rgb}{0.969, 0.969, 0.969}\color{fgcolor}\begin{figure}

{\centering \includegraphics[width=0.8\textwidth]{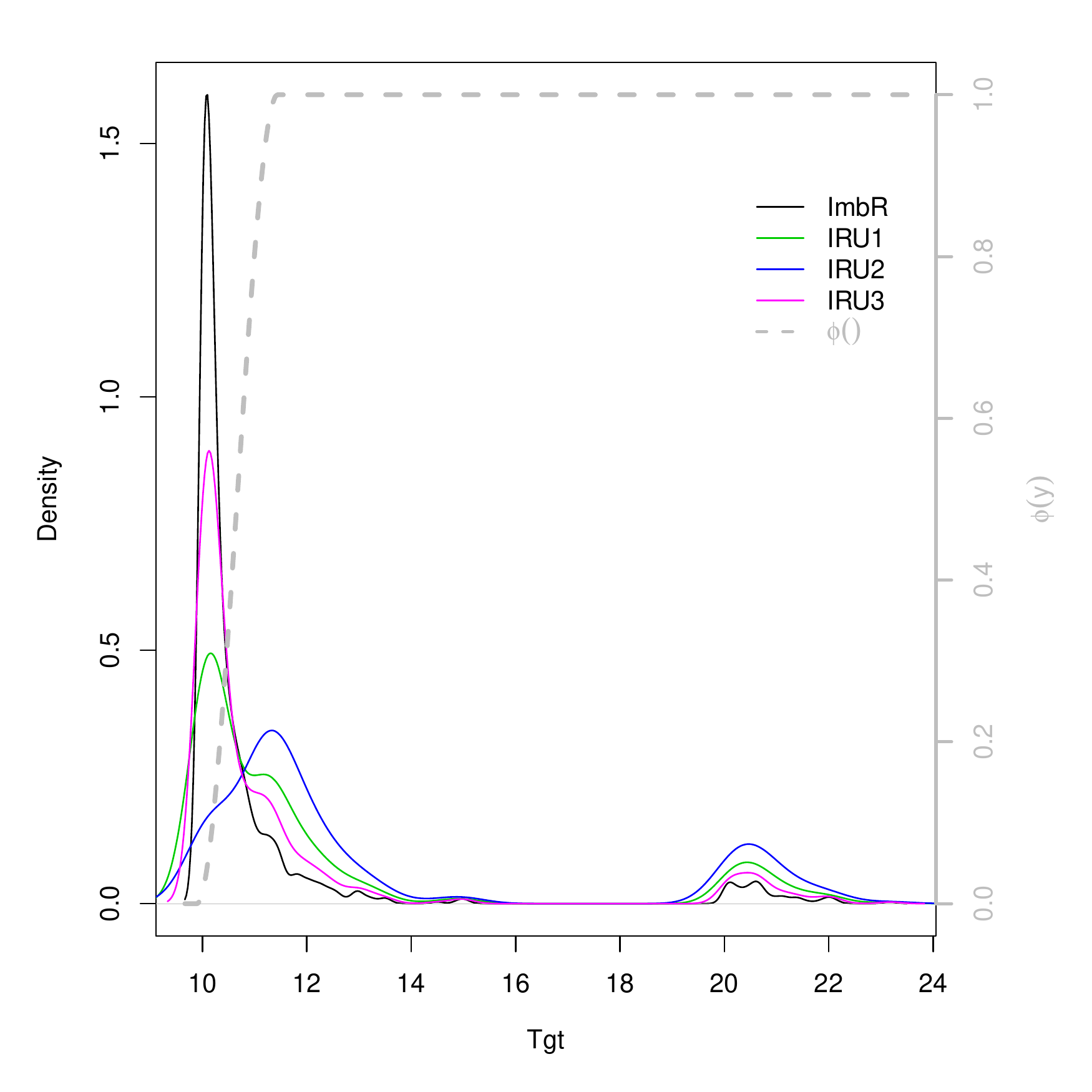} 

}

\caption[Automatic relevance function and density of the target variable in the original ImbR data and the changed data sets using Random Under-sampling strategy]{Automatic relevance function and density of the target variable in the original ImbR data and the changed data sets using Random Under-sampling strategy}\label{fig:ImbR_RU_ex1}
\end{figure}

\end{knitrout}

We can also observe the impact of the previously defined changes on the examples distribution in Figure~\ref{fig:ImbR_RU_dist}.

\begin{knitrout}\footnotesize
\definecolor{shadecolor}{rgb}{0.969, 0.969, 0.969}\color{fgcolor}\begin{figure}

{\centering \includegraphics[width=\maxwidth]{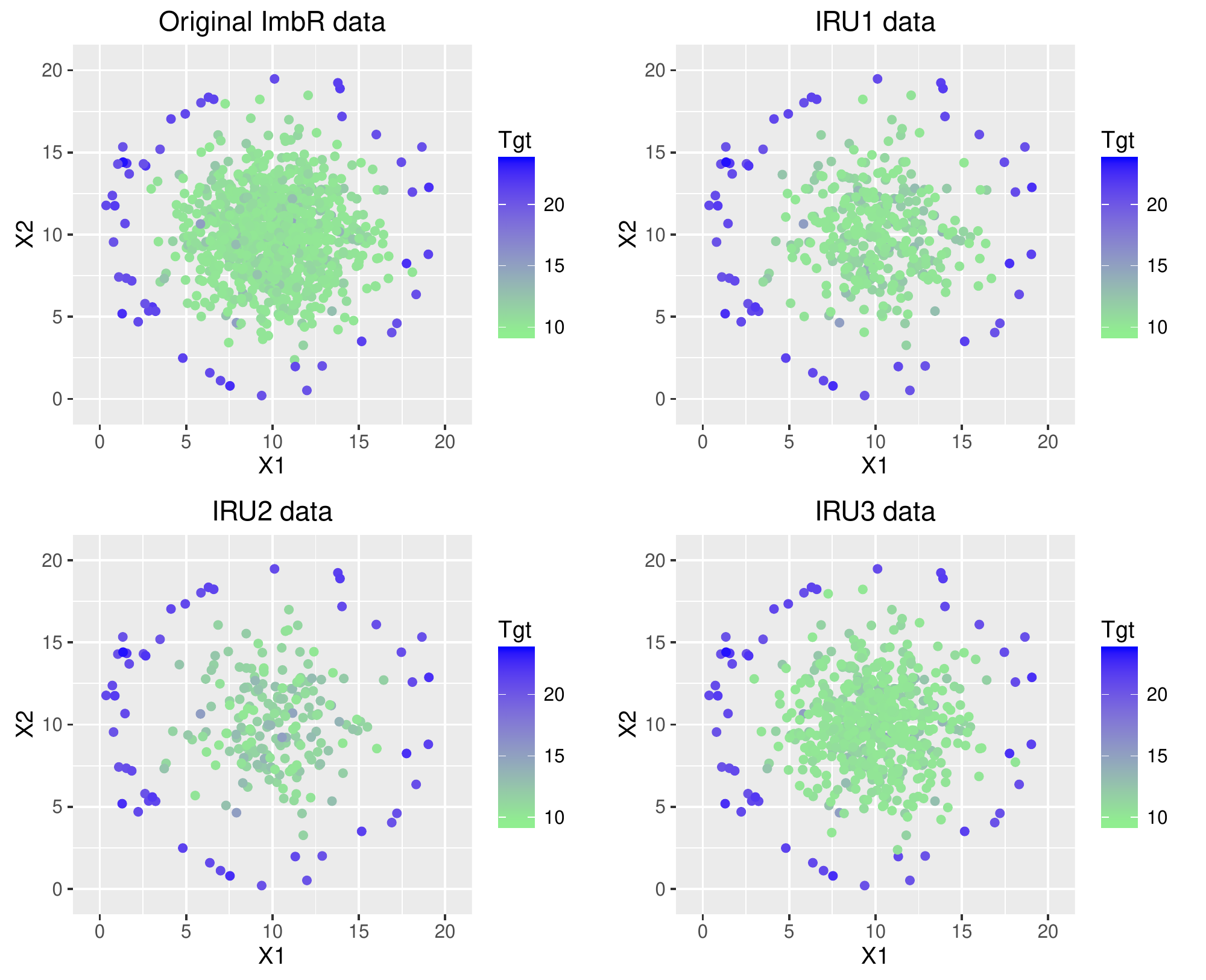} 

}

\caption[Original and changed data sets using different parameters of the random under-sampling strategy]{Original and changed data sets using different parameters of the random under-sampling strategy.}\label{fig:ImbR_RU_dist}
\end{figure}

\end{knitrout}

Let us now assume that we have some domain knowledge that leads us to considered a different relevance function. Suppose that the most relevant cases are the target varaible values close to 15. In the following example we define a new relvance function suitable for this context and apply the random under-sampling strategy to change the original data set with different parameters.

\begin{knitrout}\footnotesize
\definecolor{shadecolor}{rgb}{0.969, 0.969, 0.969}\color{fgcolor}\begin{kframe}
\begin{alltt}
\hlstd{rel} \hlkwb{<-} \hlkwd{matrix}\hlstd{(}\hlkwd{c}\hlstd{(}\hlnum{14}\hlstd{,} \hlnum{0}\hlstd{,} \hlnum{0}\hlstd{,} \hlnum{15}\hlstd{,} \hlnum{1}\hlstd{,} \hlnum{0}\hlstd{,} \hlnum{16}\hlstd{,} \hlnum{0}\hlstd{,} \hlnum{0}\hlstd{,} \hlnum{20}\hlstd{,} \hlnum{1}\hlstd{,} \hlnum{0}\hlstd{,} \hlnum{21}\hlstd{,} \hlnum{0}\hlstd{,} \hlnum{0}\hlstd{),}
              \hlkwc{ncol}\hlstd{=}\hlnum{3}\hlstd{,} \hlkwc{nrow}\hlstd{=}\hlnum{5}\hlstd{,} \hlkwc{byrow}\hlstd{=}\hlnum{TRUE}\hlstd{)}
\hlstd{dsU1} \hlkwb{<-} \hlkwd{RandUnderRegress}\hlstd{(Tgt}\hlopt{~}\hlstd{., ImbR,} \hlkwc{rel}\hlstd{=rel)}
\hlstd{dsU2} \hlkwb{<-} \hlkwd{RandUnderRegress}\hlstd{(Tgt}\hlopt{~}\hlstd{., ImbR,} \hlkwc{rel}\hlstd{=rel,} \hlkwc{C.perc}\hlstd{=}\hlkwd{list}\hlstd{(}\hlnum{0.5}\hlstd{))}
\hlstd{dsU3} \hlkwb{<-} \hlkwd{RandUnderRegress}\hlstd{(Tgt}\hlopt{~}\hlstd{., ImbR,} \hlkwc{rel}\hlstd{=rel,} \hlkwc{C.perc}\hlstd{=}\hlkwd{list}\hlstd{(}\hlnum{0.2}\hlstd{))}
\end{alltt}
\end{kframe}
\end{knitrout}

Figures~\ref{fig:I_u_RU1} and \ref{fig:I_u_RU2} show the density of the target variable in the original ImbR data and the pre-processed data sets and the examples distribution.

\begin{knitrout}\footnotesize
\definecolor{shadecolor}{rgb}{0.969, 0.969, 0.969}\color{fgcolor}\begin{figure}

{\centering \includegraphics[width=0.8\textwidth]{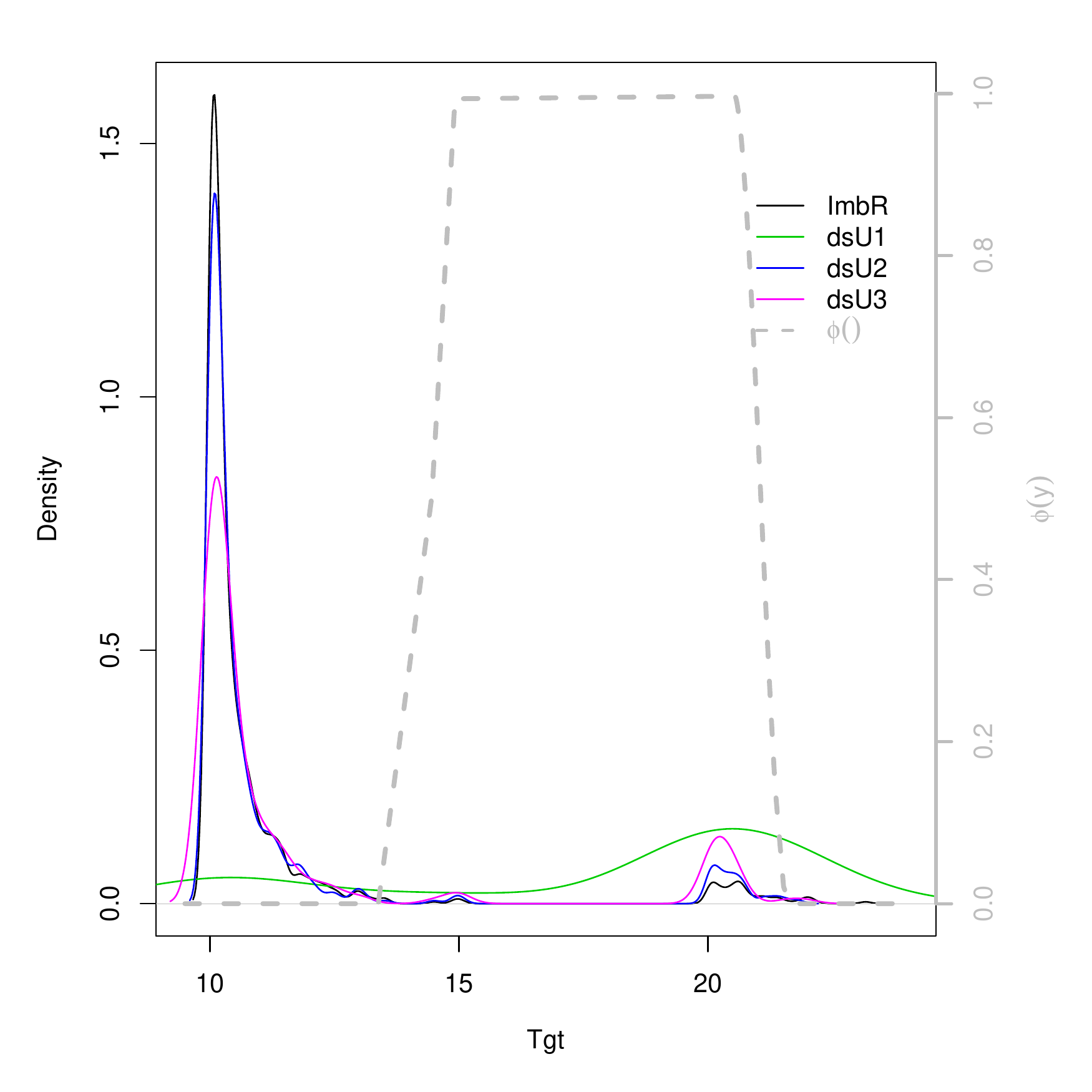} 

}

\caption[Density of the target variable in the original ImbR data and the changed data sets using Random Under-sampling strategy and a user defined relevance function]{Density of the target variable in the original ImbR data and the changed data sets using Random Under-sampling strategy and a user defined relevance function.}\label{fig:I_u_RU1}
\end{figure}

\end{knitrout}

\begin{knitrout}\footnotesize
\definecolor{shadecolor}{rgb}{0.969, 0.969, 0.969}\color{fgcolor}\begin{figure}

{\centering \includegraphics[width=\maxwidth]{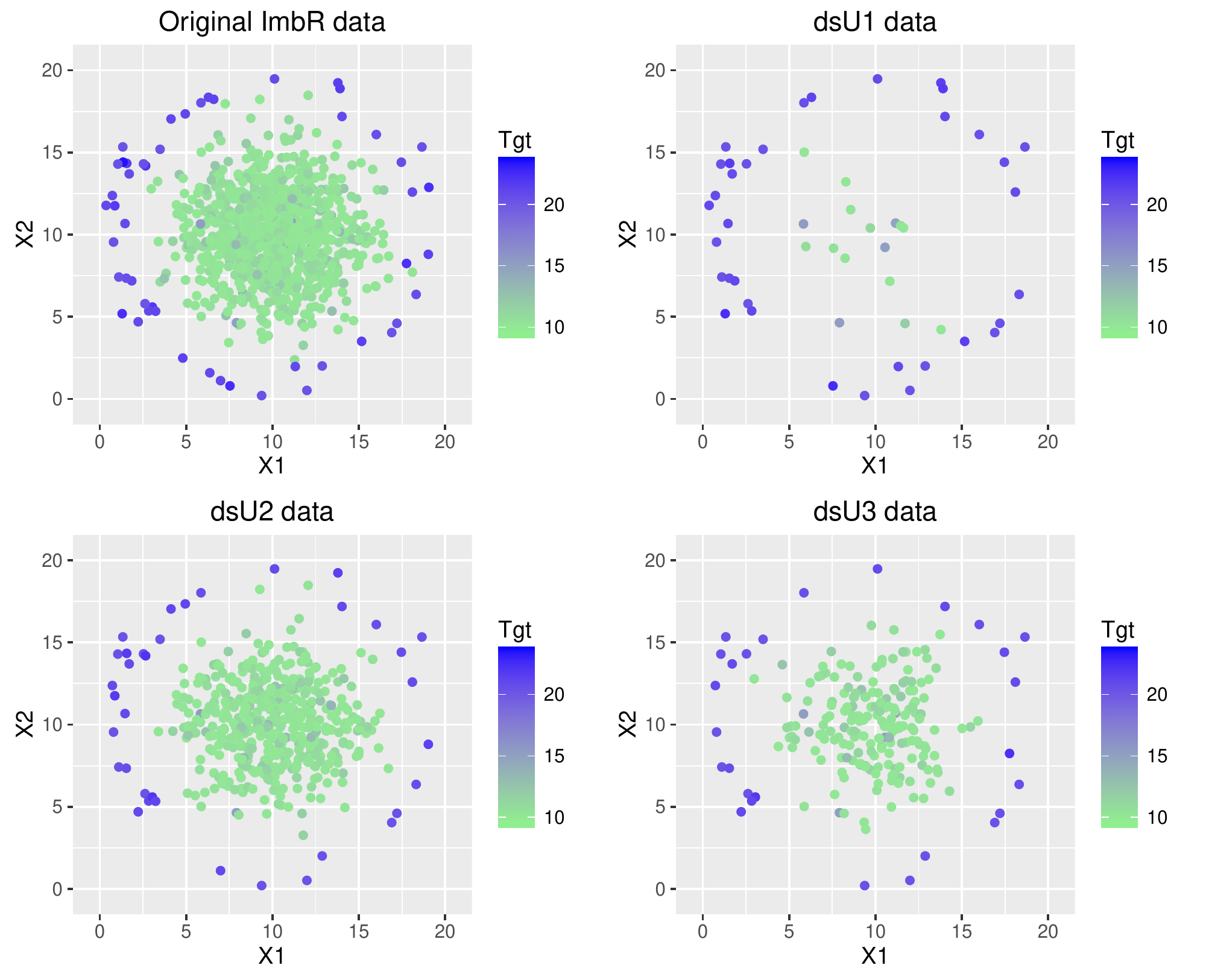} 

}

\caption[Original and changed data sets using different parameters of the random under-sampling strategy and a user defined relevance function]{Original and changed data sets using different parameters of the random under-sampling strategy and a user defined relevance function.}\label{fig:I_u_RU2}
\end{figure}

\end{knitrout}

We must highlight that this strategy entails some consequences that should not be disregarded. Namely, there can be a sever impact on the total number of examples in the modified data sets. If we are considering a large data set, possibly removing 100 points may have a negligible impact. However, if the data set is already small, then removing 100 examples may have an huge impact. 

This can be observed in the previous examples. In fact, the \texttt{C.perc} parameter must be thought carefully due to the consequences on the total number of examples. In Table \ref{tab:RUReg_table} we can check the impact of the several strategies on the data set for the two relevance functions considered (the obtained through the automatic method and the defined with a 3-column matrix).

\begin{table}[ht]
\centering
\begin{tabular}{rrrrrrrr}
  \hline
 & ImbR & IRU1 & IRU2 & IRU3 & dsU1 & dsU2 & dsU3 \\ 
  \hline
nr. examples & 1000 & 390 & 242 & 597 &  51 & 513 & 220 \\ 
   \hline
\end{tabular}
\caption{Total number of examples in each data set for different parameters of random under-sampling strategy.} 
\label{tab:RUReg_table}
\end{table}

\subsection{Random Over-sampling}\label{sec:RORegress}

The Random over-sampling method proposed is an adaptation of the Random over-sampling method proposed for classification tasks using the previously presented relevance function for utility-based regression tasks. This technique is available through \texttt{randomOverRegress} function, and is simply based on the random introduction of replicas of examples of the original data set. These replicas are only introduced in the most important ranges of the target variable, i.e., in the ranges where the relevance is above a user-defined threshold. Similarly to what happened in Random under-sampling, the user may define its own relevance function or use the automatic method provided to generate one. It is also the user responsibility to define the relevance threshold (using the \texttt{thr.rel} parameter) and the percentages of over-sampling to apply in each bump of relevance (through the \texttt{C.perc} parameter). Alternatively, the user may set the \texttt{C.perc} parameter as ``balance" or ``extreme", cases which automatically evaluate the percentages of over-sampling to apply for obtaining a new balanced data set or for inverting the frequencies of examples in the defined bumps.
In the following example we can see how to use this function.

\begin{knitrout}\footnotesize
\definecolor{shadecolor}{rgb}{0.969, 0.969, 0.969}\color{fgcolor}\begin{kframe}
\begin{alltt}
\hlcom{# using the automatic method for defining the relevance function and}
\hlcom{# the default threshold of 0.5}
\hlstd{IRO} \hlkwb{<-} \hlkwd{RandOverRegress}\hlstd{(Tgt}\hlopt{~}\hlstd{., ImbR,} \hlkwc{C.perc}\hlstd{=}\hlkwd{list}\hlstd{(}\hlnum{2.5}\hlstd{))}
\hlstd{IROBal} \hlkwb{<-} \hlkwd{RandOverRegress}\hlstd{(Tgt}\hlopt{~}\hlstd{., ImbR,} \hlkwc{C.perc}\hlstd{=}\hlstr{"balance"}\hlstd{)}
\hlstd{IROExt} \hlkwb{<-} \hlkwd{RandOverRegress}\hlstd{(Tgt}\hlopt{~}\hlstd{., ImbR,} \hlkwc{C.perc}\hlstd{=}\hlstr{"extreme"}\hlstd{)}

\hlcom{# change the relevance threshold to 0.9}
\hlstd{IRO0.9} \hlkwb{<-} \hlkwd{RandOverRegress}\hlstd{(Tgt}\hlopt{~}\hlstd{., ImbR,} \hlkwc{thr.rel}\hlstd{=}\hlnum{0.9}\hlstd{)}
\end{alltt}
\end{kframe}
\end{knitrout}

Figure \ref{fig:RO_ex1} shows the impact of this method for several parameters.

\begin{knitrout}\footnotesize
\definecolor{shadecolor}{rgb}{0.969, 0.969, 0.969}\color{fgcolor}\begin{figure}

{\centering \includegraphics[width=\maxwidth]{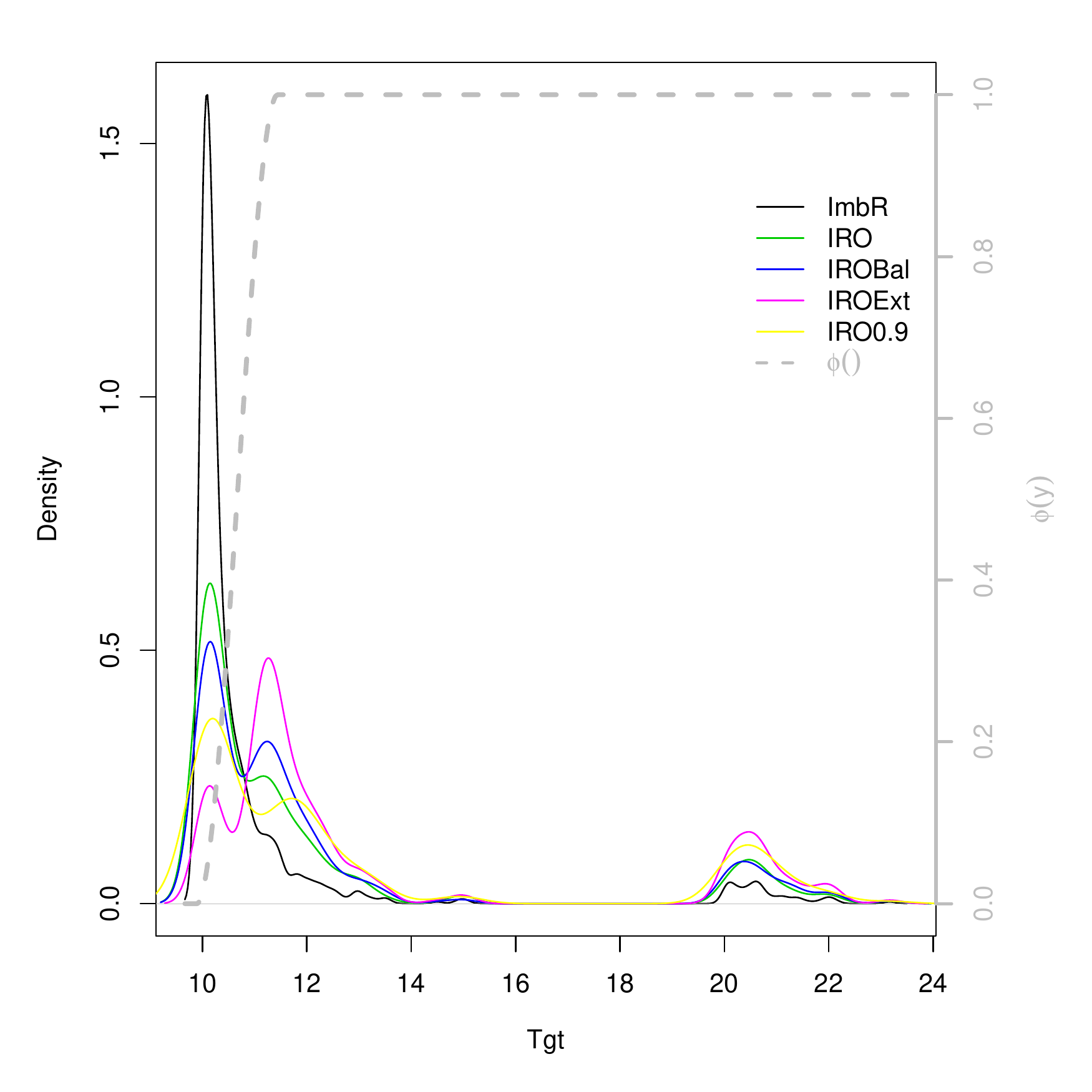} 

}

\caption[Relevance function and density of the target variable in the original and new data sets using Random over-sampling strategy]{Relevance function and density of the target variable in the original and new data sets using Random over-sampling strategy.}\label{fig:RO_ex1}
\end{figure}

\end{knitrout}

This method also carries a strong impact on the total number of examples in the modified data set. While the random Under-sampling method is able to produce a significant reduction of the data set, the random over-sampling technique will increase, sometimes drastically, the data set size. Table \ref{tab:ROReg_table} shows the impact of the previous examples on the total number of examples of used the data set.

\begin{table}[ht]
\centering
\begin{tabular}{rrrrrr}
  \hline
 & ImbR & IRO & IROBal & IROExt & IRO0.9 \\ 
  \hline
nr. examples & 1000 & 1487 & 1805 & 4323 & 1866 \\ 
   \hline
\end{tabular}
\caption{Total number of examples in each data set for different parameters of random over-sampling strategy.} 
\label{tab:ROReg_table}
\end{table}

As expected, all the data sets have an increased size. However, for the \texttt{IRO0.9} data set, the size was increased approximately 187\%. This ``side effect" must be taken into consideration when applying this technique because it may impose constraints on the used learners. We must also highlight that, although the data set size can be strongly increased, we are in fact only introducing replicas of already existing examples, and thus no new information is being inserted.

\subsection{Generation of synthetic examples by the introduction of Gaussian Noise}\label{sec:gnRegress}

The generation of synthetic examples through the introduction of small perturbations based on Gaussian Noise was a strategy proposed for classification tasks \cite{lee1999regularization, lee2000noisy}. The main idea of this strategy is to generate new synthetic examples with a desired class label, by perturbing the features of examples of that class a certain amount of the respective standard deviation. 

We have adapted this over-sampling technique to regression problems and have combined it with the random under-sampling method. To accomplish this it is required that the user defines a relevance function and a relevance threshold. The examples which have a target variable value with relevance higher than the threshold set will be over-sampled, and the remaining will be randomly under-sampled. The under-sampling strategy used is the same described in Section~\ref{sec:RURegress}. For the over-sampling strategy we use the same procedure which was described for classification tasks in Section~\ref{sec:gnClassif}. The only difference on the over-sampling method is in the target variable value. For classification tasks, the target variable value was easily assigned: it was the rare class under consideration. For regression tasks we have decided to extend the technique applied for numeric features also to the target variable. This means that the new example target variable value is obtained by a random normal perturbation of the original target value based on the target value standard deviation.

In order to use this method the user must provide a relevance function through the \texttt{rel} parameter (or use the automatic method for estimating it by setting rel to ``auto"), a threshold on the relevance (parameter \texttt{thr.rel}) and the perturbation to be used (parameter \texttt{pert}). Moreover, the user may also express using the parameter \texttt{C.perc} the percentages of over and under-sampling to apply in each bump defined, or alternatively he may set this parameter to ``balance" or ``extreme". Similarly to the behavior described in the previous techniques, setting this parameter to ``balance" or ``extreme" causes the percentages of over and under-sampling to be automatically estimated. The option ``balance" will try to distribute the examples evenly across the existing bumps while maintaining the total number of examples in the modified data set. If the choice is ``extreme" then the frequencies of the examples in the bumps will be inverted. The user can also indicate if the under-sampling process can be made with repetition of examples or not using the \texttt{repl} parameter.

We now show some examples of usage of the function \texttt{GaussNoiseRegress}.
\begin{knitrout}\footnotesize
\definecolor{shadecolor}{rgb}{0.969, 0.969, 0.969}\color{fgcolor}\begin{kframe}
\begin{alltt}
\hlcom{# relevance function estimated automatically has two bumps}
\hlcom{# defining the desired percentages of under and over-sampling to apply}
\hlstd{C.perc}\hlkwb{=}\hlkwd{list}\hlstd{(}\hlnum{0.5}\hlstd{,} \hlnum{3}\hlstd{)}
\hlcom{# define the relevance threshold}
\hlstd{thr.rel}\hlkwb{=}\hlnum{0.8}
\hlstd{mygn} \hlkwb{<-} \hlkwd{GaussNoiseRegress}\hlstd{(Tgt}\hlopt{~}\hlstd{., ImbR,} \hlkwc{thr.rel}\hlstd{=thr.rel,} \hlkwc{C.perc}\hlstd{=C.perc)}
\hlstd{gnB} \hlkwb{<-} \hlkwd{GaussNoiseRegress}\hlstd{(Tgt}\hlopt{~}\hlstd{., ImbR,} \hlkwc{thr.rel}\hlstd{=thr.rel,} \hlkwc{C.perc}\hlstd{=}\hlstr{"balance"}\hlstd{)}
\hlstd{gnE} \hlkwb{<-} \hlkwd{GaussNoiseRegress}\hlstd{(Tgt}\hlopt{~}\hlstd{., ImbR,} \hlkwc{thr.rel}\hlstd{=thr.rel,} \hlkwc{C.perc}\hlstd{=}\hlstr{"extreme"}\hlstd{)}
\end{alltt}
\end{kframe}
\end{knitrout}

Figures~\ref{fig:GN_plot1} and \ref{fig:GN_plot2} show the impact of this strategy, for the parameters considered, on the examples distribution. In Figure~\ref{fig:GN_plot2} we have binarized the data sets into rare/important ($+$) and normal/unimportant cases ($-$). We have considered the threshold of 19 on the target varaible to distinguish between the two "classes". Figure~\ref{fig:GN_plot_new} shows the true distribution of the target variable in the original ImbR data and the pre-processed data sets.

\begin{knitrout}\footnotesize
\definecolor{shadecolor}{rgb}{0.969, 0.969, 0.969}\color{fgcolor}\begin{figure}

{\centering \includegraphics[width=\maxwidth]{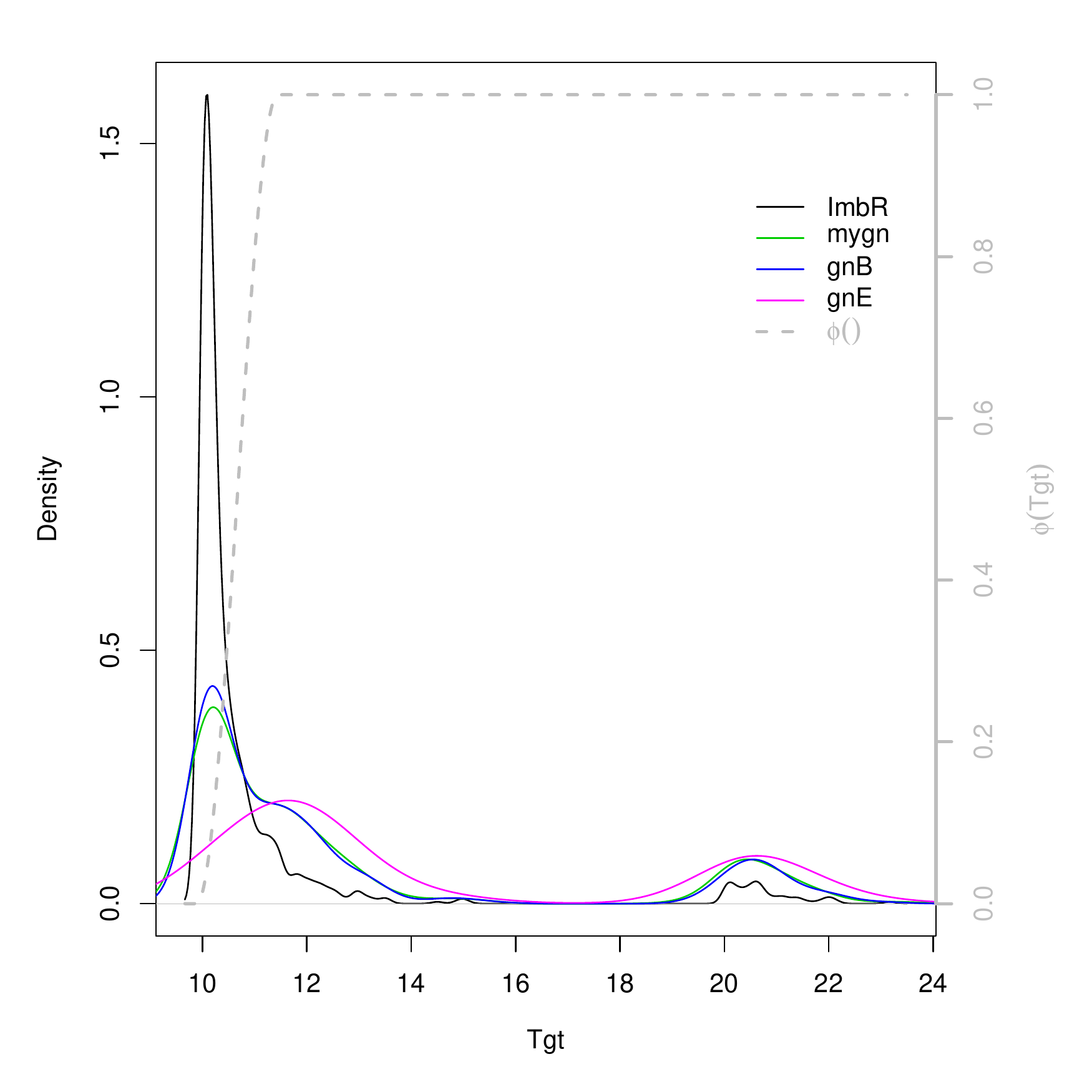} 

}

\caption[Relevance function and density of the target variable in the original and new data sets using Gaussian noise strategy]{Relevance function and density of the target variable in the original and new data sets using Gaussian noise strategy.}\label{fig:GN_plot1}
\end{figure}

\end{knitrout}

\begin{knitrout}\footnotesize
\definecolor{shadecolor}{rgb}{0.969, 0.969, 0.969}\color{fgcolor}\begin{figure}

{\centering \includegraphics[width=0.8\textwidth]{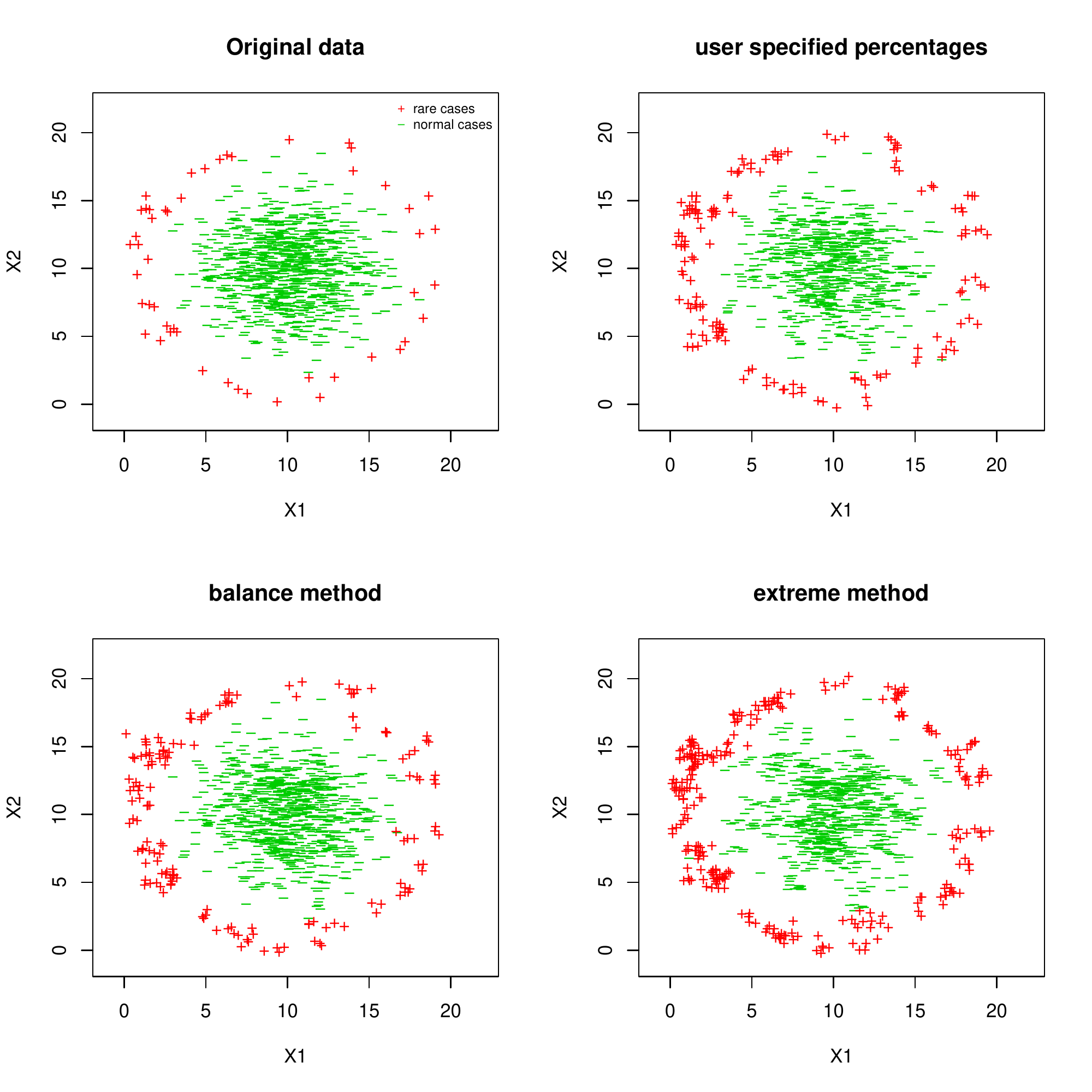} 

}

\caption[The impact of Gaussian Noise strategy in a binarized version of ImbR data considering Tgt values above 19 as rare and below 19 as normal cases]{The impact of Gaussian Noise strategy in a binarized version of ImbR data considering Tgt values above 19 as rare and below 19 as normal cases.}\label{fig:GN_plot2}
\end{figure}

\end{knitrout}

\begin{knitrout}\footnotesize
\definecolor{shadecolor}{rgb}{0.969, 0.969, 0.969}\color{fgcolor}\begin{figure}

{\centering \includegraphics[width=\maxwidth]{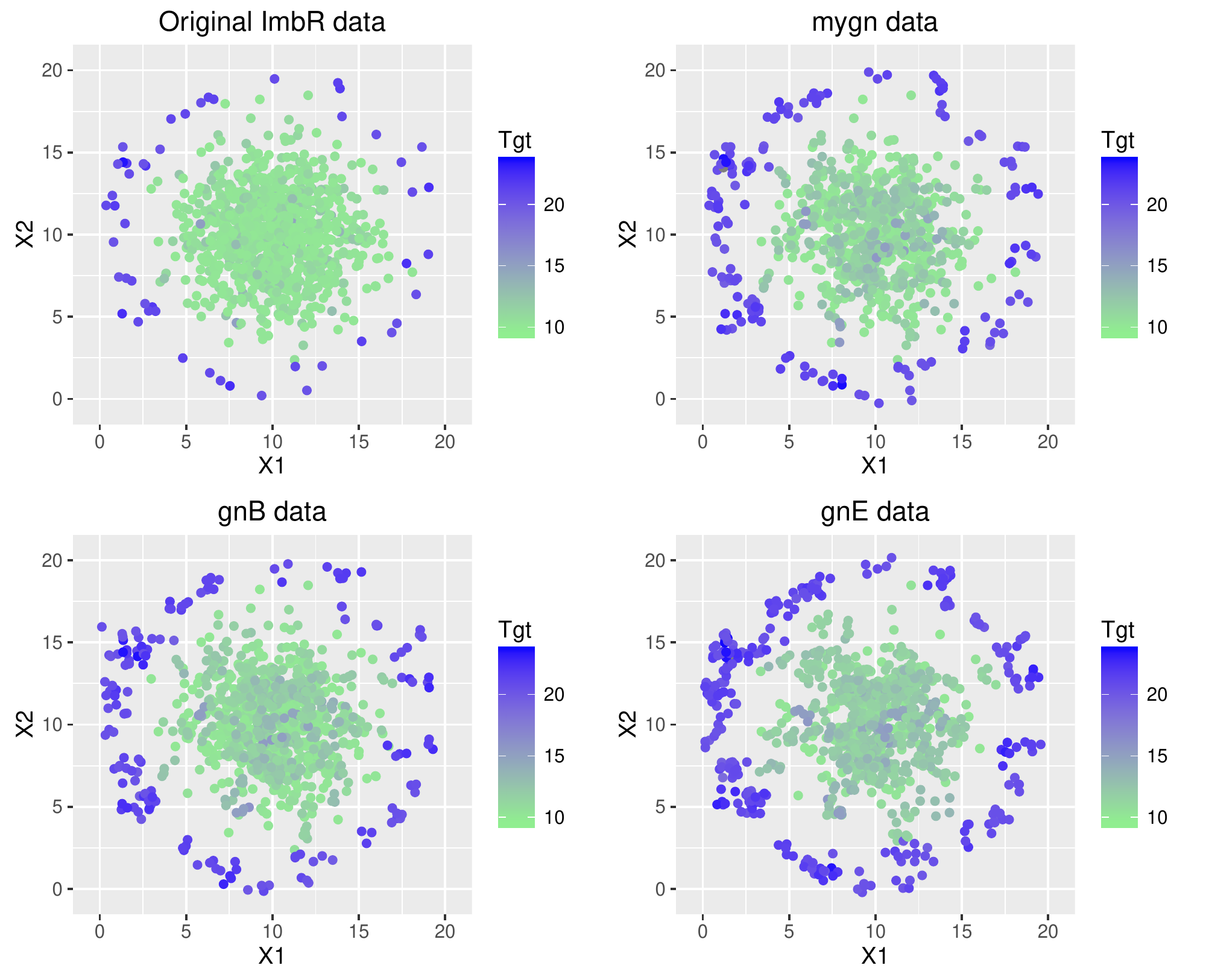} 

}

\caption[Target variable distribution on ImbR and data sets changed through Gaussian Noise strategy]{Target variable distribution on ImbR and data sets changed through Gaussian Noise strategy.}\label{fig:GN_plot_new}
\end{figure}

\end{knitrout}

In the following example we check the impact of changing the perturbation introduced.

\begin{knitrout}\footnotesize
\definecolor{shadecolor}{rgb}{0.969, 0.969, 0.969}\color{fgcolor}\begin{kframe}
\begin{alltt}
\hlcom{# the default uses the value of 0.1 for "pert" parameter}
\hlstd{gnB1} \hlkwb{<-} \hlkwd{GaussNoiseRegress}\hlstd{(Tgt}\hlopt{~}\hlstd{., ImbR,} \hlkwc{thr.rel}\hlstd{=thr.rel,} \hlkwc{C.perc}\hlstd{=}\hlstr{"balance"}\hlstd{)}

\hlcom{# try two different values for "pert" parameter}
\hlstd{gnB2} \hlkwb{<-} \hlkwd{GaussNoiseRegress}\hlstd{(Tgt}\hlopt{~}\hlstd{., ImbR,} \hlkwc{thr.rel}\hlstd{=thr.rel,} \hlkwc{C.perc}\hlstd{=}\hlstr{"balance"}\hlstd{,}
                          \hlkwc{pert}\hlstd{=}\hlnum{0.5}\hlstd{)}
\hlstd{gnB3} \hlkwb{<-} \hlkwd{GaussNoiseRegress}\hlstd{(Tgt}\hlopt{~}\hlstd{., ImbR,} \hlkwc{thr.rel}\hlstd{=thr.rel,} \hlkwc{C.perc}\hlstd{=}\hlstr{"balance"}\hlstd{,}
                          \hlkwc{pert}\hlstd{=}\hlnum{0.01}\hlstd{)}
\end{alltt}
\end{kframe}
\end{knitrout}

The impact of changing the parameter \texttt{pert} is represented in Figures~\ref{fig:GN_plot3_dist} and \ref{fig:GN_plot3}.
\begin{knitrout}\footnotesize
\definecolor{shadecolor}{rgb}{0.969, 0.969, 0.969}\color{fgcolor}\begin{figure}

{\centering \includegraphics[width=0.8\textwidth]{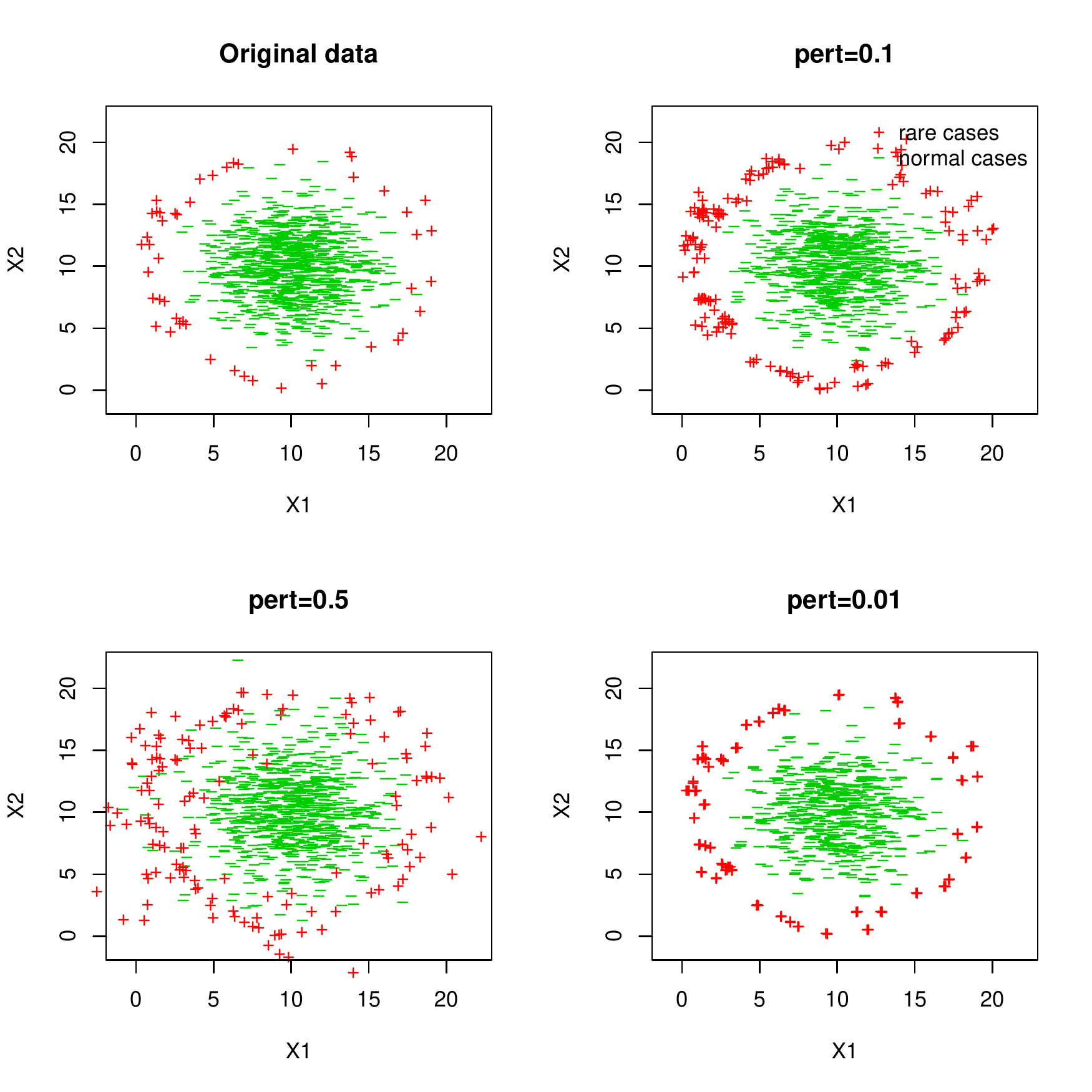} 

}

\caption[Impact of changing the pert parameter in Gaussian Noise strategy in a binarized version of ImbR data]{Impact of changing the pert parameter in Gaussian Noise strategy in a binarized version of ImbR data.}\label{fig:GN_plot3}
\end{figure}

\end{knitrout}

\begin{knitrout}\footnotesize
\definecolor{shadecolor}{rgb}{0.969, 0.969, 0.969}\color{fgcolor}\begin{figure}

{\centering \includegraphics[width=\maxwidth]{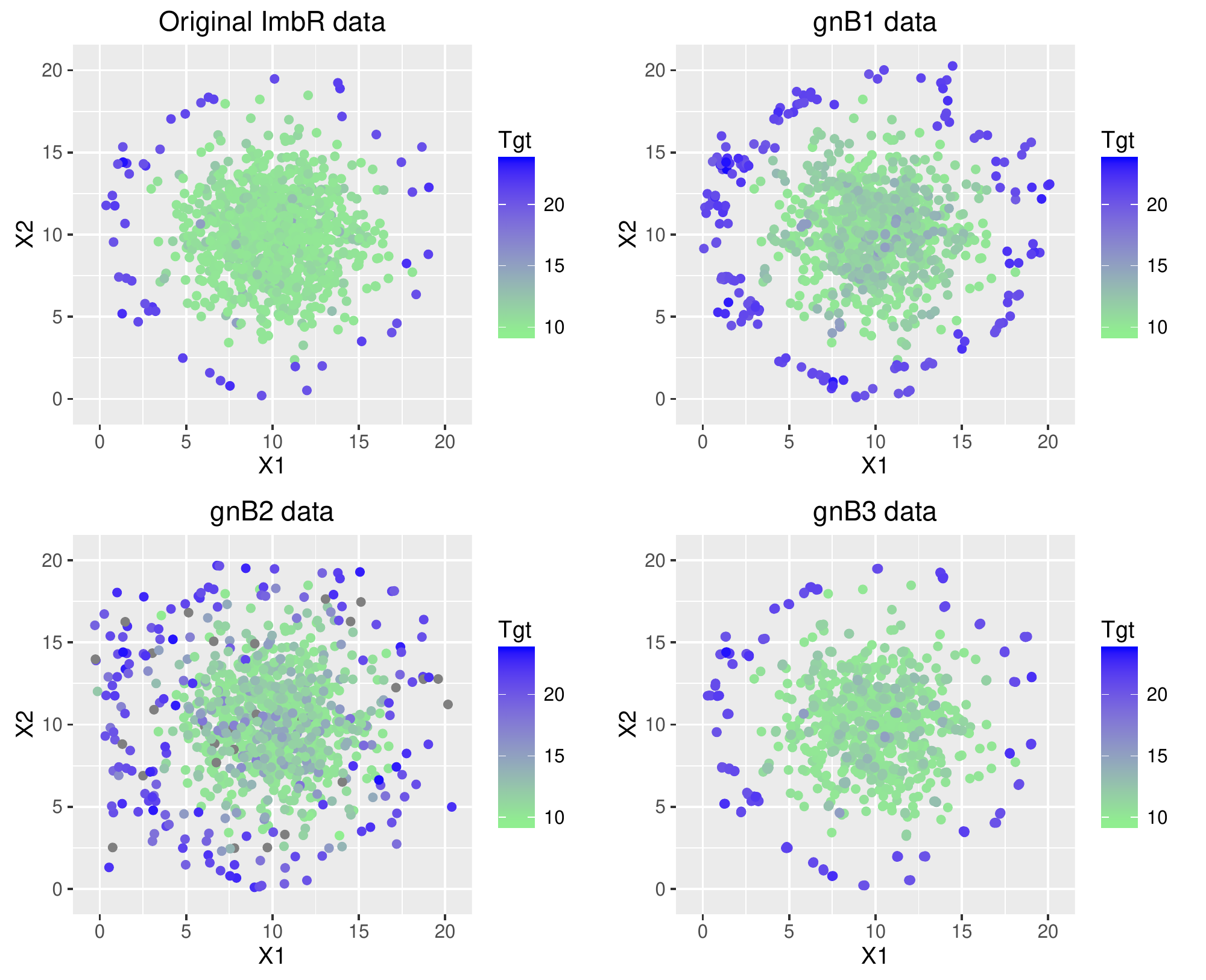} 

}

\caption[Target variable distribution on ImbR and data sets changed through Gaussian Noise strategy]{Target variable distribution on ImbR and data sets changed through Gaussian Noise strategy.}\label{fig:GN_plot3_dist}
\end{figure}

\end{knitrout}

\subsection{The SmoteR Algorithm}\label{sec:smoteR}

The SmoteR algorithm was presented in \cite{torgo2013smote}. This proposal is an adaptation for regression problems under imbalanced domains of the existing smote algorithm \cite{CBOK02} for classification tasks. As with other methods addressing regression tasks on imbalanced data distributions it is the user responsability to provide a relevance function and a relevance threshold. This function determines which are the relevant and the unimportant examples. This algorithm combines an over-sampling strategy by interpolation of examples with a random under-sampling approach. For the generation of new examples by interpolation, the same procedure proposed in smote algorithm is used. Regarding the generation of the target variable value of the new generated examples the proposed smoteR algorithm uses an weighted average of the values of target variable of the two examples used. The weights are calculated as an inverse function of the distance of the generated case to each of the two seed examples. This means that, the further away the new example is from the seed case less weight will be given for the generation of the target variable value. The random under-sampling approach is applied in the bumps containing the normal and unimportant cases. 

The smoteR algorithm is available through the \texttt{SmoteRegress} function. The user may define its own relevance function or use the automatic method, as in the previously described techniques. The user must also define the relevance threshold. 
Regarding the generation of examples it is required to specify the number of nearest neighbors to consider in smoteR algorithm. This is available through the parameter \texttt{k} and the default is set to 5. The user may then use the \texttt{C.perc} parameter to either express the percentages of under and over-sampling to use in each bump of relevance or to set which automatic method should be used for determining these percentages. Similarly to the other approaches, the automatic methods available are ``balance" and ``extreme" which estimate both where to apply the under/over-sampling and the corresponding percentages. The method ``balance" changes the examples distribution by assigning roughly the same number of examples to each bump while the ``extreme" method inverts the frequencies of each bump. Both methods approximately maintain the total number of examples. The parameter \texttt{repl} allows to select if the random under-sampling strategy is applied with repetition of examples or not. The user can also specify which distance function should be used for the nearest neighbors computation using the \texttt{dist} parameter. 

The following examples illustrate how this method can be used.

\begin{knitrout}\footnotesize
\definecolor{shadecolor}{rgb}{0.969, 0.969, 0.969}\color{fgcolor}\begin{kframe}
\begin{alltt}
\hlcom{# we will use the automatic method for defining the relevance function and will}
\hlcom{# set the relevance threshold to 0.8 }
\hlcom{# this method splits the data set in two: a first range of values normal and less}
\hlcom{# important and a second range with the interesting cases}

\hlcom{# to check this, we can plot the relevance function obtained automatically}
\hlcom{# as follows:}

\hlstd{y} \hlkwb{<-} \hlkwd{sort}\hlstd{(ImbR}\hlopt{$}\hlstd{Tgt)}
\hlstd{phiF.args} \hlkwb{<-} \hlkwd{phi.control}\hlstd{(y,} \hlkwc{method} \hlstd{=} \hlstr{"extremes"}\hlstd{,} \hlkwc{extr.type} \hlstd{=} \hlstr{"both"}\hlstd{)}
\hlstd{y.phi} \hlkwb{<-} \hlkwd{phi}\hlstd{(y,} \hlkwc{control.parms} \hlstd{= phiF.args)}

\hlcom{# plot the relevance function}
\hlkwd{plot}\hlstd{(y, y.phi,} \hlkwc{type}\hlstd{=}\hlstr{"l"}\hlstd{,}
     \hlkwc{ylab} \hlstd{=} \hlkwd{expression}\hlstd{(}\hlkwd{phi}\hlstd{(y)),} \hlkwc{xlab} \hlstd{=} \hlkwd{expression}\hlstd{(y))}

\hlcom{#add the relevance threshold to the plot}
\hlkwd{abline}\hlstd{(}\hlkwc{h} \hlstd{=} \hlnum{0.8}\hlstd{,} \hlkwc{col} \hlstd{=} \hlnum{3}\hlstd{,} \hlkwc{lty} \hlstd{=} \hlnum{2}\hlstd{)}
\end{alltt}
\end{kframe}\begin{figure}

{\centering \includegraphics[width=0.8\textwidth]{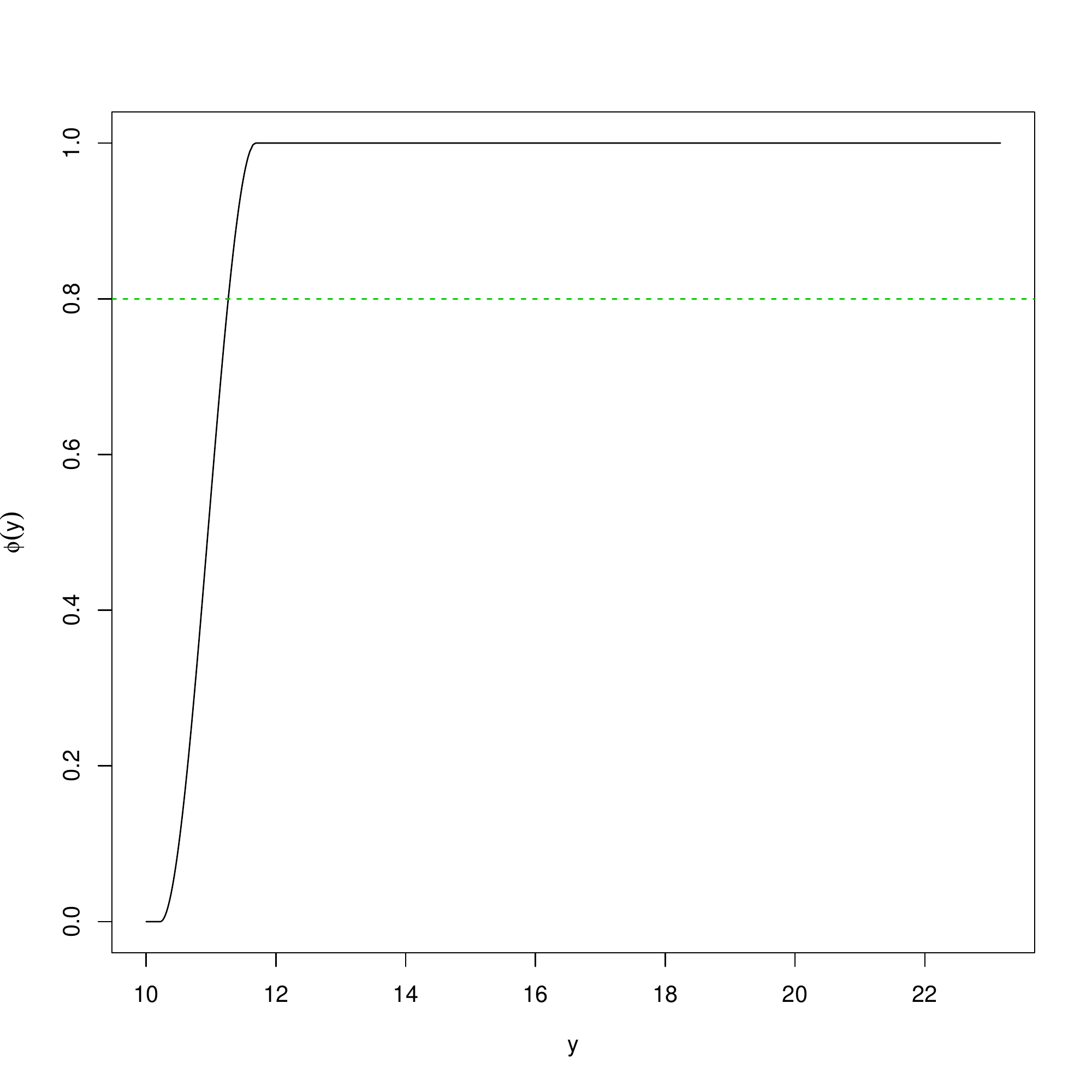} 

}

\caption[Relevance function obtained automatically for the ImbR data set]{Relevance function obtained automatically for the ImbR data set}\label{fig:smoteR_rel1}
\end{figure}

\end{knitrout}

Figure~\ref{fig:smoteR_rel1} shows that we are considering two different bumps: a first bump with the normal and less important cases and a second bump with the rare and interesting cases. Thus, to address this problem we can do the following:

\begin{knitrout}\footnotesize
\definecolor{shadecolor}{rgb}{0.969, 0.969, 0.969}\color{fgcolor}\begin{kframe}
\begin{alltt}
\hlcom{# we have two bumps: the first must be under-sampled and the second over-sampled. }
\hlcom{# Thus, we can chose the following percentages: }
\hlstd{thr.rel} \hlkwb{=} \hlnum{0.8}
\hlstd{C.perc} \hlkwb{=} \hlkwd{list}\hlstd{(}\hlnum{0.1}\hlstd{,} \hlnum{8}\hlstd{)}

\hlcom{# using these percentages and the relevance threshold of 0.8 with all}
\hlcom{# the other parameters default values}
\hlcom{# we can select any distance function }
\hlcom{# because the data set contains only numeric features}
\hlstd{mysm} \hlkwb{<-} \hlkwd{SmoteRegress}\hlstd{(Tgt}\hlopt{~}\hlstd{., ImbR,} \hlkwc{thr.rel}\hlstd{=thr.rel,} \hlkwc{dist}\hlstd{=}\hlstr{"Manhattan"}\hlstd{,} \hlkwc{C.perc}\hlstd{=C.perc)}

\hlcom{# use the automatic method for obtaining a balanced data set}
\hlstd{smB} \hlkwb{<-} \hlkwd{SmoteRegress}\hlstd{(Tgt}\hlopt{~}\hlstd{., ImbR,} \hlkwc{thr.rel}\hlstd{=thr.rel,} \hlkwc{dist}\hlstd{=}\hlstr{"Manhattan"}\hlstd{,} \hlkwc{C.perc}\hlstd{=}\hlstr{"balance"}\hlstd{)}

\hlcom{# use the automatic method for invert the frequencies of the bumps}
\hlstd{smE} \hlkwb{<-} \hlkwd{SmoteRegress}\hlstd{(Tgt}\hlopt{~}\hlstd{., ImbR,} \hlkwc{thr.rel}\hlstd{=thr.rel,} \hlkwc{dist}\hlstd{=}\hlstr{"Manhattan"}\hlstd{,} \hlkwc{C.perc}\hlstd{=}\hlstr{"extreme"}\hlstd{)}
\end{alltt}
\end{kframe}
\end{knitrout}

This strategy changes the examples distribution as shown in Figure~\ref{fig:smoteR_plot1}.

\begin{knitrout}\footnotesize
\definecolor{shadecolor}{rgb}{0.969, 0.969, 0.969}\color{fgcolor}\begin{figure}

{\centering \includegraphics[width=\maxwidth]{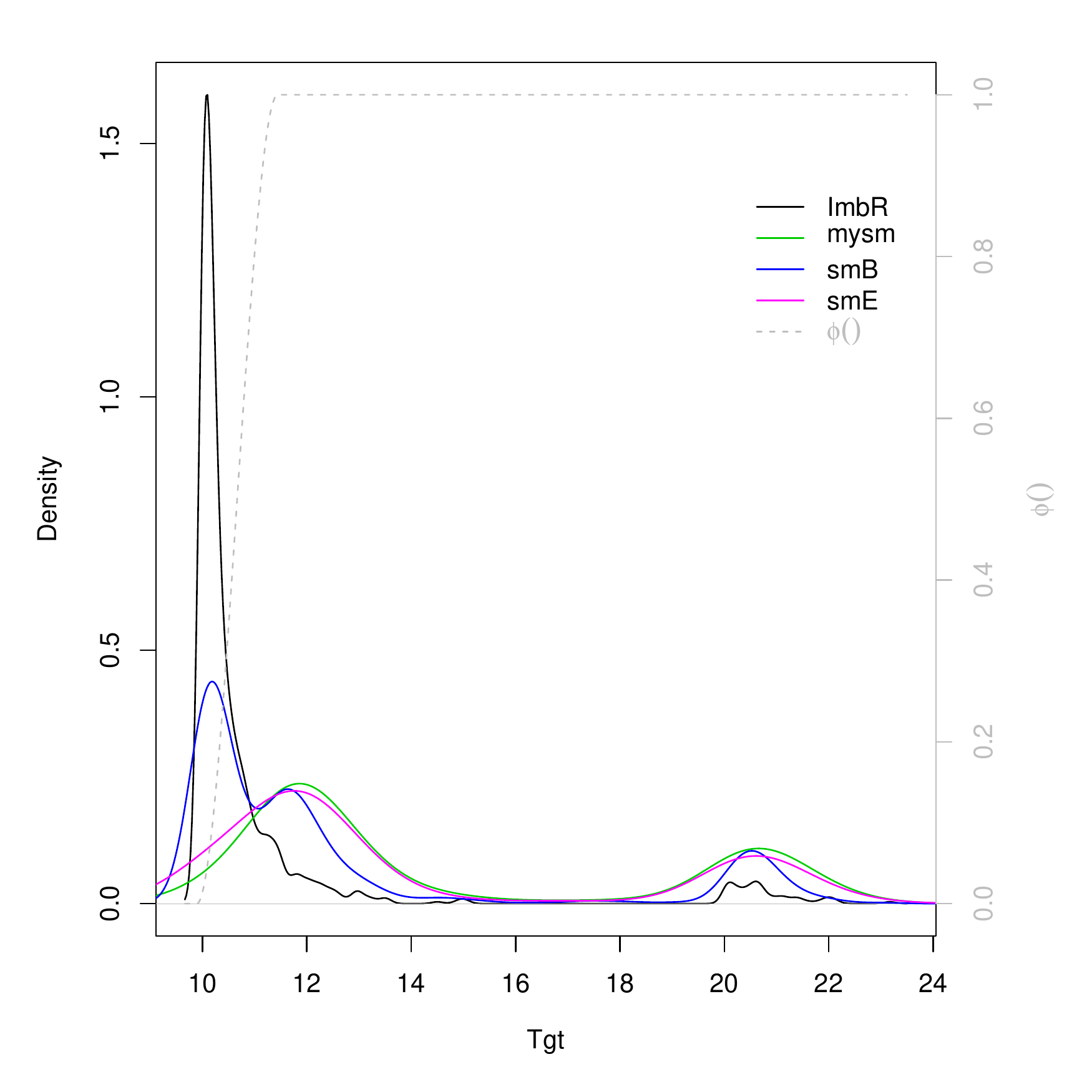} 

}

\caption[Relevance function and density of the target variable in the original and new data sets using smoteR strategy]{Relevance function and density of the target variable in the original and new data sets using smoteR strategy.}\label{fig:smoteR_plot1}
\end{figure}

\end{knitrout}

Figure~\ref{fig:SMOTE_plot3_dist} shows the Original ImbR data and the new data sets pre-processed with SmoteR strategy.

\begin{knitrout}\footnotesize
\definecolor{shadecolor}{rgb}{0.969, 0.969, 0.969}\color{fgcolor}\begin{figure}

{\centering \includegraphics[width=\maxwidth]{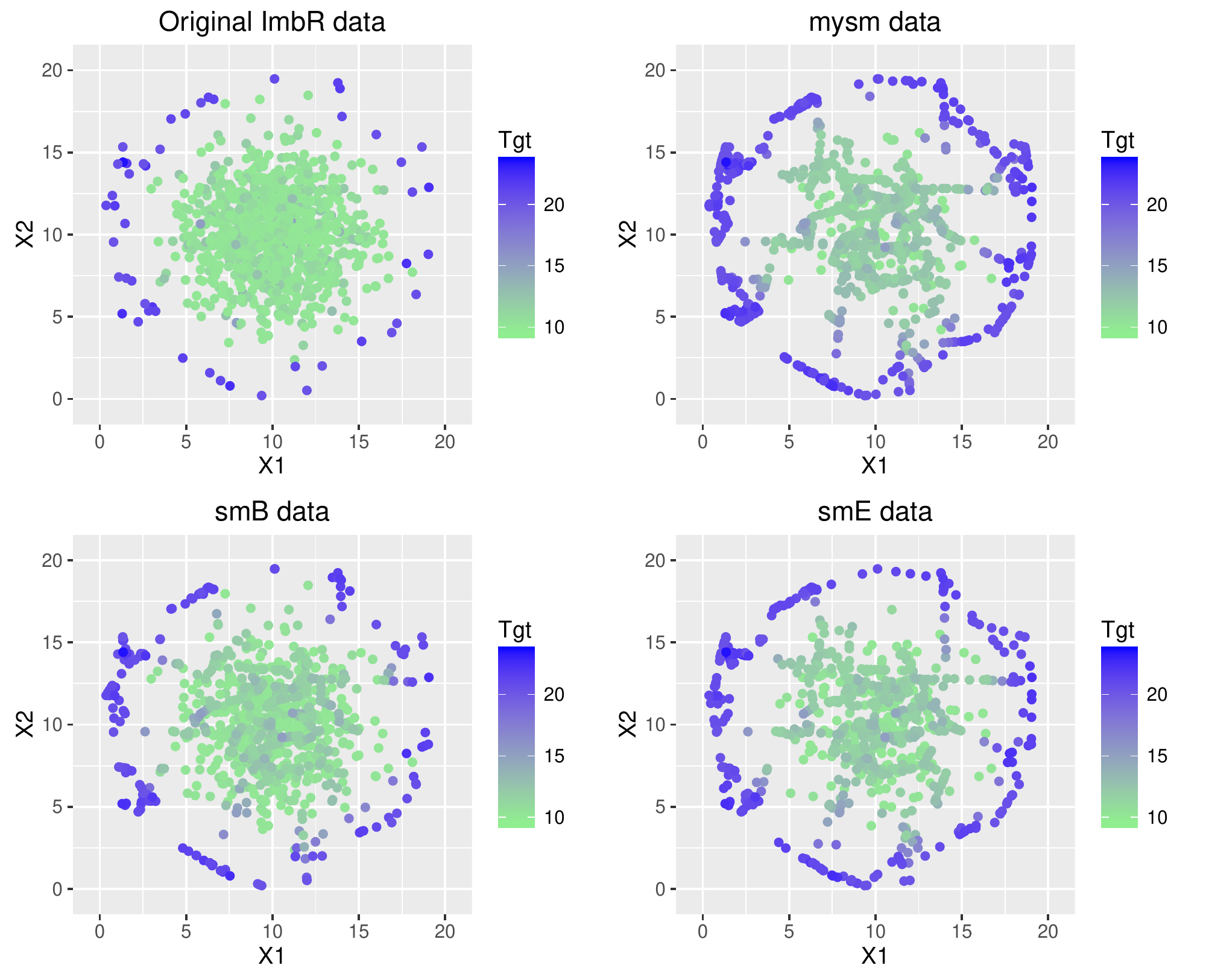} 

}

\caption[Target variable distribution on ImbR and data sets changed through SmoteR strategy]{Target variable distribution on ImbR and data sets changed through SmoteR strategy.}\label{fig:SMOTE_plot3_dist}
\end{figure}

\end{knitrout}

We can also obtain the number of examples that each bump contains. Table~\ref{tab:smoteR_1} shows the examples distribution for the considered strategies.

\begin{table}[ht]
\centering
\begin{tabular}{rrr}
  \hline
 & first bump & second bump \\ 
  \hline
ImbR & 849 & 151 \\ 
  mysm &  84 & 1208 \\ 
  smB & 499 & 500 \\ 
  smE & 169 & 845 \\ 
   \hline
\end{tabular}
\caption{Number of examples in each bump of relevance for different parameters of smoteR strategy.} 
\label{tab:smoteR_1}
\end{table}

In Figure~\ref{fig:smoteR_1bar} we can visualize the impact of these approaches on the examples distribution for each bump of relevance.

\begin{knitrout}\footnotesize
\definecolor{shadecolor}{rgb}{0.969, 0.969, 0.969}\color{fgcolor}\begin{figure}

{\centering \includegraphics[width=\maxwidth,height=0.5\textheight]{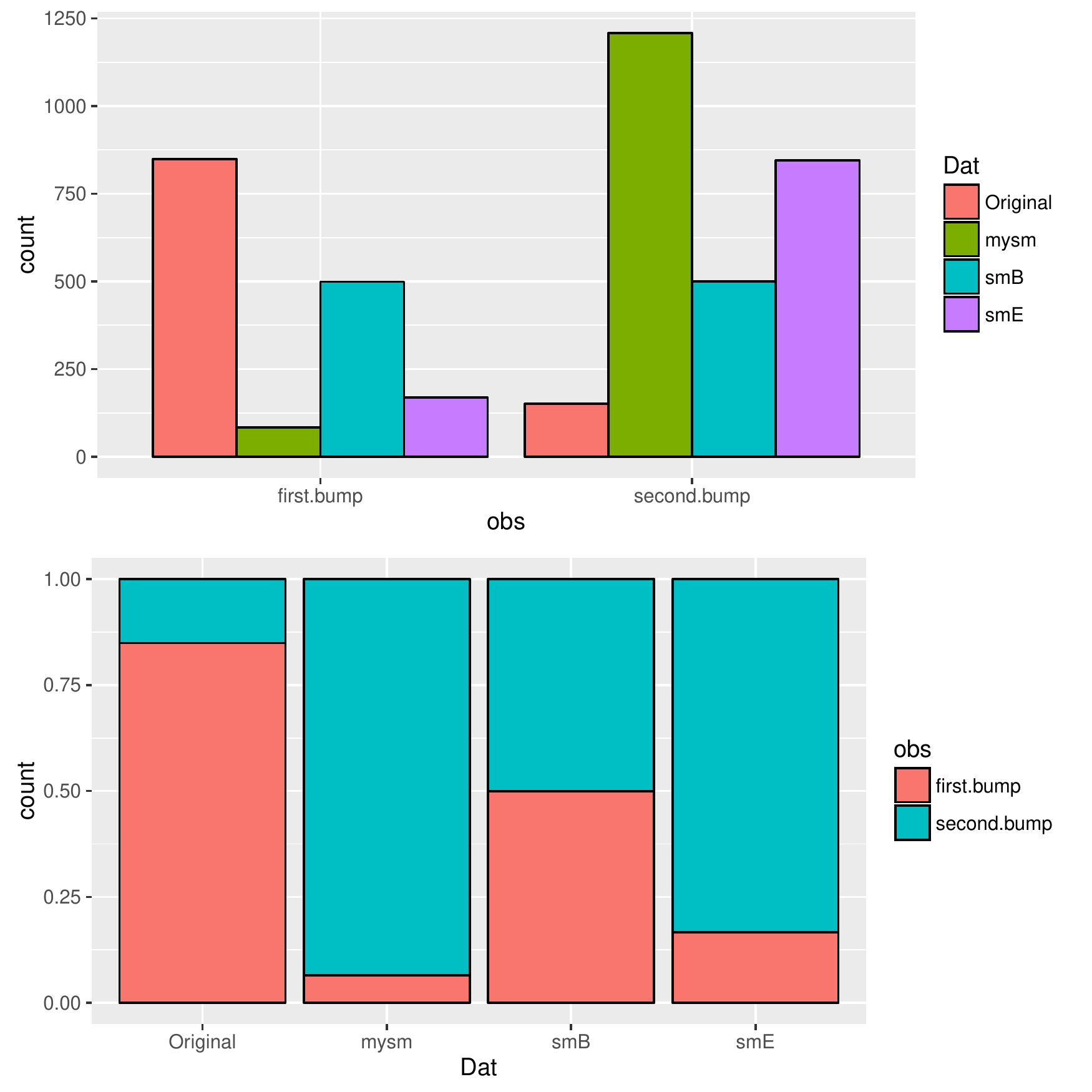} 

}

\caption[Impact in the distribution of examples for several parameters in smoteR strategy]{Impact in the distribution of examples for several parameters in smoteR strategy. }\label{fig:smoteR_1bar}
\end{figure}

\end{knitrout}

\subsection{Importance Sampling}\label{sec:ISRegress}

The Importance Sampling method is a new proposal whose main idea is to use the relevance function ($\phi()$) defined for a regression problem as a probability for resampling the examples combining over with under-sampling. This method simply removes some of the examples and includes in the data set replicas of other existing examples. There is no generation of new synthetic examples. For the over-sampling strategy, replicas of examples are introduced by selecting examples according to the relevance function defined, i.e., the higher the relevance of an example, the higher is the probability of being selected as a new replica to include. The under-sampling strategy selects examples to remove according to the function $1-\phi(y)$, i.e, the higher the relevance value of an example, the lower will be the probability of removing it.

This method includes two main behaviors which can be distinguished by the definition or not of a threshold on the relevance function. This means that, if the user decides to chose a relevance threshold the strategy will take this value into consideration with under and over-sampling being applied only on the defined bumps. However, if the user decides not to set a threshold on the relevance then over sampling and under-sampling strategies will also be applied but without a strict bound, i.e., there may be regions of the target variable values where under-sampling and over-sampling are performed together.

The strategy that depends on the definition of a relevance threshold, has the relevance bumps well defined. For these bumps, the user has several alternatives available through the \texttt{C.perc} parameter: the percentages of over and under-sampling to apply may be explicitly defined, or one of the options ``balance" or ``extreme" may be chosen. These last two option for the \texttt{C.perc} parameter allow to estimate the under and over-sampling percentages automatically. The option ``balance" allows to obtain  a balanced data set across the different existing bumps. The ``extreme" option will produce a new data set with the examples frequencies in the bumps inverted. In this setting, there is no range of the target variable where both under and over-sampling techniques are applied. 

As previously mentioned, there is the possibility of not defining a relevance threshold, and simply use the relevance function to decide which examples should be replicated and which should be removed. In this case, the user does not set a threshold on the relevance, but he can define the importance that over and under-sampling should have. In this case, the \texttt{C.perc} parameter is ignored and two other parameters(\texttt{U} and \texttt{O}) are considered instead. The parameters \texttt{U} and \texttt{O} allow the user to define (in a $[0,1] scale$) the importance that the under/over-sampling have, i.e, these parameters assign a weight to the two methods. The higher is \texttt{O} parameter, the higher is the number of replicas selected. In a similar way, the higher is \texttt{U} parameter the higher is the number of examples removed.

The function \texttt{ImpSampRegress} allows the use of Importance Sampling strategy. Some examples on how to use this approach are provided next.

\begin{knitrout}\footnotesize
\definecolor{shadecolor}{rgb}{0.969, 0.969, 0.969}\color{fgcolor}\begin{kframe}
\begin{alltt}
\hlcom{# relevance function estimated automatically has two bumps}
\hlcom{# using the strategy with threshold definition}
\hlstd{C.perc}\hlkwb{=}\hlkwd{list}\hlstd{(}\hlnum{0.2}\hlstd{,}\hlnum{6}\hlstd{)}
\hlstd{myIS} \hlkwb{<-} \hlkwd{ImpSampRegress}\hlstd{(Tgt}\hlopt{~}\hlstd{., ImbR,} \hlkwc{thr.rel}\hlstd{=}\hlnum{0.8}\hlstd{,}\hlkwc{C.perc}\hlstd{=C.perc)}
\hlstd{ISB} \hlkwb{<-} \hlkwd{ImpSampRegress}\hlstd{(Tgt}\hlopt{~}\hlstd{., ImbR,} \hlkwc{thr.rel}\hlstd{=}\hlnum{0.8}\hlstd{,} \hlkwc{C.perc}\hlstd{=}\hlstr{"balance"}\hlstd{)}
\hlstd{ISE} \hlkwb{<-} \hlkwd{ImpSampRegress}\hlstd{(Tgt}\hlopt{~}\hlstd{., ImbR,} \hlkwc{thr.rel}\hlstd{=}\hlnum{0.8}\hlstd{,} \hlkwc{C.perc}\hlstd{=}\hlstr{"extreme"}\hlstd{)}
\end{alltt}
\end{kframe}
\end{knitrout}

Figures \ref{fig:IS_plot1} and \ref{fig:IS_new_plot3} show the impact on the density and distribution of the examples for the new data sets obtained with Importance Sampling strategy. Figure~\ref{fig:IS_plot2} shows a binarized version of the previous data sets considering the value 19 as the threshold between the rare/important cases and the normal/unimportant cases. 

\begin{knitrout}\footnotesize
\definecolor{shadecolor}{rgb}{0.969, 0.969, 0.969}\color{fgcolor}\begin{figure}

{\centering \includegraphics[width=\maxwidth]{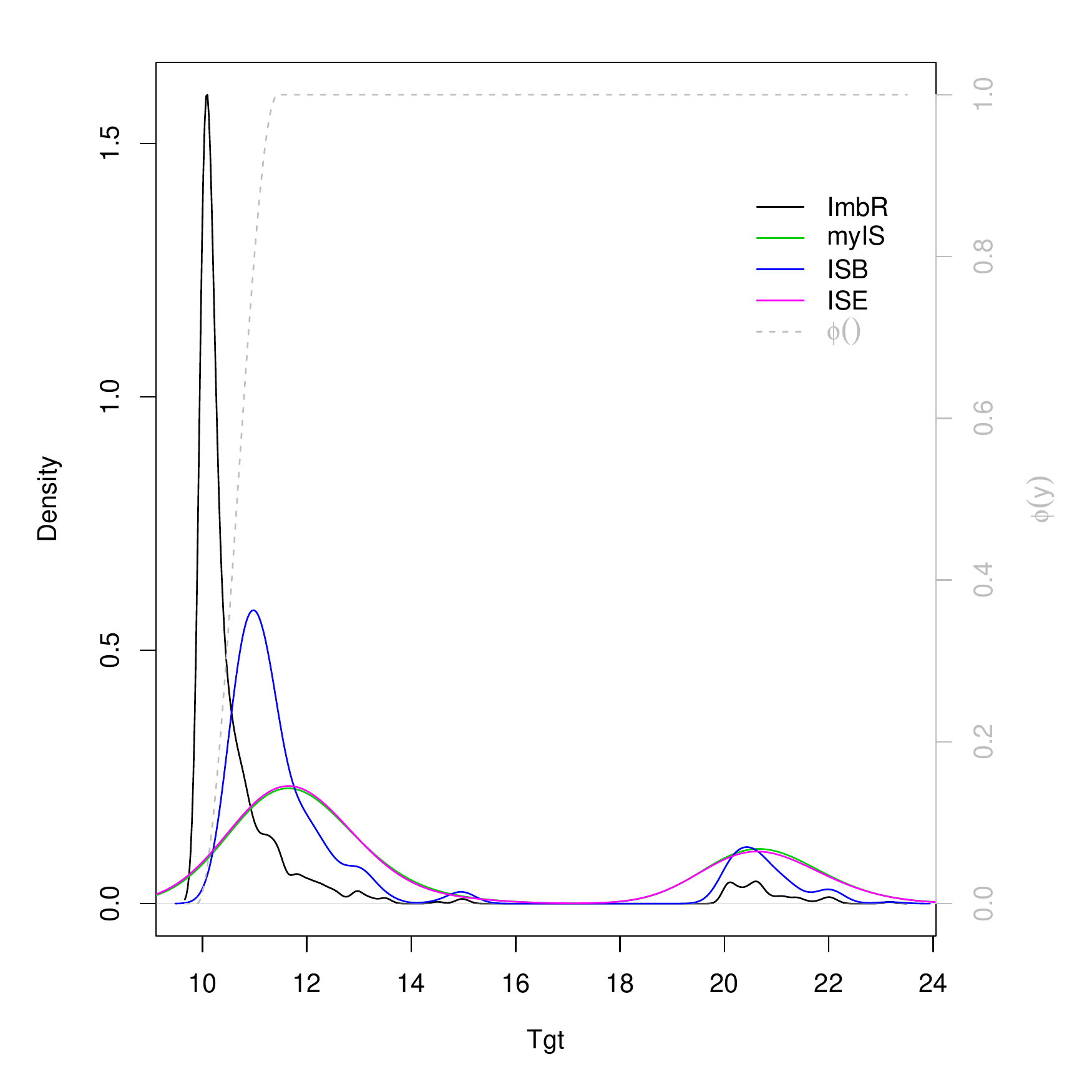} 

}

\caption[Relevance function and density of the target variable in the original and new data sets using Importance Sampling strategy]{Relevance function and density of the target variable in the original and new data sets using Importance Sampling strategy.}\label{fig:IS_plot1}
\end{figure}

\end{knitrout}

\begin{knitrout}\footnotesize
\definecolor{shadecolor}{rgb}{0.969, 0.969, 0.969}\color{fgcolor}\begin{figure}

{\centering \includegraphics[width=\maxwidth]{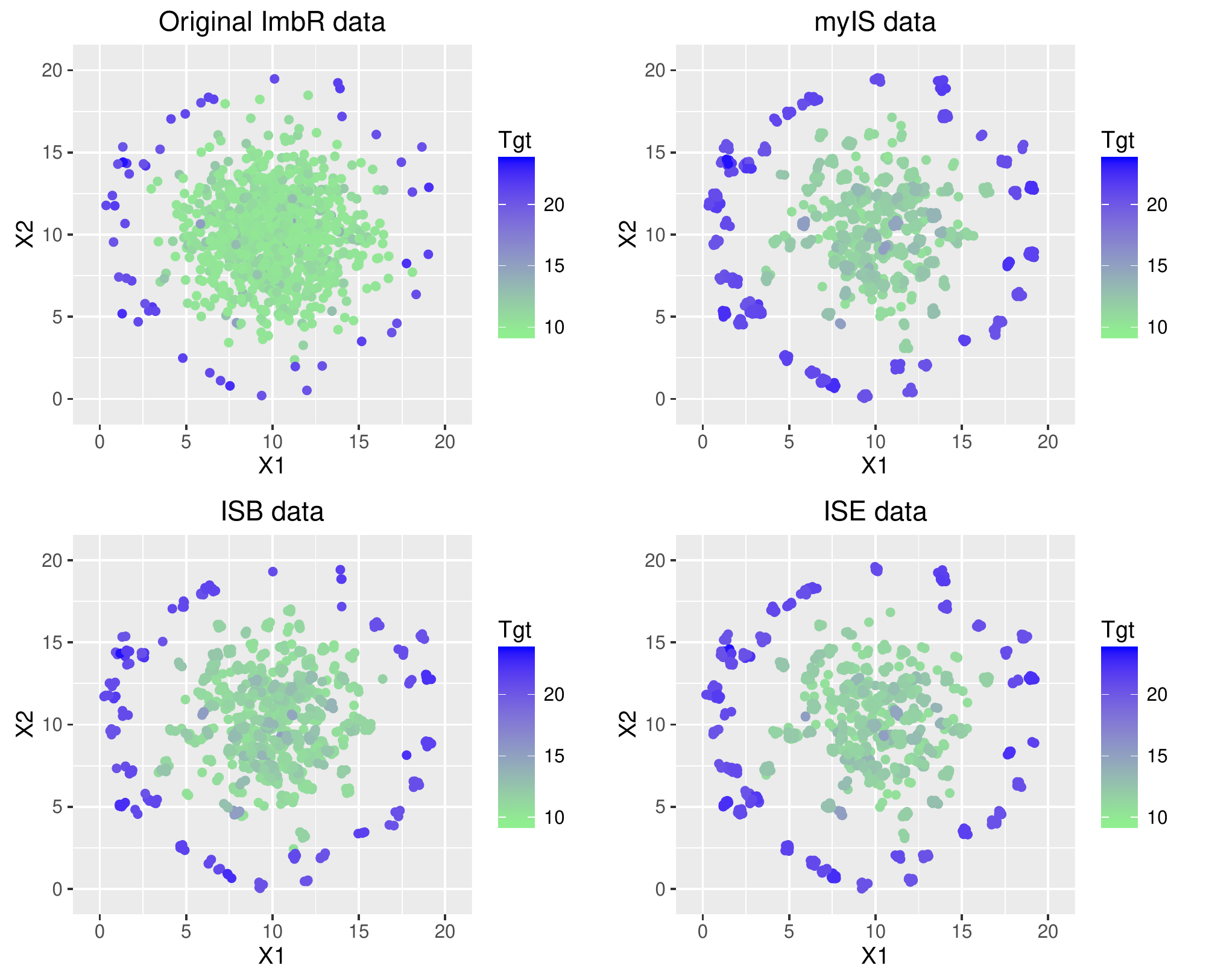} 

}

\caption[Target variable distribution on ImbR and data sets changed through Importance Sampling strategy]{Target variable distribution on ImbR and data sets changed through Importance Sampling strategy.}\label{fig:IS_new_plot3}
\end{figure}

\end{knitrout}

\begin{knitrout}\footnotesize
\definecolor{shadecolor}{rgb}{0.969, 0.969, 0.969}\color{fgcolor}\begin{figure}

{\centering \includegraphics[width=0.8\textwidth]{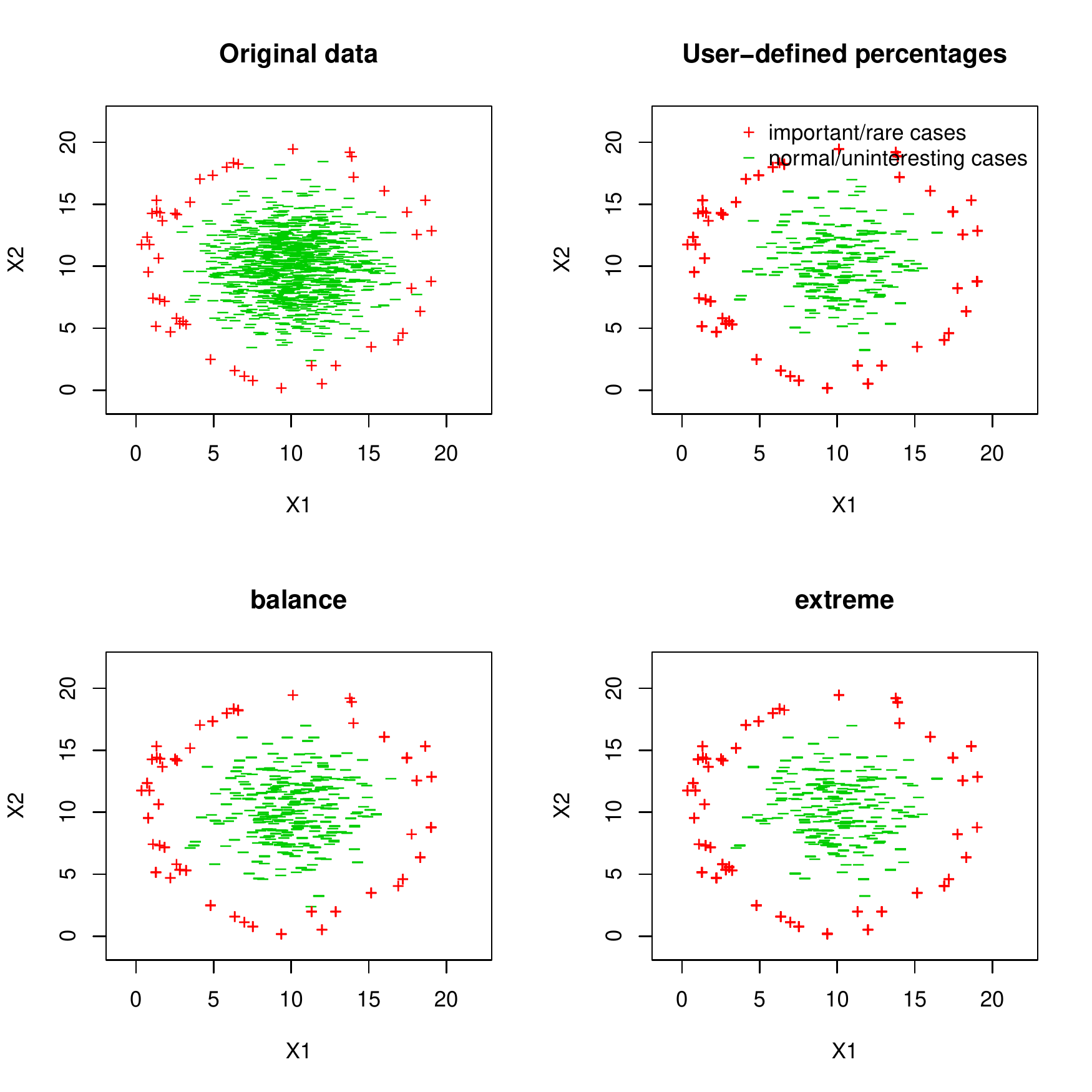} 

}

\caption[Impact of Importance Sampling strategy in a binarized version of the data sets considering the value 19 as the threshold between rare and normal cases]{Impact of Importance Sampling strategy in a binarized version of the data sets considering the value 19 as the threshold between rare and normal cases.}\label{fig:IS_plot2}
\end{figure}

\end{knitrout}

We now provide some examples of the use of this strategy without the definition of a relevance threshold.

\begin{knitrout}\footnotesize
\definecolor{shadecolor}{rgb}{0.969, 0.969, 0.969}\color{fgcolor}\begin{kframe}
\begin{alltt}
\hlcom{# relevance function is also estimated automatically}
\hlcom{# the default is not to use a relevance threshold and to assign equal }
\hlcom{# importance to under and over-sampling, i.e., U=0.5 and O=0.5}
\hlstd{ISD} \hlkwb{<-} \hlkwd{ImpSampRegress}\hlstd{(Tgt}\hlopt{~}\hlstd{., ImbR)}
\hlstd{IS1} \hlkwb{<-} \hlkwd{ImpSampRegress}\hlstd{(Tgt}\hlopt{~}\hlstd{., ImbR,} \hlkwc{U}\hlstd{=}\hlnum{0.9}\hlstd{,} \hlkwc{O}\hlstd{=}\hlnum{0.2}\hlstd{)}
\hlstd{IS2} \hlkwb{<-} \hlkwd{ImpSampRegress}\hlstd{(Tgt}\hlopt{~}\hlstd{., ImbR,} \hlkwc{U}\hlstd{=}\hlnum{0.5}\hlstd{,} \hlkwc{O}\hlstd{=}\hlnum{0.8}\hlstd{)}
\end{alltt}
\end{kframe}
\end{knitrout}

Figures \ref{fig:IS_plot3} and \ref{fig:IS_plot4} show the impact on the density and distribution of the examples for the new data sets obtained with Importance Sampling strategy.
\begin{knitrout}\footnotesize
\definecolor{shadecolor}{rgb}{0.969, 0.969, 0.969}\color{fgcolor}\begin{figure}

{\centering \includegraphics[width=\maxwidth]{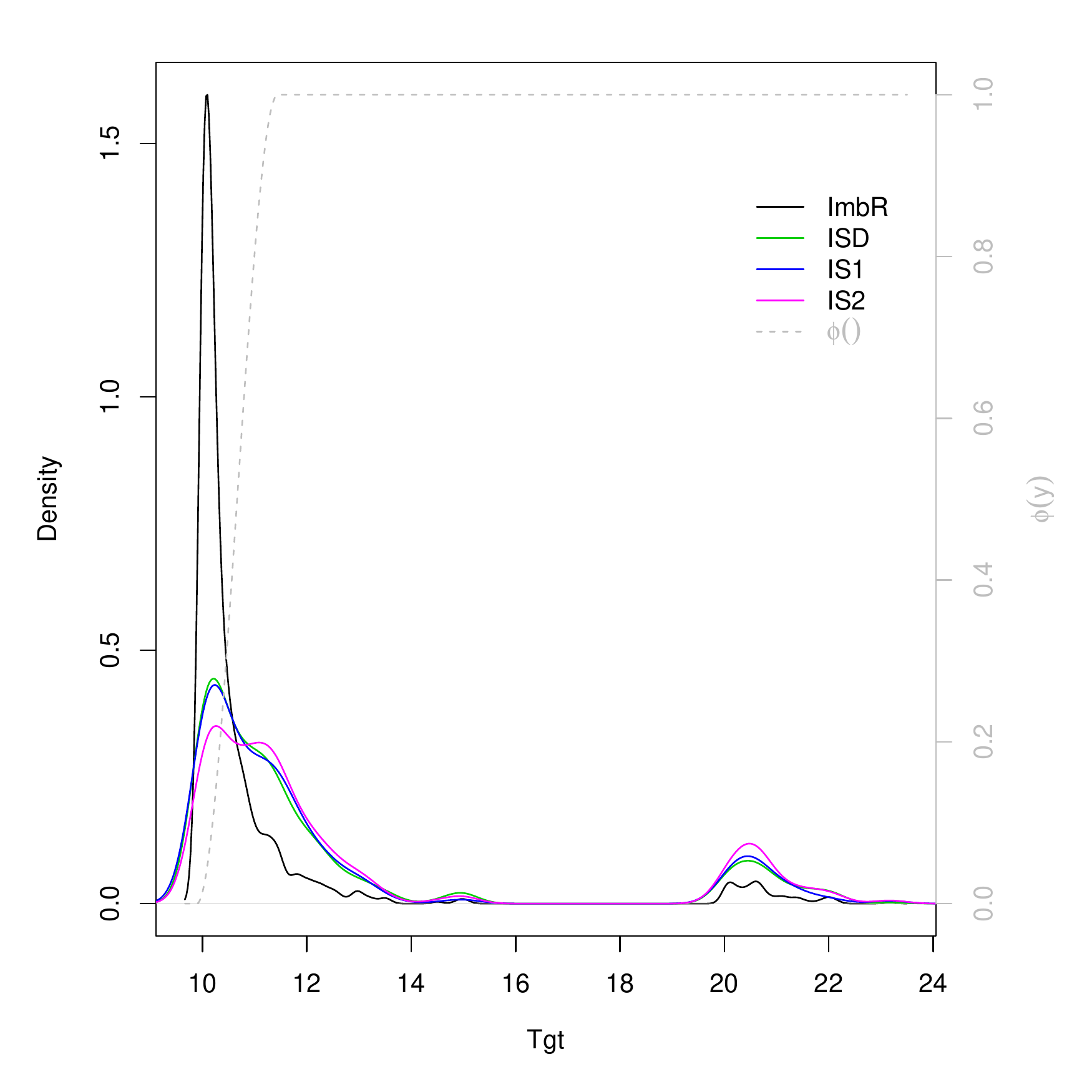} 

}

\caption[Relevance function and density of the target variable in the original and new data sets using Importance Sampling strategy]{Relevance function and density of the target variable in the original and new data sets using Importance Sampling strategy.}\label{fig:IS_plot3}
\end{figure}

\end{knitrout}

\begin{knitrout}\footnotesize
\definecolor{shadecolor}{rgb}{0.969, 0.969, 0.969}\color{fgcolor}\begin{figure}

{\centering \includegraphics[width=0.8\textwidth]{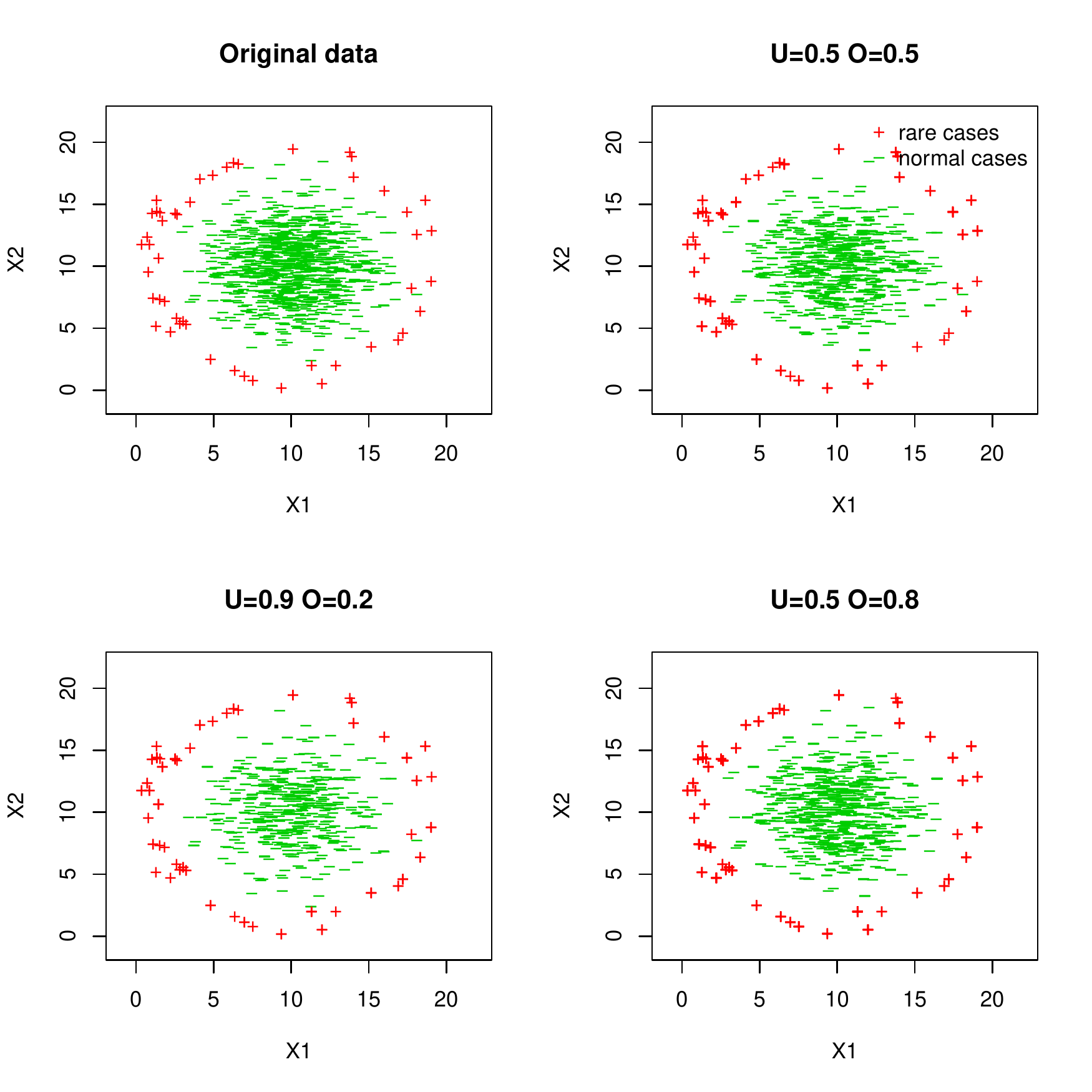} 

}

\caption[Impact of Importance Sampling strategy in a binarized version of the data sets with the value 19 as the threshold between rare and normal cases]{Impact of Importance Sampling strategy in a binarized version of the data sets with the value 19 as the threshold between rare and normal cases.}\label{fig:IS_plot4}
\end{figure}

\end{knitrout}

\begin{knitrout}\footnotesize
\definecolor{shadecolor}{rgb}{0.969, 0.969, 0.969}\color{fgcolor}\begin{figure}

{\centering \includegraphics[width=\maxwidth]{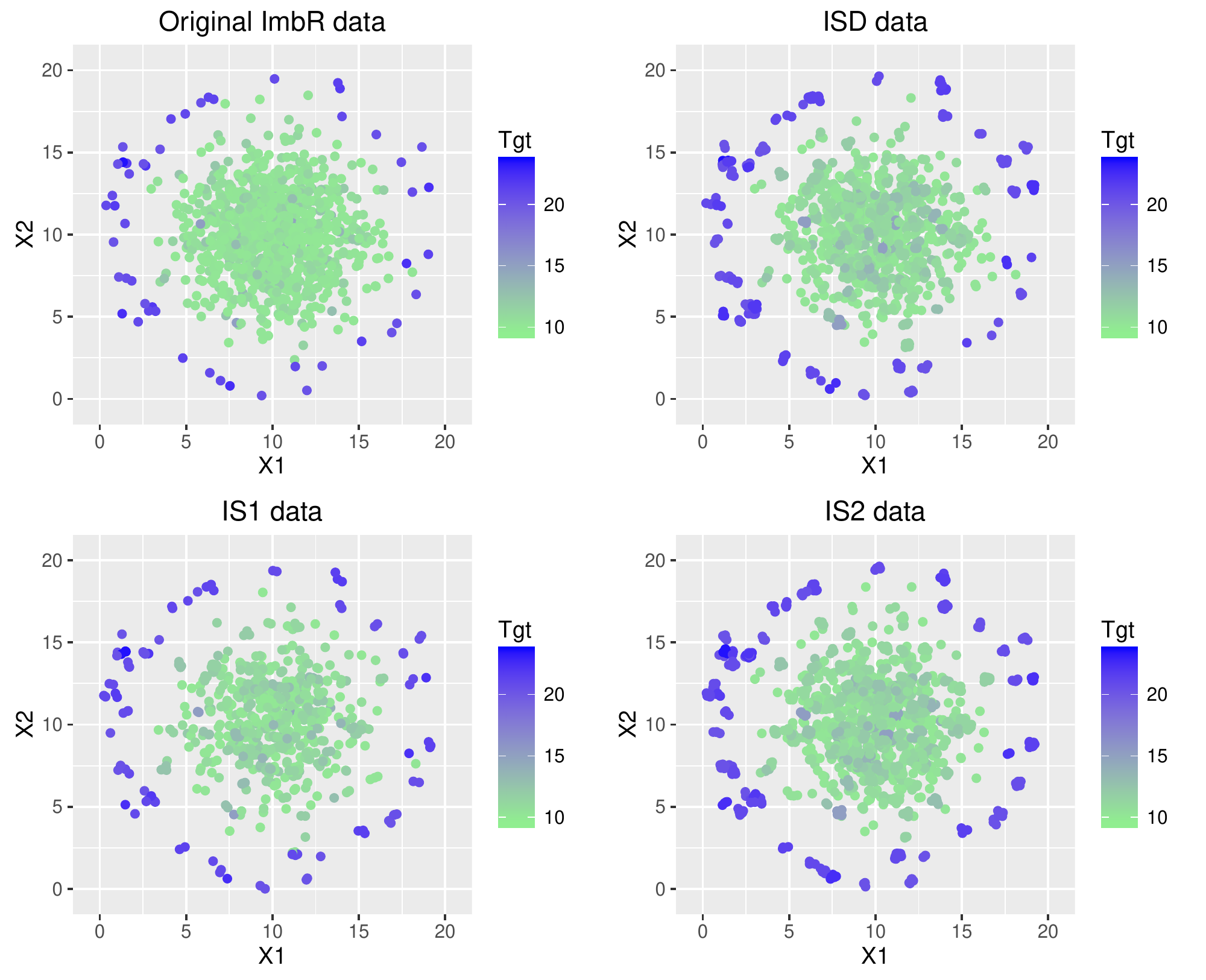} 

}

\caption[Target variable distribution on ImbR and data sets changed through Importance Sampling strategy]{Target variable distribution on ImbR and data sets changed through Importance Sampling strategy.}\label{fig:IS_new_plot4}
\end{figure}

\end{knitrout}

% ====================================================================
\section{Distance Functions}\label{sec:distFunc}

In this section we briefly explain the different distance functions implemented, which can be used for calculating the neighbors of the examples along several strategies for classification or regression tasks.
The implementation of these functions was motivated by the inclusion in \UBL of several methods which depend on the nearest neighbors computation. Although several efficient tools exist for evaluating the nearest neighbors, they are mostly limited to the use of the Euclidean distance. In this context, restricting the user to the use of the Euclidean distance can be a limitation, namely because several data sets include nominal features which can and should also be considered in the neighbors computation. In fact, all the features contained in the data set, whether nominal or numeric, should be taken into account when computing the nearest neighbors. Thus, in order to avoid the restriction of computing nearest neighbors based only on the data set numeric features we have implemented several possible measures which can be used for data sets containing only nominal or numeric features or simultaneously both types. By the implementation of several distance functions, we aim at providing an increased flexibility for computing the nearest neighbors while ensuring that no feature information is wasted.

Several distance measures exist which can deal only with numeric or nominal features or can integrate both types in the distance evaluation. Distance functions such as \texttt{Canberra}, \texttt{Euclidean} or \texttt{Chebyshev} are able to deal solely with numeric attributes while the \texttt{Overlap} measure handles only nominal features. Other measures such as \texttt{HEOM} or \texttt{HVDM} try to use both types of features.

We now briefly describe the distance functions implemented in this package. We begin with the distance functions suitable for data sets with only numeric features. Let us suppose $x$ and $y$ are two examples of a data set with m features. The well-known Euclidean distance can be computed as shown in Equation \ref{eq:Eucl}. The Manhattan distance, also known as city-block distance or taxicab metric, may be calculated with Equation \ref{eq:Manhat}. 

\begin{equation}\label{eq:Eucl}
D(x,y)=\sqrt{ \sum_{i=1}^{m}(x_i-y_i)^2 }
\end{equation}

\begin{equation}\label{eq:Manhat}
D(x,y)=\sum_{i=1}^{m}|x_i-y_i|
\end{equation}
A generalization of these distance functions is obtained with the Minkowsky distance (cf. Equation \ref{eq:Minkowsky}). In this case, by setting $r$ to 1 or 2 we can obtain respectively the Manhattan and Euclidean distance functions.
\begin{equation}\label{eq:Minkowsky}
D(x,y)=\left( \sum_{i=1}^{m}|x_i-y_i|^r\right) ^{\frac{1}{r}}
\end{equation}

The Canberra distance, defined in Equation \ref{eq:Canberra}, and the Chebyshev distance (Equation \ref{eq:Chebychev}) are also functions which can be applied to evaluate the distance between examples described only by numeric features.

\begin{equation}\label{eq:Canberra}
D(x,y)= \sum_{i=1}^{m}\frac{|x_i-y_i|}{|x_i|+|y_i|}
\end{equation}

\begin{equation}\label{eq:Chebychev}
D(x,y)=\max_{i=1}^{m}|x_i-y_i|
\end{equation}

All the previous distance functions can be used in \UBL for computing the nearest neighbors. After selecting an appropriate approach to apply on a data set, it is only necessary to set the parameter \texttt{dist} of the approach to the desired distance function and the \texttt{p} parameter if it is a Minkowsky distance. We illustrate this in the next example.

\begin{knitrout}\footnotesize
\definecolor{shadecolor}{rgb}{0.969, 0.969, 0.969}\color{fgcolor}\begin{kframe}
\begin{alltt}
\hlstd{dat} \hlkwb{<-} \hlstd{iris[}\hlopt{-}\hlkwd{c}\hlstd{(}\hlnum{91}\hlopt{:}\hlnum{125}\hlstd{),]}
\hlcom{# using the default of smote to invert the frequencies of the data set}
\hlkwd{set.seed}\hlstd{(}\hlnum{123}\hlstd{)}
\hlstd{sm.Eu} \hlkwb{<-} \hlkwd{SmoteClassif}\hlstd{(Species}\hlopt{~}\hlstd{., dat,} \hlkwc{dist}\hlstd{=}\hlstr{"Euclidean"}\hlstd{,}
                      \hlkwc{C.perc}\hlstd{=}\hlstr{"extreme"}\hlstd{,} \hlkwc{k}\hlstd{=}\hlnum{3}\hlstd{)}
\hlkwd{set.seed}\hlstd{(}\hlnum{123}\hlstd{)}
\hlstd{sm.Man1} \hlkwb{<-} \hlkwd{SmoteClassif}\hlstd{(Species}\hlopt{~}\hlstd{., dat,} \hlkwc{dist}\hlstd{=}\hlstr{"Manhattan"}\hlstd{,}
                        \hlkwc{C.perc}\hlstd{=}\hlstr{"extreme"}\hlstd{,} \hlkwc{k}\hlstd{=}\hlnum{3}\hlstd{)}
\hlkwd{set.seed}\hlstd{(}\hlnum{123}\hlstd{)}
\hlstd{sm.Man2} \hlkwb{<-} \hlkwd{SmoteClassif}\hlstd{(Species}\hlopt{~}\hlstd{., dat,} \hlkwc{dist}\hlstd{=}\hlstr{"p-norm"}\hlstd{,} \hlkwc{p}\hlstd{=}\hlnum{1}\hlstd{,}
                        \hlkwc{C.perc}\hlstd{=}\hlstr{"extreme"}\hlstd{,} \hlkwc{k}\hlstd{=}\hlnum{3}\hlstd{)}
\hlkwd{set.seed}\hlstd{(}\hlnum{123}\hlstd{)}
\hlstd{sm.5norm} \hlkwb{<-} \hlkwd{SmoteClassif}\hlstd{(Species}\hlopt{~}\hlstd{., dat,} \hlkwc{dist}\hlstd{=}\hlstr{"p-norm"}\hlstd{,} \hlkwc{p}\hlstd{=}\hlnum{5}\hlstd{,}
                         \hlkwc{C.perc}\hlstd{=}\hlstr{"extreme"}\hlstd{,} \hlkwc{k}\hlstd{=}\hlnum{3}\hlstd{)}
\hlkwd{set.seed}\hlstd{(}\hlnum{123}\hlstd{)}
\hlstd{sm.Cheb} \hlkwb{<-} \hlkwd{SmoteClassif}\hlstd{(Species}\hlopt{~}\hlstd{., dat,} \hlkwc{dist}\hlstd{=}\hlstr{"Chebyshev"}\hlstd{,}
                        \hlkwc{C.perc}\hlstd{=}\hlstr{"extreme"}\hlstd{,} \hlkwc{k}\hlstd{=}\hlnum{3}\hlstd{)}
\hlkwd{set.seed}\hlstd{(}\hlnum{123}\hlstd{)}
\hlstd{sm.Canb} \hlkwb{<-} \hlkwd{SmoteClassif}\hlstd{(Species}\hlopt{~}\hlstd{., dat,} \hlkwc{dist}\hlstd{=}\hlstr{"Canberra"}\hlstd{,}
                        \hlkwc{C.perc}\hlstd{=}\hlstr{"extreme"}\hlstd{,} \hlkwc{k}\hlstd{=}\hlnum{3}\hlstd{)}
\end{alltt}
\end{kframe}
\end{knitrout}

The impact of using these distance functions with smote strategy can be visualized in Figure \ref{fig:dist_num}.

\begin{knitrout}\footnotesize
\definecolor{shadecolor}{rgb}{0.969, 0.969, 0.969}\color{fgcolor}\begin{figure}

{\centering \includegraphics[width=\maxwidth]{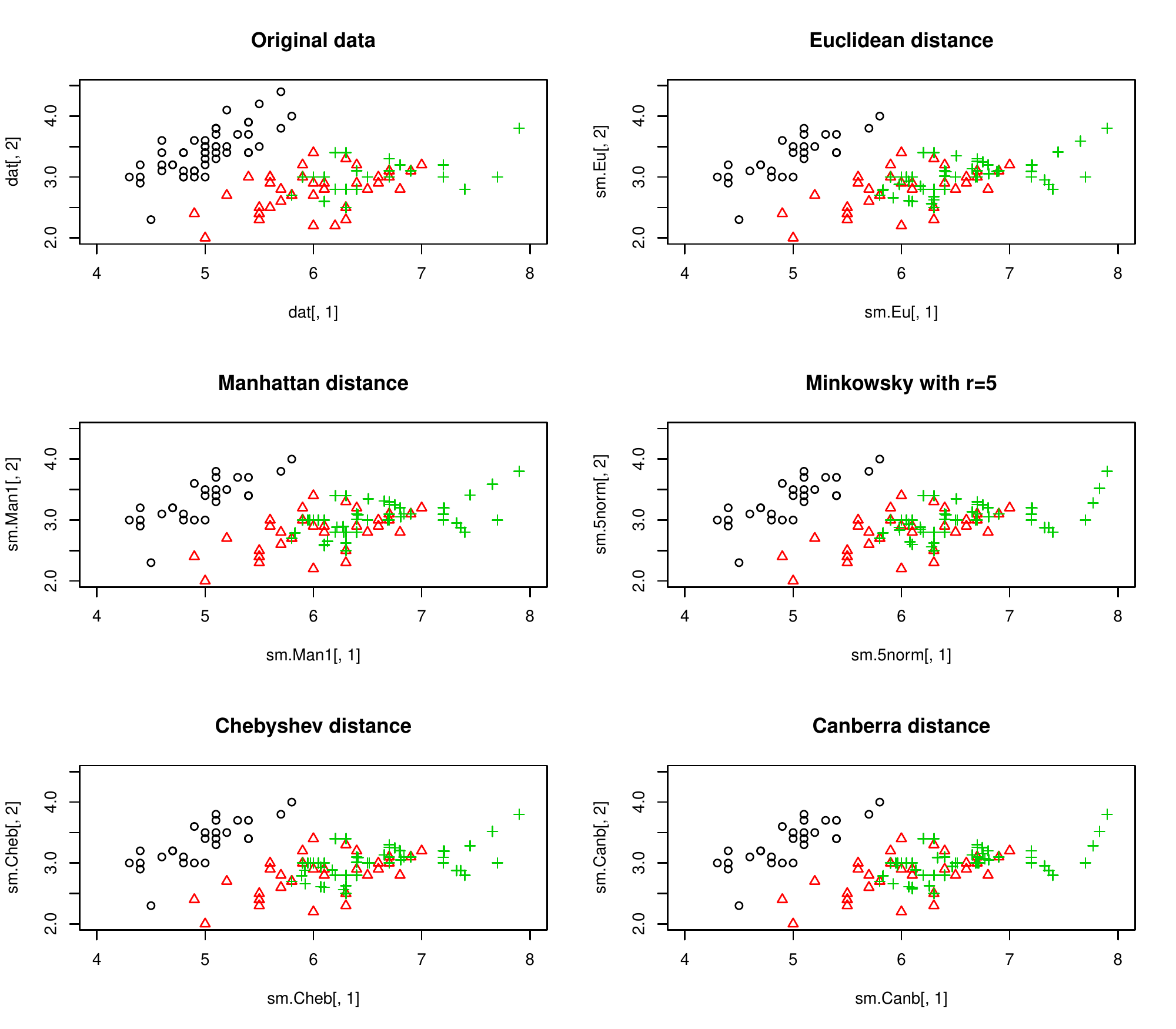} 

}

\caption[Impact of using different distance functions with smote strategy]{Impact of using different distance functions with smote strategy.}\label{fig:dist_num}
\end{figure}

\end{knitrout}

All the previously described metrics do not perform any type of normalization. This step, if wanted, should be performed previously by the user.

Regarding nominal attributes, a distance function which is suitable for handling this type of variables is the overlap measure, which is defined in Equation \ref{eq:overlap}.

\begin{equation}\label{eq:overlap}
overlap(x,y) = \begin{cases} 1 &\mbox{if } x \neq y \\
0 & \mbox{if } x = y. \end{cases} 
\end{equation}

This distance function can be used in strategies that require the computation of nearest neighbors as follows:

\begin{knitrout}\footnotesize
\definecolor{shadecolor}{rgb}{0.969, 0.969, 0.969}\color{fgcolor}\begin{kframe}
\begin{alltt}
\hlcom{# build a data set with all nominal features}
\hlkwd{library}\hlstd{(DMwR)}
\hlkwd{data}\hlstd{(algae)}
\hlstd{clean.algae} \hlkwb{<-} \hlstd{algae[}\hlkwd{complete.cases}\hlstd{(algae),}\hlnum{1}\hlopt{:}\hlnum{3}\hlstd{]}

\hlcom{# speed is considered the target class}
\hlkwd{summary}\hlstd{(clean.algae)}
\end{alltt}
\begin{verbatim}
##     season       size       speed   
##  autumn:36   large :42   high  :76  
##  spring:48   medium:83   low   :31  
##  summer:43   small :59   medium:77  
##  winter:57
\end{verbatim}
\begin{alltt}
\hlstd{ndat1} \hlkwb{<-} \hlkwd{ENNClassif}\hlstd{(speed}\hlopt{~}\hlstd{., clean.algae,} \hlkwc{dist}\hlstd{=}\hlstr{"Overlap"}\hlstd{,}  \hlkwc{Cl}\hlstd{=}\hlkwd{c}\hlstd{(}\hlstr{"high"}\hlstd{,} \hlstr{"medium"}\hlstd{))}
\hlstd{ndat2} \hlkwb{<-} \hlkwd{ENNClassif}\hlstd{(speed}\hlopt{~}\hlstd{., clean.algae,} \hlkwc{dist}\hlstd{=}\hlstr{"Overlap"}\hlstd{,}  \hlkwc{Cl}\hlstd{=}\hlstr{"all"}\hlstd{)}

\hlcom{#all the smaller classes are the most important}
\hlstd{ndat3} \hlkwb{<-} \hlkwd{NCLClassif}\hlstd{(speed}\hlopt{~}\hlstd{., clean.algae,} \hlkwc{dist}\hlstd{=}\hlstr{"Overlap"}\hlstd{,}  \hlkwc{Cl}\hlstd{=}\hlstr{"smaller"}\hlstd{)}
\hlcom{# the most important classes are "high" and "low"}
\hlstd{ndat4} \hlkwb{<-} \hlkwd{NCLClassif}\hlstd{(speed}\hlopt{~}\hlstd{., clean.algae,} \hlkwc{dist}\hlstd{=}\hlstr{"Overlap"}\hlstd{,}  \hlkwc{Cl}\hlstd{=}\hlkwd{c}\hlstd{(}\hlstr{"high"}\hlstd{,} \hlstr{"low"}\hlstd{))}

\hlstd{ndat5} \hlkwb{<-} \hlkwd{SmoteClassif}\hlstd{(speed}\hlopt{~}\hlstd{., clean.algae,} \hlkwc{dist}\hlstd{=}\hlstr{"Overlap"}\hlstd{,} \hlkwc{C.perc}\hlstd{=}\hlstr{"balance"}\hlstd{)}
\end{alltt}
\end{kframe}
\end{knitrout}

Figure \ref{fig:dist_overlap} shows the impact of using the overlap distance function, with several different strategies, on a data set consisting of only nominal variables.
\begin{knitrout}\footnotesize
\definecolor{shadecolor}{rgb}{0.969, 0.969, 0.969}\color{fgcolor}\begin{figure}

{\centering \includegraphics[width=\maxwidth]{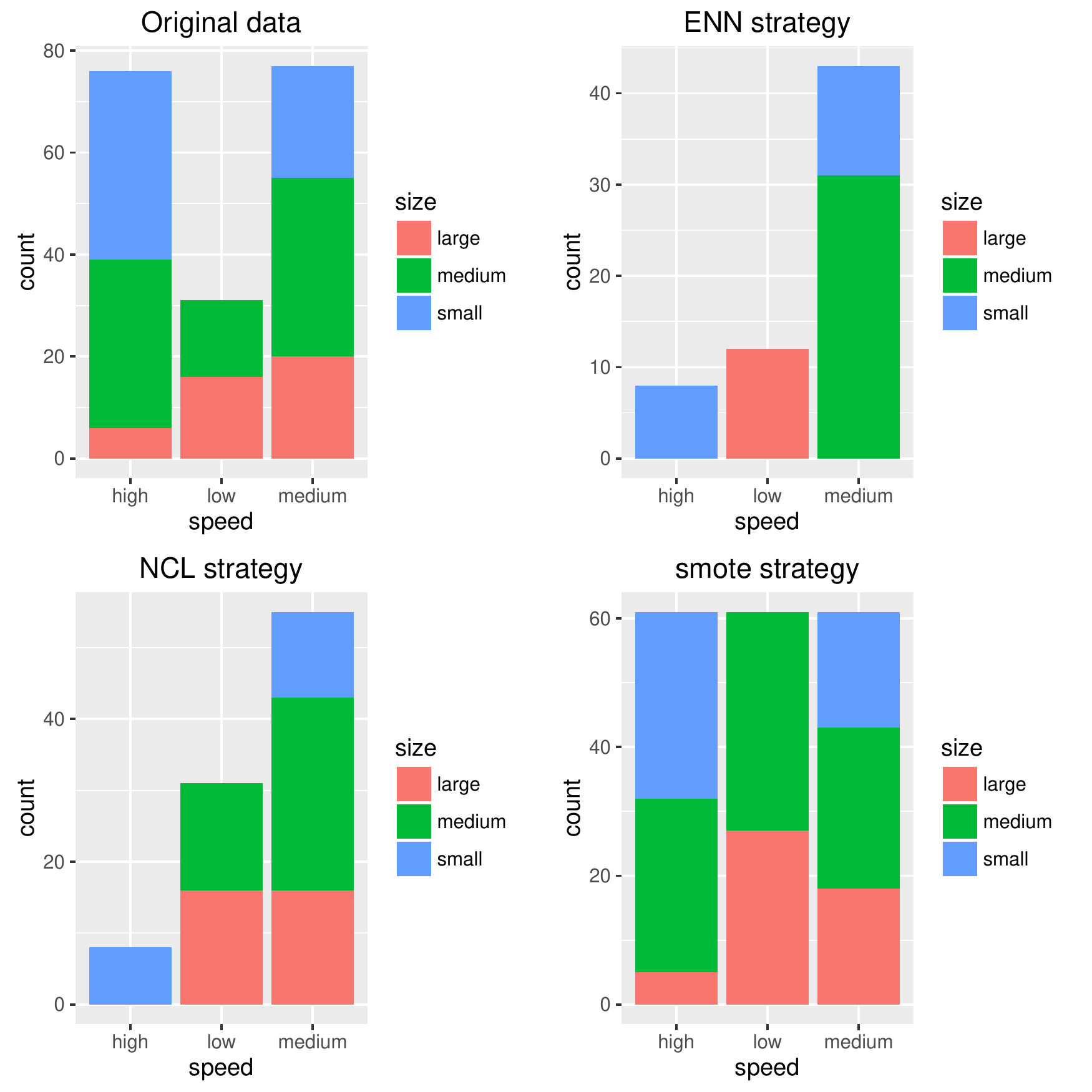} 

}

\caption[Using Overlap distance function with different strategies on a data set with only nominal features]{Using Overlap distance function with different strategies on a data set with only nominal features.}\label{fig:dist_overlap}
\end{figure}

\end{knitrout}

To evaluate the distance between examples described by nominal and numeric variables a simple adaptation of the previous distance functions can be performed. The Heterogeneous Euclidean-Overlap Metric function (HEOM) is a popular solution for these situations. Equations \ref{eq:HEOM} and \ref{eq:auxHEOM} describe how this distance is computed.

\begin{equation}\label{eq:HEOM}
HEOM(x,y)= \sqrt{\sum_{a=1}^{m}d_a^2(x_a,y_a) }
\end{equation}

\begin{equation}\label{eq:auxHEOM}
\mbox{where  } d_a(x,y)= \begin{cases} 1 & \mbox{if } x \vee y \mbox{ are unknown, else} \\
overlap(x,y) & \mbox{if } a \mbox{ is nominal, else} \\
\frac{|x-y|}{range_a}
\end{cases}
\end{equation}
\noindent where $range_a=max_a-min_a$

\begin{knitrout}\footnotesize
\definecolor{shadecolor}{rgb}{0.969, 0.969, 0.969}\color{fgcolor}\begin{kframe}
\begin{alltt}
\hlcom{# build a data set with nominal and numeric features}
\hlkwd{library}\hlstd{(DMwR)}
\hlkwd{data}\hlstd{(algae)}
\hlstd{clean.algae} \hlkwb{<-} \hlstd{algae[}\hlkwd{complete.cases}\hlstd{(algae),}\hlnum{1}\hlopt{:}\hlnum{5}\hlstd{]}

\hlcom{# speed is the target class}
\hlkwd{summary}\hlstd{(clean.algae)}
\end{alltt}
\begin{verbatim}
##     season       size       speed         mxPH            mnO2       
##  autumn:36   large :42   high  :76   Min.   :7.000   Min.   : 1.500  
##  spring:48   medium:83   low   :31   1st Qu.:7.777   1st Qu.: 7.675  
##  summer:43   small :59   medium:77   Median :8.100   Median : 9.750  
##  winter:57                           Mean   :8.078   Mean   : 9.019  
##                                      3rd Qu.:8.400   3rd Qu.:10.700  
##                                      Max.   :9.500   Max.   :13.400
\end{verbatim}
\begin{alltt}
\hlstd{enn} \hlkwb{<-} \hlkwd{ENNClassif}\hlstd{(speed}\hlopt{~}\hlstd{., clean.algae,} \hlkwc{dist}\hlstd{=}\hlstr{"HEOM"}\hlstd{,}  \hlkwc{Cl}\hlstd{=}\hlstr{"all"}\hlstd{,} \hlkwc{k}\hlstd{=}\hlnum{5}\hlstd{)[[}\hlnum{1}\hlstd{]]}
\hlcom{#consider all the smaller classes as the most important}
\hlstd{ncl} \hlkwb{<-} \hlkwd{NCLClassif}\hlstd{(speed}\hlopt{~}\hlstd{., clean.algae,} \hlkwc{dist}\hlstd{=}\hlstr{"HEOM"}\hlstd{,}  \hlkwc{Cl}\hlstd{=}\hlstr{"smaller"}\hlstd{)}
\hlstd{sm} \hlkwb{<-} \hlkwd{SmoteClassif}\hlstd{(speed}\hlopt{~}\hlstd{., clean.algae,} \hlkwc{dist}\hlstd{=}\hlstr{"HEOM"}\hlstd{,} \hlkwc{C.perc}\hlstd{=}\hlstr{"balance"}\hlstd{)}
\end{alltt}
\end{kframe}
\end{knitrout}

In Figure \ref{fig:dist_heom} we can observe the impact of using the HEOM distance function with several strategies.
\begin{knitrout}\footnotesize
\definecolor{shadecolor}{rgb}{0.969, 0.969, 0.969}\color{fgcolor}\begin{figure}

{\centering \includegraphics[width=\maxwidth]{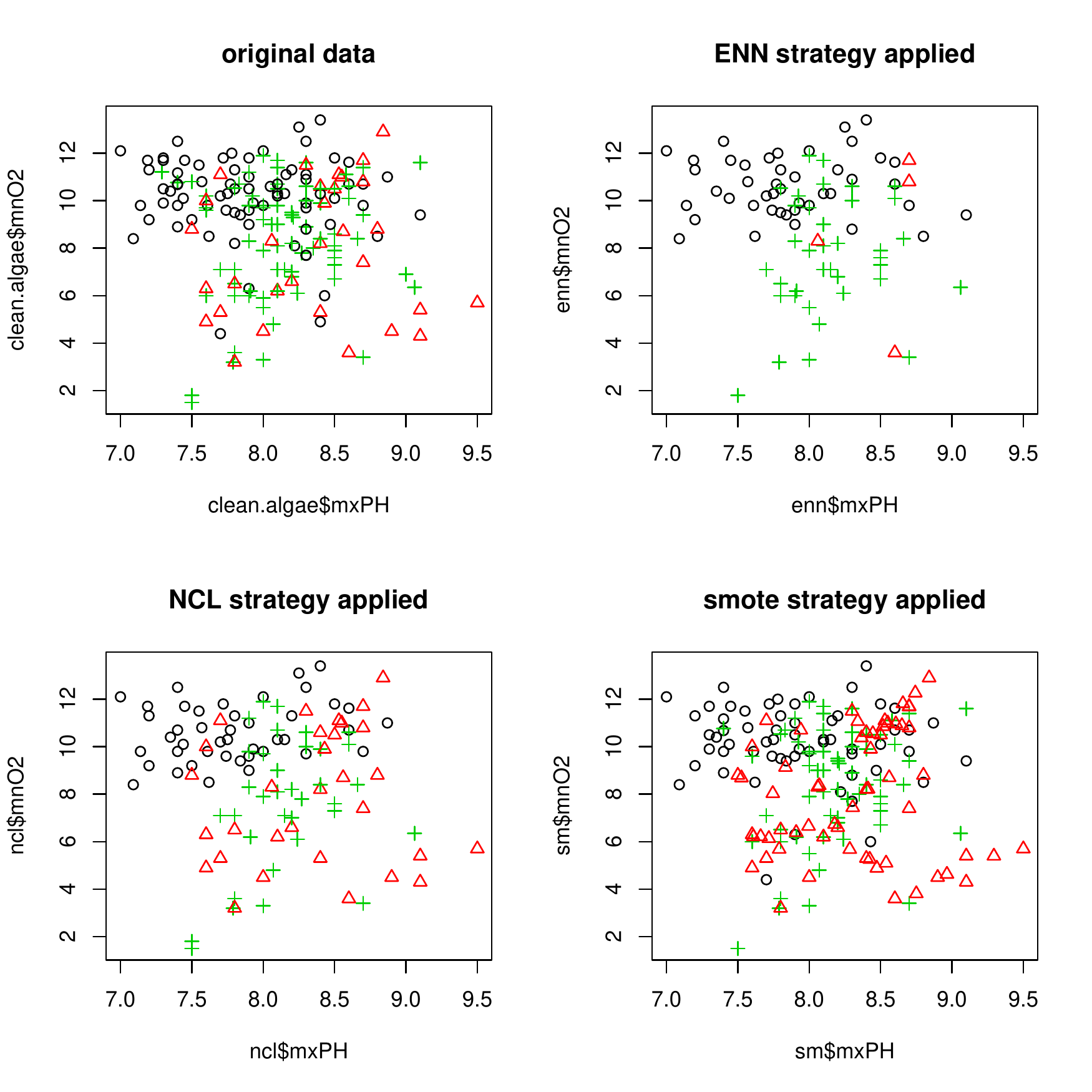} 

}

\caption[Using HEOM distance function with different strategies on a data set with both nominal and numeric features]{Using HEOM distance function with different strategies on a data set with both nominal and numeric features.}\label{fig:dist_heom}
\end{figure}

\end{knitrout}

Other proposals, such as the Heterogeneous Value Difference Metric (HVDM), were tested for handling both nominal and numeric features. The HVDM uses the notion of Value Distance Metric (VDM) which was introduced by \cite{stanfill1986toward} to address the distance computation with nominal variables. The VDM metric is described in Equation \ref{eq:VDM}.

\begin{equation}\label{eq:VDM}
VDM_a(x,y)= \sum_{c=1}^{C}  \left\lvert \frac{N_{a,x,c}}{N_{a,x}}-\frac{N_{a,y,c}}{N_{a,y}}\right\rvert ^q
\end{equation}
where, \\
$a$ is the nominal attribute under consideration;\\
$C$ is the number of classes existing on the data set;\\
$q$ is a constant;\\
$N_{a,x,c}$ represents the number of examples which have value $x$ for the feature $a$ and class label $c$;\\
$N_{a,x}$ is the number of examples that have value $x$ for the feature $a$.\\

The HVDM distance function was proposed by \cite{wilson1997improved} and its definition, presented in Equations \ref{eq:HVDM}and \ref{eq:auxHVDM}, is similar to the HEOM.

\begin{equation}\label{eq:HVDM}
HVDM(x,y)= \sqrt{\sum_{a=1}^{m}d_a^2(x_a,y_a) }
\end{equation}

\begin{equation}\label{eq:auxHVDM}
\mbox{where  } d_a(x,y)= \begin{cases} 1 & \mbox{if } x \vee y \mbox{ are unknown, otherwise} \\
norm-vdm_a(x,y) & \mbox{if } a \mbox{ is nominal} \\
norm-diff_a(x,y) & \mbox{if } a \mbox{ is numeric} \\
\end{cases}
\end{equation}

The HVDM distance function uses a normalized version of the absolute value of the difference between two examples for the numeric attributes (Equation \ref{eq:HVDMnum}) and uses for the nominal attributes an also normalized version of the VDM measure for the nominal attributes (Equation \ref{eq:HVDMnom}) .

\begin{equation}\label{eq:HVDMnom}
norm-vdm_a(x,y)=\sqrt{VDM_a(x,y)}=\sqrt{\sum_{c=1}^{C}  \left\lvert \frac{N_{a,x,c}}{N_{a,x}}-\frac{N_{a,y,c}}{N_{a,y}}\right\rvert ^2}
\end{equation}

\begin{equation}\label{eq:HVDMnum}
norm-diff_a(x,y)= \frac{|x-y|}{4\sigma_a}
\end{equation}

Regarding Equation \ref{eq:HVDMnom}, several normalization of the VDM measure were proposed and tested in \cite{wilson1997improved}. The version presented here and implemented in \UBL was the one that achieved the best performance. We also highlight that the distance function proposed for the numeric attributes uses a different normalization which relies on the standard deviation of of each attribute $\sigma_a$.

The HVDM distance can be used simply by setting the \texttt{dist} parameter to ``HVDM". Although it is a function suitable for both nominal and numeric, if the data set provided contains only one type of attributes only the corresponding distance will be used.

\begin{knitrout}\footnotesize
\definecolor{shadecolor}{rgb}{0.969, 0.969, 0.969}\color{fgcolor}\begin{kframe}
\begin{alltt}
\hlcom{# build a data set with both nominal and numeric features}
\hlkwd{library}\hlstd{(DMwR)}
\hlkwd{data}\hlstd{(algae)}
\hlstd{clean.algae} \hlkwb{<-} \hlstd{algae[}\hlkwd{complete.cases}\hlstd{(algae),}\hlkwd{c}\hlstd{(}\hlnum{1}\hlopt{:}\hlnum{6}\hlstd{)]}

\hlcom{# speed is considered the target class}
\hlkwd{summary}\hlstd{(clean.algae)}
\end{alltt}
\begin{verbatim}
##     season       size       speed         mxPH            mnO2              Cl        
##  autumn:36   large :42   high  :76   Min.   :7.000   Min.   : 1.500   Min.   :  0.80  
##  spring:48   medium:83   low   :31   1st Qu.:7.777   1st Qu.: 7.675   1st Qu.: 11.85  
##  summer:43   small :59   medium:77   Median :8.100   Median : 9.750   Median : 35.08  
##  winter:57                           Mean   :8.078   Mean   : 9.019   Mean   : 44.88  
##                                      3rd Qu.:8.400   3rd Qu.:10.700   3rd Qu.: 58.52  
##                                      Max.   :9.500   Max.   :13.400   Max.   :391.50
\end{verbatim}
\begin{alltt}
\hlstd{dat1} \hlkwb{<-} \hlkwd{SmoteClassif}\hlstd{(speed}\hlopt{~}\hlstd{., clean.algae,} \hlkwc{dist}\hlstd{=}\hlstr{"HVDM"}\hlstd{,} \hlkwc{C.perc}\hlstd{=}\hlstr{"extreme"}\hlstd{)}

\hlstd{dat2} \hlkwb{<-} \hlkwd{NCLClassif}\hlstd{(speed}\hlopt{~}\hlstd{., clean.algae,} \hlkwc{k}\hlstd{=}\hlnum{3}\hlstd{,} \hlkwc{dist}\hlstd{=}\hlstr{"HVDM"}\hlstd{,} \hlkwc{Cl}\hlstd{=}\hlstr{"smaller"}\hlstd{)}

\hlstd{dat3} \hlkwb{<-} \hlkwd{TomekClassif}\hlstd{(speed}\hlopt{~}\hlstd{., clean.algae,} \hlkwc{dist}\hlstd{=}\hlstr{"HVDM"}\hlstd{,} \hlkwc{Cl}\hlstd{=}\hlstr{"all"}\hlstd{,} \hlkwc{rem}\hlstd{=}\hlstr{"both"}\hlstd{)}
\end{alltt}
\end{kframe}
\end{knitrout}

Figure \ref{fig:dist_HVDM} shows the result of applying HVDM distance function for several different strategies, on a data set consisting of numeric and nominal features.
\begin{knitrout}\footnotesize
\definecolor{shadecolor}{rgb}{0.969, 0.969, 0.969}\color{fgcolor}\begin{figure}

{\centering \includegraphics[width=\maxwidth]{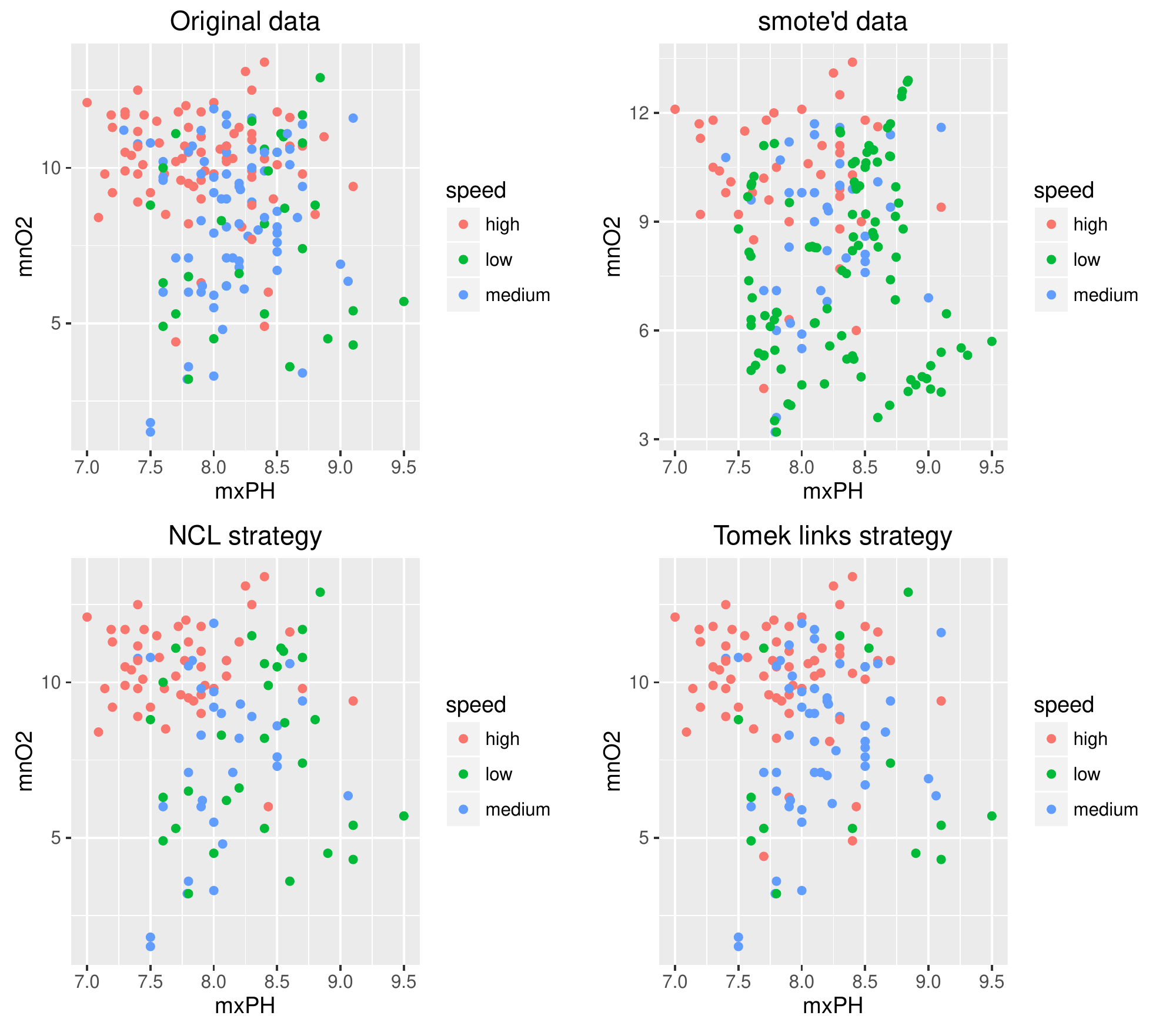} 

}

\caption[Using HVDM distance function with different strategies]{Using HVDM distance function with different strategies.}\label{fig:dist_HVDM}
\end{figure}

\end{knitrout}

In Figure \ref{fig:dist_HVDM2} the impact of smote strategy applied with different distance functions on a data set can be observed.

\begin{knitrout}\footnotesize
\definecolor{shadecolor}{rgb}{0.969, 0.969, 0.969}\color{fgcolor}\begin{figure}

{\centering \includegraphics[width=\maxwidth]{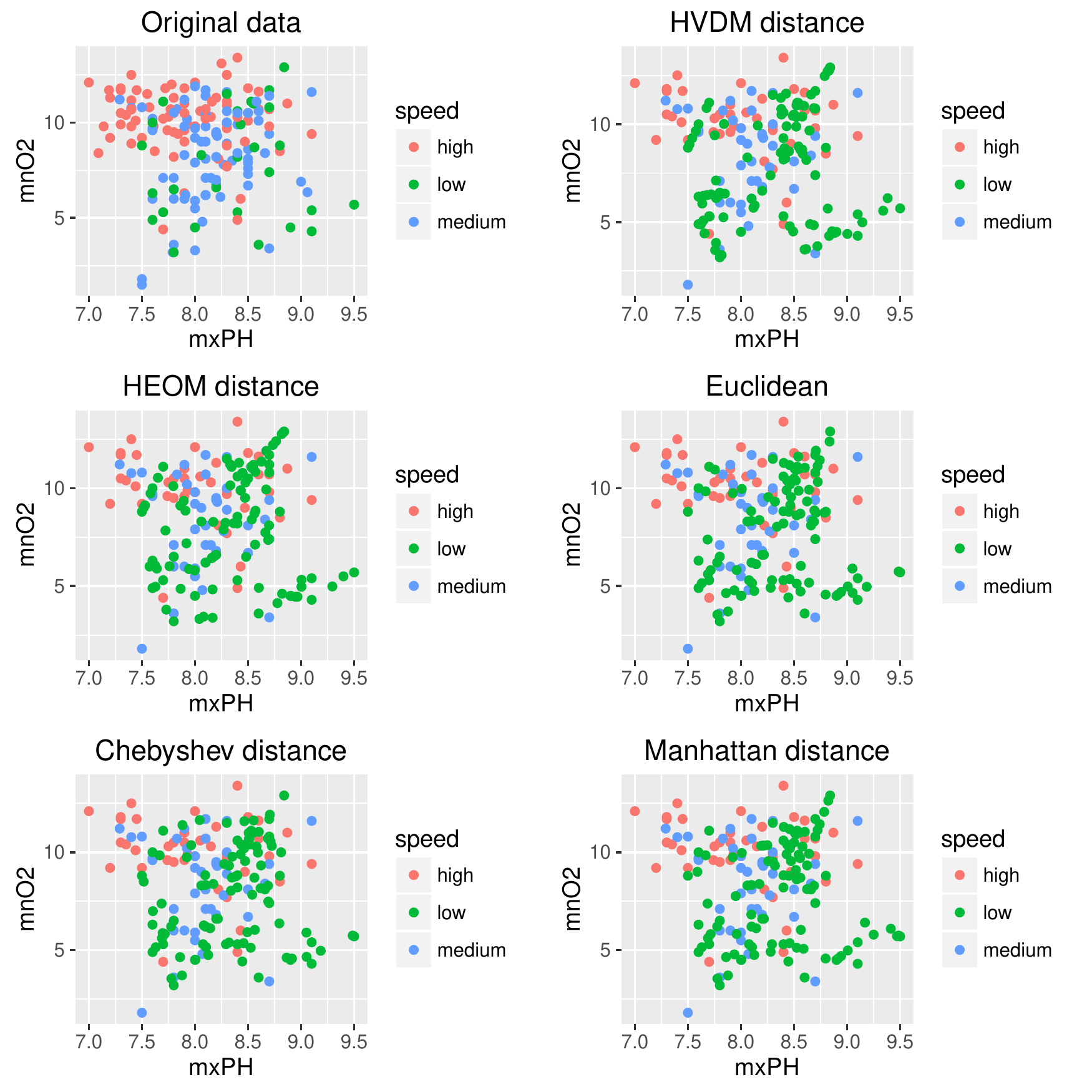} 

}

\caption[Using different distance functions with  smote strategy]{Using different distance functions with  smote strategy.}\label{fig:dist_HVDM2}
\end{figure}

\end{knitrout}

% =======================================================
\section{Conclusions}\label{sec:conc}

We have presented \pUBL that aims at dealing with utility-based predictive tasks. This package offers several methods for multiclass and regression problems. The approaches implemented are pre-processing methods for changing the target variable distribution. This change in the data set is performed with the goal of incorporating the user preference bias. The use of pre-processing methods that change the original data set force the learning algorithms to focus on the most relevant cases for the user.

The existing strategies for dealing with utility-based problems as a pre-processing step present the advantage of allowing the use of any standard learning algorithm without changing it. Moreover, these methods do not compromise the interpretability of the models used. As possible disadvantages we must point the difficulty of determining the ideal distribution of the domain. In fact, a perfectly balanced distribution of examples is not always the solution that provides the best results.

\UBLp is a versatile tool for tackling problems at a pre-processing level that have some information regarding the domain. This package extends some methods previously developed for binary classification to a multiclass setting and also allows to deal with regression problems with multiple important regions. It offers to the user the possibility of manually defining how the data set should be changed for a selected pre-processing approach. Moreover, for each implemented approach it also enables the use of automatic methods for estimating the changes to apply. These automatic methods use the original domain disribution for assigning more importance to the least represented examples, a commom setting when learning from imbalanced domains.

\newpage

\bibliographystyle{alpha}
\bibliography{UBL}
\end{document}